\documentclass[%
 reprint,
 amsmath,amssymb,
 aps,
]{revtex4-2}
\usepackage{caption}
\usepackage{ragged2e}
\usepackage{xcolor}
\usepackage{graphicx}
\usepackage{dcolumn}
\usepackage{bm}

\begin{document}

\preprint{APS/123-QED}

\title{Exciton photoemission from a ground state of a solid Ta$_2$Pd$_3$Te$_5$}

\author{Siwon Lee$^{1,2}$, Kyung-Hwan Jin$^{2,3}$, SeongJin Kwon$^{1,2}$, Hyunjin Jung$^{1,2}$, Choongjae Won$^{4,5}$, Sang-Wook Cheong$^{5,6}$, Gil Young Cho$^{1,2}$, Jaeyoung Kim$^{2}$, Han Woong Yeom$^{1,2}$}%
\email[Email Address: ]{yeom@postech.ac.kr}
\affiliation{1 Department of Physics, Pohang University of Science and Technology (POSTECH), Pohang 37673, Republic of Korea\\
2 Center for Artificial Low Dimensional Electronic Systems, Institute for Basic Science (IBS), Pohang 37673, Republic of Korea\\
3 Department of Physics and Research Institute for Materials and Energy Sciences, Jeonbuk National University, Jeonju, 54896 South Korea\\
4 Max Planck POSTECH/Korea Research Initiative, Center for Complex Phase of Materials, Pohang, 37673, Republic of Korea\\
5 Laboratory of Pohang Emergent Materials, Pohang Accelerator Laboratory, Pohang, 37673, Republic of Korea\\
6 Rutgers Center for emergent Materials and Department of Physics and Astronomy, Rutgers University, NJ, USA}

\date{\today}

\begin{abstract}
Excitons are bosonic quasiparticles with a variety of applications in optoelectronics, photosynthesis, and dissipationless informatics, and their lifetime can become sufficiently long to form a quantum condensate. While exciton condensation has been predicted to occur as a ground state of a solid, so called an excitonic insulator, whose material realization has been elusive. 
Here we report the observation of direct photoemission signals from excitons in a ground state of a very recent excitonic insulator candidate Ta$_2$Pd$_3$Te$_5$ below its metal-insulator transition temperature using orbital-selective angle-resolved photoemission spectroscopy. It is confirmed that the excitons have a lower energy than the valence band maximum to possibly drive the phase transition.
This measurement further discloses the size and the unusual odd parity of the exciton wave function. The present finding opens an avenue toward applications of coherent excitons in solid systems and searching for exotic quantum phases of exciton condensates.
\end{abstract}

\maketitle



Excitons represent bound states of electrons and holes formed by Coulomb attraction, which play important roles in diverse photoexcitation processes from semiconductor optical properties and photovoltaics to photosynthesis \cite{van2000photosynthetic,brixner2005two,kuznetsova2010all,menke2013tailored,romero2014quantum,ross2014electrically,furchi2014photovoltaic,ye2015monolayer,romero2017quantum,mueller2018exciton}.
An exciton can also be spontaneously formed as a ground state of a solid when the exciton binding energy exceeds the band gap; a single-electron band gap is replaced by a manybody excitonic gap below a transition temperature  \textit{T}$_c$           
 and excitons form a quantum condensate.
This was predicted to occur in a semimetal and a small-gap semiconductor with poorly-screened Coulomb interaction to minimize the band gap with maximized Coulomb attraction \cite{jerome1967excitonic,halperin1968possible, wakisaka2009excitonic,pillo2000photoemission, seki2014excitonic, sugawara2016unconventional, kogar2017signatures, song2023signatures, gao2023evidence, gao2024observation,kwan2021theory,jia2022evidence,sun2022evidence,que2024gate}.
The evidence of the exciton condensation, a transition into an excitonic insulator, was recently reported in monolayers of transition metal dichalcogenides \cite{gao2023evidence,gao2024observation,kwan2021theory,jia2022evidence,sun2022evidence,que2024gate}.
However, the confirmation of an excitonic insulator remains elusive in a 3D crystalline form of solids in spite of substantial recent efforts devoted particularly to transition metal chalcogenide materials such as TiSe$_2$ \cite{rossnagel2002charge,porer2014non,wegner2020evidence,jeong2024dichotomy} and Ta$_2$NiSe$_5$ \cite{watson2020band,windgatter2021common,lu2021evolution,kaneko2013orthorhombic,mazza2020nature}. 
The major challenge has been the inevitable coupling of electronic and lattice degrees of freedom, which would induce substantial structural changes in a putative excitonic gap opening. 
The major driving force of the transition thus becomes ambiguous with both excitonic and structural contributions entangled \cite{rossnagel2002charge,porer2014non,wegner2020evidence,watson2020band,windgatter2021common,lu2021evolution,kaneko2013orthorhombic,mazza2020nature,jeong2024dichotomy}. 

On the other hand, recent works demonstrated that excitons manifest themselves as a characteristic photoemission signal \cite{scholes2006excitons,cui2014transient,byrnes2014exciton,tanimura2019dynamics,madeo2020directly,man2021experimental,karni2022structure,mori2023spin,pareek2024driving}. 
When it absorbs an incoming photon of sufficient energy, an exciton emits a photoelectron leaving a hole within the solid (Fig. 1a). 
The exciton photoelectron spectral function is given as a convolution of the exciton wave function and that of conduction band electrons \cite{rustagi2018photoemission} as well established in pump-probe photoemission experiments for excited state excitons within band gaps of semiconductors \cite{scholes2006excitons,cui2014transient,byrnes2014exciton,tanimura2019dynamics,madeo2020directly,man2021experimental,karni2022structure,mori2023spin,pareek2024driving}.
While applying a similar method is not trivial in an excitonic insulator due to an energy overlap of excitons with valence bands, a recent work successfully isolated the characteristic exciton photoemission signal in Ta$_2$NiSe$_5$ utilizing orbital selection rules of photoemission \cite{fukutani2021detecting}. 
However, very intriguingly, the exciton photoemission signal was observed only above \textit{T}$_c$. This was interpreted to indicate the spontaneous formation of many incoherent excitons above \textit{T}$_c$ and, although not completely clear yet, the exciton photoemission signal may be suppressed by the strong exciton-lattice interaction below \textit{T}$_c$. 
This made the confirmation of excitons in an insulating ground state of a solid as well as the excitonic insulator nature of Ta$_2$NiSe$_5$ elusive.

In this context, we focus on another, very recent, excitonic insulator candidate material Ta$_2$Pd$_3$Te$_5$ \cite {hossain2023discovery, huang2024evidence, yao2024excitonic, jiang2024surface,zhang2024spontaneous}.
While Ta$_2$Pd$_3$Te$_5$ has a similar band structure and a similar gap-opening transition (\textit{T}$_c$ = 365 K) with Ta$_2$NiSe$_5$, its structural distortion is undetectably small in x-ray diffraction \cite{hossain2023discovery,huang2024evidence,zhang2024spontaneous} with only signatures of symmetry breaking in electron diffraction \cite{huang2024evidence,zhang2024spontaneous} and scanning tunneling microscopy (See Fig. S1 in Ref. \cite{sup}). 
This unique property positions Ta$_2$Pd$_3$Te$_5$ as an unprecedentedly promising candidate for an excitonic insulator with the excitonic coupling dominating over the electron-lattice coupling. 
Nevertheless, the clear evidence for excitons within the insulating phase has not been available.
Here, we identify the exciton photoemission signal from the insulating phase of this materials using ARPES, confirming the existence of excitons in a ground state of a crystal. 
The detailed properties of the excitons formed, which cannot be probed by other experimental probes, are investigated to reveal the size and the unusual odd parity of excitons in this material. 



Ta$_2$Pd$_3$Te$_5$ is a layered material and its single layer is composed of unit cells with multiple Ta and Pd chains (Fig. 1a).
The chain structure gives rise to semimetallic band dispersions along the chains and flat bands across the chains. 
Its butterfly-shape valence bands along $k_x$ (Fig. 1b) come from Pd $d_{xz}$ and Te $p_{x}$ orbitals and the parabolic conduction band from Ta $d_{z^2}$ electrons  \cite{huang2024evidence,yao2024excitonic,wang2021observation}. 
These bands degenerate at the center of the Brillouin zone (BZ) ($\Gamma$) to make a zero-gap semimetal in the normal state, which are well reproduced in the band calculations using density functional theory (DFT) \cite{huang2024evidence,jiang2024surface,zhang2024spontaneous}. 
Due to the mirror symmetry inherent in the atomic and orbital structures,
each band exhibits a pronounced polarization dependence (see Figs. S4 and S5 in Ref. \cite{sup} for detail) \cite{fukutani2021detecting}. 
The valence (conduction) band spectral weight is suppressed in ARPES spectra taken at $yz(x)$-polarization shown in Fig. 2c (Fig. 2d) along $\Gamma-X$ due to their orbital characters [see also Fig. S5a (S5b) in Ref. \cite{sup} for detail]. 
A metal-insulator transition starts from 365 K as the valence (conduction) bands move gradually to a lower energy (higher above the Fermi level) as shown in Fig. 2 (see also Fig. S6 and S7 in Ref. \cite{sup} for detail). 
The band gap becomes obvious below 300 K beyond the energy broadening of ARPES spectra (Fig. 2c and f). 
Temperature-dependent band dispersions are fully consistent with the recent ARPES works \cite{hossain2023discovery,huang2024evidence,zhang2024spontaneous}. 
The structural investigation within those works further found a structural symmetry breaking with marginal distortions. The band gap expected in the DFT calculations is, however, in the order of a few meV so that an extra manybody interaction is needed to explain the observed band gap of $\sim$ 100 meV (see Fig. S8 in Ref. \cite{sup} for detail).
The existence of any extra charge density wave is excluded in the structural studies \cite{huang2024evidence,zhang2024spontaneous} and the scanning tunneling microscopy measurement (See Fig. S1 in Ref. \cite{sup}). 
Moreover, the strong correlation effect is not likely since the gap forms from a highly dispersing Dirac band. 

The ARPES spectra in Fig. 1c combine those obtained along high-symmetry lines of the BZ, $\Gamma-Y$ and $\Gamma-X$, in the insulating phase at 80 K.
One can immediately notice a strong and unusual photoemission signal at the $\Gamma$ point below the valence band maximum located at -0.08 eV (schematics in Fig. 1b and see Fig. S9 in Ref. \cite{sup} for more details).
This signal deviates largley from the valence band dispersions, both calculated and measured ones, shown in Figs. 2a, c, and d \cite{huang2024evidence,yao2024excitonic,wang2021observation}. 
Moreover, it disappears completely in the second BZ as shown in Fig. 1e, where the bare valence band signals persist. 
Note also that the valence band dispersion changes dramatically from $\Gamma-X$ to $\Gamma-Y$ while the spectral feature at $\Gamma$ is consistent. 
It is thus unambiguous that this signal does not come from valence bands. 
To quantify this unusual spectral function, we subtracted the ARPES spectra of the first BZ by the bare valence band spectra obtained in the second BZ, which is given on the right panel of Fig. 1e (see Fig. S9 in Ref. \cite{sup} for detail). 
The spectral function subtracted is largely isotropic in momentum but slightly squeezed along the $\Gamma-Y$ direction (Fig. 1c and see Fig. S9 in Ref. \cite{sup} for details).
The unusual spectral shape in energy and momentum is not compatible with spectra features due to electron-phonon interactions such as polaronic sidebands and shake-up replica of the valence band \cite{chen2015observation,kang2018holstein,riley2018crossover,antonius2020polaron,caruso2021two,kang2021band}.
The fluctuation- or temperature-induced spectral broadening cannot explain this spectral distribution, since it is consistent down to a very low temperature of 10 K (see Fig. S12  in Ref. \cite{sup} for detail).
We further excluded the phonon broadening of the valence band by quantitatively simulating the spectral function using a two-band Bethe-Salpeter equation and phonon-induced replica bands (see Fig. S10, S11 in Ref. \cite{sup} for detail).
It was also reported that the top of the valence band has no k$_z$ dispersion, which rules out any substantial final state effect to broaden the spectra at $\Gamma$ \cite{wang2023robust}.


However, this egg-shape photoemission signal is remarkably similar to the spectral function of the exciton photoemission observed previously \cite{scholes2006excitons,cui2014transient,byrnes2014exciton,tanimura2019dynamics,madeo2020directly,man2021experimental,karni2022structure,mori2023spin,pareek2024driving,rustagi2018photoemission,fukutani2021detecting,christiansen2019theory}. 
This signal can be simulated quantitatively well by the exciton photoemission formula (Fig. 1d and Fig. 3d) with the exciton wavefunction and the valence band dispersions at the $\Gamma$ point \cite{rustagi2018photoemission, fukutani2021detecting}.
The exciton wavefunction was extracted by fitting the momentum distribution curves (MDCs, Figs. 3a and 3b) of the subtracted spectra along both $\Gamma-Y$ (left) and $\Gamma-X$ (right) (see the tables in Ref. \cite{sup} for the fitting parameters). 
The wavefunction corresponds to an 1$s$ wavefunction, which exhibits a slightly larger Bohr radius along the $\Gamma-X$ direction (approximately 15.1 \AA) than along $\Gamma-Y$ (approximately 13.6 \AA) due to the marginal anisotropy mentioned above. 
The anisotropic shape of the exciton seems to reflect the anisotropic band structure \cite{fukutani2021detecting}.  
The angular distribution of the exciton wavefunction (Fig. 3d) is consistent with the recent result of tight-binding calculations of excitons in a Ta$_2$Pd$_3$Te$_5$ monolayer \cite{yao2024excitonic} while the calculation predicts two times larger exciton radius. 
The energy distribution curves (EDCs) at $\Gamma$ are fitted by Voigt functions as illustrated in Fig. 3c, which shows an energy broadening of approximately 0.1 eV.  
This reflects the scattering of excitons, which is thought to be due mainly to phonons \cite{ni2017real}. 



The temperature-dependent ARPES measurements shown in Fig. 2 (see also Figs. S12 and S13 in Ref. \cite{sup} for detail) show that the exciton photoemission signal remains robust below \textit{T}$_c$ (Fig. 2a).
However, the excitonic signal moves systematically to a lower energy (Figs. 2a and 2e) with rising temperature, which is consistently lower than the valence band top.
Namely, the exciton binding energy surpasses the band gap enough to drive the metal-insulator transition in the whole temperature range below \textit{T}$_c$. 
In addition, a gradual spectral weight transfer is observed from the valence band to the exciton photoemission feature during the transition (see also Fig. S14 in Ref. \cite{sup} for detail).
In clear contrast, the exciton photoemission in Ta$_2$NiSe$_5$ was observed only above \textit{T}$_c$ \cite{fukutani2021detecting}. 
This suggests that the observed excitons in Ta$_2$NiSe$_5$ are incoherent excitons preformed above the condensation temperature and the excitonic transition is near the crossover of a Bose-Einstein condensation \cite{lu2017zero,okazaki2018photo,werdehausen2018photo,huang2024evidence,zhang2024spontaneous}.
The present experimental setup did not allow to reach a temperature above 400 K so that the existence of preformed excitons in Ta$_2$Pd$_3$Te$_5$ is not clear.  
The reason for the loss of the exciton photoemission below \textit{T}$_c$ in Ta$_2$NiSe$_5$ is not understood yet but may be related to the strong electron-lattice and exciton-lattice interactions \cite{kaneko2013orthorhombic,mazza2020nature,fukutani2021detecting,chen2023role,chen2023anomalous}.


A further contrasting property of the exciton photoemission can be found in its polarization dependence, which is dictated by the photoemission dipole selection rule.
The exciton photoemission signal is a product of the \textit{s-}wave-like exciton envelope and the photoemission matrix element from the conduction band orbital \cite{rustagi2018photoemission,fukutani2021detecting}.
The exciton feature in Ta$_2$NiSe$_5$ was found to have even parity for both mirror planes as selectively probed by \textit{z}-polarized photons \cite{fukutani2021detecting}.
In contrast, the present exciton photoemission signal is clearly observed for \textit{x}-polarized photons but suppressed in the \textit{yz} polarization (Fig. 2c), indicating a contribution of an odd parity wave function along the \textit{yz} plane (see Figs. S4 and S15 in Ref. \cite{sup} for detail). 
We note that the only odd parity wave function involved here is the valence Pd $d_{xy}$ orbital (Fig. 2d), which is not expected to contribute to the photoemission from a conventional excited-state exciton within a band gap.  
However, the strong mixing of the valence and conduction bands near $\Gamma$ is natural in an excitonic insulator \cite{jerome1967excitonic,huang2024evidence,zhang2024spontaneous}, which allows the valence band orbital to contribute to the exciton photoemission matrix element. 
The interband hybridization is allowed by or induces the structural symmetry breaking mentioned above.
Then, the distinct parity of exciton spectral functions in Ta$_2$NiSe$_5$ and the present crystal is also straightforwardly explained by the difference in the orbital parity of their valence orbitals; Ta$_2$NiSe$_5$'s valence band has even parity for both x and y axes.

In summary, we detected a characteristic photoemission signal from excitons in an emerging insulating state of a 3D crystalline solid, Ta$_2$Pd$_3$Te$_5$, which evidences the excitonic insulator nature of the insulating state. 
The exciton binding energy was tracked throughout the gradual gap opening transition and the symmetry and the envelope of exciton wave function were determined. 
The nearly distortion-less nature of the transition and the highest \textit{T}$_c$ makes this system an ideal platform to exploit exciton condensates in a bulk solid. 
The symmetry-broken, odd-parity, nature of the excitons observed may lead to discovering exotic manybody quantum states \cite{wang2019prediction, kozii2015odd} under the non-trivial topological nature of the present system \cite{guo2021quantum,wang2021observation,wang2023robust,li2024interfering,yu2024observation}.

This work was supported by the Institute for Basic Science (Grant No. IBS-R014-D1). K.-H.J. was supported by Global-Learning \& Academic research institution for Master's·PhD students, and Postdocs (LAMP) Program of the National Research Foundation of Korea (NRF) funded by the Ministry of Education (Grant No. RS-2024-00443714). C.W. was supported by NRF research grant (No. RS-2022-NR068223).

\appendix

\section{Crystal growth}
Single crystals of Ta$_2$Pd$_3$Te$_5$ were synthesized using the self-flux method. A mixture of Ta (powder, 99.999 \%), Pd (powder, 99.9999 \%), and Te (lump, 99.9999 \%) in a molar ratio of 2.0 : 4.5 : 7.5 (Ta : Pd : Te) was prepared in an argon-filled glove box. The mixture was placed in an alumina crucible, sealed in an evacuated quartz tube, and heated to 950 $^{\circ}\mathrm{C}$ over a period of 10 hours. It was then maintained at this temperature for 2 days. Following this, the tube was cooled to 800  $^{\circ}\mathrm{C}$ at a rate of 0.5 $^{\circ}\mathrm{C}$/h. The crystalline quality was confirmed by x-ray diffraction and spectroscopy (see Figs. S2 and S3 in Ref. \cite{sup} for detail).\nocite{zhang2011precise}\\

\section{ARPES experiment}
All ARPES experiments were performed using the DA-30L electron analyzer (Scienta Omicron, Sweden) and a laser of photon energy 11 eV (Oxide, Japan). The typical energy resolution was better than 0.6 meV. Ta$_2$Pd$_3$Te$_5$ single crystal was cleaved at 75-80 K in ultra high vacuum ($\sim$ 10$^{-11}$ torr) before measurements.\\

\section{DFT calculation}
The first-principles calculations were performed within the framework of the density functional theory (DFT) using the Perdew-Burke-Ernzerhof-type generalized gradient approximation for the exchange-correlation functional, as implemented in the Vienna ab initio simulation package \cite{kresse1996efficient,perdew1996generalized}. In the self-consistent process, all the calculations were carried out using the kinetic energy cutoff of 520 eV on a 15$\times$5$\times$4 Monkhorst-Pack k-point mesh for the Ta$_2$Pd$_3$Te$_5$ bulk and a 15$\times$5$\times$1 for the slab calculations. The spin–orbit coupling effect was included in the self-consistent electronic structure calculation. The lattice constants were taken from experiments \cite{wang2021observation}, but the atoms in the unit cell are fully relaxed with the force cutoff of 0.01 eV \AA$^{-1}$. The electronic self-consistent iteration was converged to $10^{-5}$ eV precision of the total energy.\\

\bibliography{apssamp}

\begin{thebibliography}{74}%
\makeatletter
\providecommand \@ifxundefined [1]{%
 \@ifx{#1\undefined}
}%
\providecommand \@ifnum [1]{%
 \ifnum #1\expandafter \@firstoftwo
 \else \expandafter \@secondoftwo
 \fi
}%
\providecommand \@ifx [1]{%
 \ifx #1\expandafter \@firstoftwo
 \else \expandafter \@secondoftwo
 \fi
}%
\providecommand \natexlab [1]{#1}%
\providecommand \enquote  [1]{``#1''}%
\providecommand \bibnamefont  [1]{#1}%
\providecommand \bibfnamefont [1]{#1}%
\providecommand \citenamefont [1]{#1}%
\providecommand \href@noop [0]{\@secondoftwo}%
\providecommand \href [0]{\begingroup \@sanitize@url \@href}%
\providecommand \@href[1]{\@@startlink{#1}\@@href}%
\providecommand \@@href[1]{\endgroup#1\@@endlink}%
\providecommand \@sanitize@url [0]{\catcode `\\12\catcode `\$12\catcode `\&12\catcode `\#12\catcode `\^12\catcode `\_12\catcode `\%12\relax}%
\providecommand \@@startlink[1]{}%
\providecommand \@@endlink[0]{}%
\providecommand \url  [0]{\begingroup\@sanitize@url \@url }%
\providecommand \@url [1]{\endgroup\@href {#1}{\urlprefix }}%
\providecommand \urlprefix  [0]{URL }%
\providecommand \Eprint [0]{\href }%
\providecommand \doibase [0]{https://doi.org/}%
\providecommand \selectlanguage [0]{\@gobble}%
\providecommand \bibinfo  [0]{\@secondoftwo}%
\providecommand \bibfield  [0]{\@secondoftwo}%
\providecommand \translation [1]{[#1]}%
\providecommand \BibitemOpen [0]{}%
\providecommand \bibitemStop [0]{}%
\providecommand \bibitemNoStop [0]{.\EOS\space}%
\providecommand \EOS [0]{\spacefactor3000\relax}%
\providecommand \BibitemShut  [1]{\csname bibitem#1\endcsname}%
\let\auto@bib@innerbib\@empty
\bibitem [{\citenamefont {Van~Amerongen}\ \emph {et~al.}(2000)\citenamefont {Van~Amerongen}, \citenamefont {Van~Grondelle} \emph {et~al.}}]{van2000photosynthetic}%
  \BibitemOpen
  \bibfield  {author} {\bibinfo {author} {\bibfnamefont {H.}~\bibnamefont {Van~Amerongen}}, \bibinfo {author} {\bibfnamefont {R.}~\bibnamefont {Van~Grondelle}}, \emph {et~al.},\ }\href@noop {} {\emph {\bibinfo {title} {Photosynthetic excitons}}}\ (\bibinfo  {publisher} {World Scientific},\ \bibinfo {year} {2000})\BibitemShut {NoStop}%
\bibitem [{\citenamefont {Brixner}\ \emph {et~al.}(2005)\citenamefont {Brixner}, \citenamefont {Stenger}, \citenamefont {Vaswani}, \citenamefont {Cho}, \citenamefont {Blankenship},\ and\ \citenamefont {Fleming}}]{brixner2005two}%
  \BibitemOpen
  \bibfield  {author} {\bibinfo {author} {\bibfnamefont {T.}~\bibnamefont {Brixner}}, \bibinfo {author} {\bibfnamefont {J.}~\bibnamefont {Stenger}}, \bibinfo {author} {\bibfnamefont {H.~M.}\ \bibnamefont {Vaswani}}, \bibinfo {author} {\bibfnamefont {M.}~\bibnamefont {Cho}}, \bibinfo {author} {\bibfnamefont {R.~E.}\ \bibnamefont {Blankenship}},\ and\ \bibinfo {author} {\bibfnamefont {G.~R.}\ \bibnamefont {Fleming}},\ }\bibfield  {title} {\bibinfo {title} {Two-dimensional spectroscopy of electronic couplings in photosynthesis},\ }\href@noop {} {\bibfield  {journal} {\bibinfo  {journal} {Nature}\ }\textbf {\bibinfo {volume} {434}},\ \bibinfo {pages} {625} (\bibinfo {year} {2005})}\BibitemShut {NoStop}%
\bibitem [{\citenamefont {Kuznetsova}\ \emph {et~al.}(2010)\citenamefont {Kuznetsova}, \citenamefont {Remeika}, \citenamefont {High}, \citenamefont {Hammack}, \citenamefont {Butov}, \citenamefont {Hanson},\ and\ \citenamefont {Gossard}}]{kuznetsova2010all}%
  \BibitemOpen
  \bibfield  {author} {\bibinfo {author} {\bibfnamefont {Y.}~\bibnamefont {Kuznetsova}}, \bibinfo {author} {\bibfnamefont {M.}~\bibnamefont {Remeika}}, \bibinfo {author} {\bibfnamefont {A.}~\bibnamefont {High}}, \bibinfo {author} {\bibfnamefont {A.}~\bibnamefont {Hammack}}, \bibinfo {author} {\bibfnamefont {L.}~\bibnamefont {Butov}}, \bibinfo {author} {\bibfnamefont {M.}~\bibnamefont {Hanson}},\ and\ \bibinfo {author} {\bibfnamefont {A.}~\bibnamefont {Gossard}},\ }\bibfield  {title} {\bibinfo {title} {All-optical excitonic transistor},\ }\href@noop {} {\bibfield  {journal} {\bibinfo  {journal} {Optics letters}\ }\textbf {\bibinfo {volume} {35}},\ \bibinfo {pages} {1587} (\bibinfo {year} {2010})}\BibitemShut {NoStop}%
\bibitem [{\citenamefont {Menke}\ \emph {et~al.}(2013)\citenamefont {Menke}, \citenamefont {Luhman},\ and\ \citenamefont {Holmes}}]{menke2013tailored}%
  \BibitemOpen
  \bibfield  {author} {\bibinfo {author} {\bibfnamefont {S.~M.}\ \bibnamefont {Menke}}, \bibinfo {author} {\bibfnamefont {W.~A.}\ \bibnamefont {Luhman}},\ and\ \bibinfo {author} {\bibfnamefont {R.~J.}\ \bibnamefont {Holmes}},\ }\bibfield  {title} {\bibinfo {title} {Tailored exciton diffusion in organic photovoltaic cells for enhanced power conversion efficiency},\ }\href@noop {} {\bibfield  {journal} {\bibinfo  {journal} {Nature materials}\ }\textbf {\bibinfo {volume} {12}},\ \bibinfo {pages} {152} (\bibinfo {year} {2013})}\BibitemShut {NoStop}%
\bibitem [{\citenamefont {Romero}\ \emph {et~al.}(2014)\citenamefont {Romero}, \citenamefont {Augulis}, \citenamefont {Novoderezhkin}, \citenamefont {Ferretti}, \citenamefont {Thieme}, \citenamefont {Zigmantas},\ and\ \citenamefont {Van~Grondelle}}]{romero2014quantum}%
  \BibitemOpen
  \bibfield  {author} {\bibinfo {author} {\bibfnamefont {E.}~\bibnamefont {Romero}}, \bibinfo {author} {\bibfnamefont {R.}~\bibnamefont {Augulis}}, \bibinfo {author} {\bibfnamefont {V.~I.}\ \bibnamefont {Novoderezhkin}}, \bibinfo {author} {\bibfnamefont {M.}~\bibnamefont {Ferretti}}, \bibinfo {author} {\bibfnamefont {J.}~\bibnamefont {Thieme}}, \bibinfo {author} {\bibfnamefont {D.}~\bibnamefont {Zigmantas}},\ and\ \bibinfo {author} {\bibfnamefont {R.}~\bibnamefont {Van~Grondelle}},\ }\bibfield  {title} {\bibinfo {title} {Quantum coherence in photosynthesis for efficient solar-energy conversion},\ }\href@noop {} {\bibfield  {journal} {\bibinfo  {journal} {Nature physics}\ }\textbf {\bibinfo {volume} {10}},\ \bibinfo {pages} {676} (\bibinfo {year} {2014})}\BibitemShut {NoStop}%
\bibitem [{\citenamefont {Ross}\ \emph {et~al.}(2014)\citenamefont {Ross}, \citenamefont {Klement}, \citenamefont {Jones}, \citenamefont {Ghimire}, \citenamefont {Yan}, \citenamefont {Mandrus}, \citenamefont {Taniguchi}, \citenamefont {Watanabe}, \citenamefont {Kitamura}, \citenamefont {Yao} \emph {et~al.}}]{ross2014electrically}%
  \BibitemOpen
  \bibfield  {author} {\bibinfo {author} {\bibfnamefont {J.~S.}\ \bibnamefont {Ross}}, \bibinfo {author} {\bibfnamefont {P.}~\bibnamefont {Klement}}, \bibinfo {author} {\bibfnamefont {A.~M.}\ \bibnamefont {Jones}}, \bibinfo {author} {\bibfnamefont {N.~J.}\ \bibnamefont {Ghimire}}, \bibinfo {author} {\bibfnamefont {J.}~\bibnamefont {Yan}}, \bibinfo {author} {\bibfnamefont {D.}~\bibnamefont {Mandrus}}, \bibinfo {author} {\bibfnamefont {T.}~\bibnamefont {Taniguchi}}, \bibinfo {author} {\bibfnamefont {K.}~\bibnamefont {Watanabe}}, \bibinfo {author} {\bibfnamefont {K.}~\bibnamefont {Kitamura}}, \bibinfo {author} {\bibfnamefont {W.}~\bibnamefont {Yao}}, \emph {et~al.},\ }\bibfield  {title} {\bibinfo {title} {Electrically tunable excitonic light-emitting diodes based on monolayer \uppercase{WS}e$_2$ p--n junctions},\ }\href@noop {} {\bibfield  {journal} {\bibinfo  {journal} {Nature nanotechnology}\ }\textbf {\bibinfo {volume} {9}},\ \bibinfo {pages} {268} (\bibinfo {year} {2014})}\BibitemShut {NoStop}%
\bibitem [{\citenamefont {Furchi}\ \emph {et~al.}(2014)\citenamefont {Furchi}, \citenamefont {Pospischil}, \citenamefont {Libisch}, \citenamefont {Burgdorfer},\ and\ \citenamefont {Mueller}}]{furchi2014photovoltaic}%
  \BibitemOpen
  \bibfield  {author} {\bibinfo {author} {\bibfnamefont {M.~M.}\ \bibnamefont {Furchi}}, \bibinfo {author} {\bibfnamefont {A.}~\bibnamefont {Pospischil}}, \bibinfo {author} {\bibfnamefont {F.}~\bibnamefont {Libisch}}, \bibinfo {author} {\bibfnamefont {J.}~\bibnamefont {Burgdorfer}},\ and\ \bibinfo {author} {\bibfnamefont {T.}~\bibnamefont {Mueller}},\ }\bibfield  {title} {\bibinfo {title} {Photovoltaic effect in an electrically tunable van der waals heterojunction},\ }\href@noop {} {\bibfield  {journal} {\bibinfo  {journal} {Nano letters}\ }\textbf {\bibinfo {volume} {14}},\ \bibinfo {pages} {4785} (\bibinfo {year} {2014})}\BibitemShut {NoStop}%
\bibitem [{\citenamefont {Ye}\ \emph {et~al.}(2015)\citenamefont {Ye}, \citenamefont {Wong}, \citenamefont {Lu}, \citenamefont {Ni}, \citenamefont {Zhu}, \citenamefont {Chen}, \citenamefont {Wang},\ and\ \citenamefont {Zhang}}]{ye2015monolayer}%
  \BibitemOpen
  \bibfield  {author} {\bibinfo {author} {\bibfnamefont {Y.}~\bibnamefont {Ye}}, \bibinfo {author} {\bibfnamefont {Z.~J.}\ \bibnamefont {Wong}}, \bibinfo {author} {\bibfnamefont {X.}~\bibnamefont {Lu}}, \bibinfo {author} {\bibfnamefont {X.}~\bibnamefont {Ni}}, \bibinfo {author} {\bibfnamefont {H.}~\bibnamefont {Zhu}}, \bibinfo {author} {\bibfnamefont {X.}~\bibnamefont {Chen}}, \bibinfo {author} {\bibfnamefont {Y.}~\bibnamefont {Wang}},\ and\ \bibinfo {author} {\bibfnamefont {X.}~\bibnamefont {Zhang}},\ }\bibfield  {title} {\bibinfo {title} {Monolayer excitonic laser},\ }\href@noop {} {\bibfield  {journal} {\bibinfo  {journal} {Nature Photonics}\ }\textbf {\bibinfo {volume} {9}},\ \bibinfo {pages} {733} (\bibinfo {year} {2015})}\BibitemShut {NoStop}%
\bibitem [{\citenamefont {Romero}\ \emph {et~al.}(2017)\citenamefont {Romero}, \citenamefont {Novoderezhkin},\ and\ \citenamefont {Van~Grondelle}}]{romero2017quantum}%
  \BibitemOpen
  \bibfield  {author} {\bibinfo {author} {\bibfnamefont {E.}~\bibnamefont {Romero}}, \bibinfo {author} {\bibfnamefont {V.~I.}\ \bibnamefont {Novoderezhkin}},\ and\ \bibinfo {author} {\bibfnamefont {R.}~\bibnamefont {Van~Grondelle}},\ }\bibfield  {title} {\bibinfo {title} {Quantum design of photosynthesis for bio-inspired solar-energy conversion},\ }\href@noop {} {\bibfield  {journal} {\bibinfo  {journal} {Nature}\ }\textbf {\bibinfo {volume} {543}},\ \bibinfo {pages} {355} (\bibinfo {year} {2017})}\BibitemShut {NoStop}%
\bibitem [{\citenamefont {Mueller}\ and\ \citenamefont {Malic}(2018)}]{mueller2018exciton}%
  \BibitemOpen
  \bibfield  {author} {\bibinfo {author} {\bibfnamefont {T.}~\bibnamefont {Mueller}}\ and\ \bibinfo {author} {\bibfnamefont {E.}~\bibnamefont {Malic}},\ }\bibfield  {title} {\bibinfo {title} {Exciton physics and device application of two-dimensional transition metal dichalcogenide semiconductors},\ }\href@noop {} {\bibfield  {journal} {\bibinfo  {journal} {npj 2D Materials and Applications}\ }\textbf {\bibinfo {volume} {2}},\ \bibinfo {pages} {29} (\bibinfo {year} {2018})}\BibitemShut {NoStop}%
\bibitem [{\citenamefont {J{\'e}rome}\ \emph {et~al.}(1967)\citenamefont {J{\'e}rome}, \citenamefont {Rice},\ and\ \citenamefont {Kohn}}]{jerome1967excitonic}%
  \BibitemOpen
  \bibfield  {author} {\bibinfo {author} {\bibfnamefont {D.}~\bibnamefont {J{\'e}rome}}, \bibinfo {author} {\bibfnamefont {T.}~\bibnamefont {Rice}},\ and\ \bibinfo {author} {\bibfnamefont {W.}~\bibnamefont {Kohn}},\ }\bibfield  {title} {\bibinfo {title} {Excitonic insulator},\ }\href@noop {} {\bibfield  {journal} {\bibinfo  {journal} {Physical Review}\ }\textbf {\bibinfo {volume} {158}},\ \bibinfo {pages} {462} (\bibinfo {year} {1967})}\BibitemShut {NoStop}%
\bibitem [{\citenamefont {Halperin}\ and\ \citenamefont {Rice}(1968)}]{halperin1968possible}%
  \BibitemOpen
  \bibfield  {author} {\bibinfo {author} {\bibfnamefont {B.}~\bibnamefont {Halperin}}\ and\ \bibinfo {author} {\bibfnamefont {T.}~\bibnamefont {Rice}},\ }\bibfield  {title} {\bibinfo {title} {Possible anomalies at a semimetal-semiconductor transistion},\ }\href@noop {} {\bibfield  {journal} {\bibinfo  {journal} {Reviews of Modern Physics}\ }\textbf {\bibinfo {volume} {40}},\ \bibinfo {pages} {755} (\bibinfo {year} {1968})}\BibitemShut {NoStop}%
\bibitem [{\citenamefont {Wakisaka}\ \emph {et~al.}(2009)\citenamefont {Wakisaka}, \citenamefont {Sudayama}, \citenamefont {Takubo}, \citenamefont {Mizokawa}, \citenamefont {Arita}, \citenamefont {Namatame}, \citenamefont {Taniguchi}, \citenamefont {Katayama}, \citenamefont {Nohara},\ and\ \citenamefont {Takagi}}]{wakisaka2009excitonic}%
  \BibitemOpen
  \bibfield  {author} {\bibinfo {author} {\bibfnamefont {Y.}~\bibnamefont {Wakisaka}}, \bibinfo {author} {\bibfnamefont {T.}~\bibnamefont {Sudayama}}, \bibinfo {author} {\bibfnamefont {K.}~\bibnamefont {Takubo}}, \bibinfo {author} {\bibfnamefont {T.}~\bibnamefont {Mizokawa}}, \bibinfo {author} {\bibfnamefont {M.}~\bibnamefont {Arita}}, \bibinfo {author} {\bibfnamefont {H.}~\bibnamefont {Namatame}}, \bibinfo {author} {\bibfnamefont {M.}~\bibnamefont {Taniguchi}}, \bibinfo {author} {\bibfnamefont {N.}~\bibnamefont {Katayama}}, \bibinfo {author} {\bibfnamefont {M.}~\bibnamefont {Nohara}},\ and\ \bibinfo {author} {\bibfnamefont {H.}~\bibnamefont {Takagi}},\ }\bibfield  {title} {\bibinfo {title} {Excitonic insulator state in \uppercase{T}a$_2$\uppercase{N}i\uppercase{S}e$_5$ probed by photoemission spectroscopy},\ }\href@noop {} {\bibfield  {journal} {\bibinfo  {journal} {Physical review letters}\ }\textbf {\bibinfo {volume} {103}},\ \bibinfo {pages} {026402} (\bibinfo {year} {2009})}\BibitemShut {NoStop}%
\bibitem [{\citenamefont {Pillo}\ \emph {et~al.}(2000)\citenamefont {Pillo}, \citenamefont {Hayoz}, \citenamefont {Berger}, \citenamefont {L{\'e}vy}, \citenamefont {Schlapbach},\ and\ \citenamefont {Aebi}}]{pillo2000photoemission}%
  \BibitemOpen
  \bibfield  {author} {\bibinfo {author} {\bibfnamefont {T.}~\bibnamefont {Pillo}}, \bibinfo {author} {\bibfnamefont {J.}~\bibnamefont {Hayoz}}, \bibinfo {author} {\bibfnamefont {H.}~\bibnamefont {Berger}}, \bibinfo {author} {\bibfnamefont {F.}~\bibnamefont {L{\'e}vy}}, \bibinfo {author} {\bibfnamefont {L.}~\bibnamefont {Schlapbach}},\ and\ \bibinfo {author} {\bibfnamefont {P.}~\bibnamefont {Aebi}},\ }\bibfield  {title} {\bibinfo {title} {Photoemission of bands above the fermi level: The excitonic insulator phase transition in 1\uppercase{T}- \uppercase{T}i\uppercase{S}e$_2$},\ }\href@noop {} {\bibfield  {journal} {\bibinfo  {journal} {Physical Review B}\ }\textbf {\bibinfo {volume} {61}},\ \bibinfo {pages} {16213} (\bibinfo {year} {2000})}\BibitemShut {NoStop}%
\bibitem [{\citenamefont {Seki}\ \emph {et~al.}(2014)\citenamefont {Seki}, \citenamefont {Wakisaka}, \citenamefont {Kaneko}, \citenamefont {Toriyama}, \citenamefont {Konishi}, \citenamefont {Sudayama}, \citenamefont {Saini}, \citenamefont {Arita}, \citenamefont {Namatame}, \citenamefont {Taniguchi} \emph {et~al.}}]{seki2014excitonic}%
  \BibitemOpen
  \bibfield  {author} {\bibinfo {author} {\bibfnamefont {K.}~\bibnamefont {Seki}}, \bibinfo {author} {\bibfnamefont {Y.}~\bibnamefont {Wakisaka}}, \bibinfo {author} {\bibfnamefont {T.}~\bibnamefont {Kaneko}}, \bibinfo {author} {\bibfnamefont {T.}~\bibnamefont {Toriyama}}, \bibinfo {author} {\bibfnamefont {T.}~\bibnamefont {Konishi}}, \bibinfo {author} {\bibfnamefont {T.}~\bibnamefont {Sudayama}}, \bibinfo {author} {\bibfnamefont {N.}~\bibnamefont {Saini}}, \bibinfo {author} {\bibfnamefont {M.}~\bibnamefont {Arita}}, \bibinfo {author} {\bibfnamefont {H.}~\bibnamefont {Namatame}}, \bibinfo {author} {\bibfnamefont {M.}~\bibnamefont {Taniguchi}}, \emph {et~al.},\ }\bibfield  {title} {\bibinfo {title} {Excitonic \uppercase{B}ose-\uppercase{E}instein condensation in \uppercase{T}a$_2$\uppercase{N}i\uppercase{S}e$_5$ above room temperature},\ }\href@noop {} {\bibfield  {journal} {\bibinfo  {journal} {Physical Review B}\ }\textbf {\bibinfo {volume} {90}},\ \bibinfo {pages} {155116} (\bibinfo {year}
  {2014})}\BibitemShut {NoStop}%
\bibitem [{\citenamefont {Sugawara}\ \emph {et~al.}(2016)\citenamefont {Sugawara}, \citenamefont {Nakata}, \citenamefont {Shimizu}, \citenamefont {Han}, \citenamefont {Hitosugi}, \citenamefont {Sato},\ and\ \citenamefont {Takahashi}}]{sugawara2016unconventional}%
  \BibitemOpen
  \bibfield  {author} {\bibinfo {author} {\bibfnamefont {K.}~\bibnamefont {Sugawara}}, \bibinfo {author} {\bibfnamefont {Y.}~\bibnamefont {Nakata}}, \bibinfo {author} {\bibfnamefont {R.}~\bibnamefont {Shimizu}}, \bibinfo {author} {\bibfnamefont {P.}~\bibnamefont {Han}}, \bibinfo {author} {\bibfnamefont {T.}~\bibnamefont {Hitosugi}}, \bibinfo {author} {\bibfnamefont {T.}~\bibnamefont {Sato}},\ and\ \bibinfo {author} {\bibfnamefont {T.}~\bibnamefont {Takahashi}},\ }\bibfield  {title} {\bibinfo {title} {Unconventional charge-density-wave transition in monolayer 1\uppercase{T}-\uppercase{T}i\uppercase{S}e$_2$},\ }\href@noop {} {\bibfield  {journal} {\bibinfo  {journal} {ACS nano}\ }\textbf {\bibinfo {volume} {10}},\ \bibinfo {pages} {1341} (\bibinfo {year} {2016})}\BibitemShut {NoStop}%
\bibitem [{\citenamefont {Kogar}\ \emph {et~al.}(2017)\citenamefont {Kogar}, \citenamefont {Rak}, \citenamefont {Vig}, \citenamefont {Husain}, \citenamefont {Flicker}, \citenamefont {Joe}, \citenamefont {Venema}, \citenamefont {MacDougall}, \citenamefont {Chiang}, \citenamefont {Fradkin} \emph {et~al.}}]{kogar2017signatures}%
  \BibitemOpen
  \bibfield  {author} {\bibinfo {author} {\bibfnamefont {A.}~\bibnamefont {Kogar}}, \bibinfo {author} {\bibfnamefont {M.~S.}\ \bibnamefont {Rak}}, \bibinfo {author} {\bibfnamefont {S.}~\bibnamefont {Vig}}, \bibinfo {author} {\bibfnamefont {A.~A.}\ \bibnamefont {Husain}}, \bibinfo {author} {\bibfnamefont {F.}~\bibnamefont {Flicker}}, \bibinfo {author} {\bibfnamefont {Y.~I.}\ \bibnamefont {Joe}}, \bibinfo {author} {\bibfnamefont {L.}~\bibnamefont {Venema}}, \bibinfo {author} {\bibfnamefont {G.~J.}\ \bibnamefont {MacDougall}}, \bibinfo {author} {\bibfnamefont {T.~C.}\ \bibnamefont {Chiang}}, \bibinfo {author} {\bibfnamefont {E.}~\bibnamefont {Fradkin}}, \emph {et~al.},\ }\bibfield  {title} {\bibinfo {title} {Signatures of exciton condensation in a transition metal dichalcogenide},\ }\href@noop {} {\bibfield  {journal} {\bibinfo  {journal} {Science}\ }\textbf {\bibinfo {volume} {358}},\ \bibinfo {pages} {1314} (\bibinfo {year} {2017})}\BibitemShut {NoStop}%
\bibitem [{\citenamefont {Song}\ \emph {et~al.}(2023)\citenamefont {Song}, \citenamefont {Jia}, \citenamefont {Xiong}, \citenamefont {Wang}, \citenamefont {Jiang}, \citenamefont {Huang}, \citenamefont {Hwang}, \citenamefont {Li}, \citenamefont {Hwang}, \citenamefont {Liu} \emph {et~al.}}]{song2023signatures}%
  \BibitemOpen
  \bibfield  {author} {\bibinfo {author} {\bibfnamefont {Y.}~\bibnamefont {Song}}, \bibinfo {author} {\bibfnamefont {C.}~\bibnamefont {Jia}}, \bibinfo {author} {\bibfnamefont {H.}~\bibnamefont {Xiong}}, \bibinfo {author} {\bibfnamefont {B.}~\bibnamefont {Wang}}, \bibinfo {author} {\bibfnamefont {Z.}~\bibnamefont {Jiang}}, \bibinfo {author} {\bibfnamefont {K.}~\bibnamefont {Huang}}, \bibinfo {author} {\bibfnamefont {J.}~\bibnamefont {Hwang}}, \bibinfo {author} {\bibfnamefont {Z.}~\bibnamefont {Li}}, \bibinfo {author} {\bibfnamefont {C.}~\bibnamefont {Hwang}}, \bibinfo {author} {\bibfnamefont {Z.}~\bibnamefont {Liu}}, \emph {et~al.},\ }\bibfield  {title} {\bibinfo {title} {Signatures of the exciton gas phase and its condensation in monolayer 1\uppercase{T}-\uppercase{Z}r\uppercase{T}e$_2$},\ }\href@noop {} {\bibfield  {journal} {\bibinfo  {journal} {Nature communications}\ }\textbf {\bibinfo {volume} {14}},\ \bibinfo {pages} {1116} (\bibinfo {year} {2023})}\BibitemShut {NoStop}%
\bibitem [{\citenamefont {Gao}\ \emph {et~al.}(2023)\citenamefont {Gao}, \citenamefont {Chan}, \citenamefont {Wang}, \citenamefont {Zhang}, \citenamefont {Jinxu}, \citenamefont {Cui}, \citenamefont {Yang}, \citenamefont {Liu}, \citenamefont {Shen}, \citenamefont {Sun} \emph {et~al.}}]{gao2023evidence}%
  \BibitemOpen
  \bibfield  {author} {\bibinfo {author} {\bibfnamefont {Q.}~\bibnamefont {Gao}}, \bibinfo {author} {\bibfnamefont {Y.-h.}\ \bibnamefont {Chan}}, \bibinfo {author} {\bibfnamefont {Y.}~\bibnamefont {Wang}}, \bibinfo {author} {\bibfnamefont {H.}~\bibnamefont {Zhang}}, \bibinfo {author} {\bibfnamefont {P.}~\bibnamefont {Jinxu}}, \bibinfo {author} {\bibfnamefont {S.}~\bibnamefont {Cui}}, \bibinfo {author} {\bibfnamefont {Y.}~\bibnamefont {Yang}}, \bibinfo {author} {\bibfnamefont {Z.}~\bibnamefont {Liu}}, \bibinfo {author} {\bibfnamefont {D.}~\bibnamefont {Shen}}, \bibinfo {author} {\bibfnamefont {Z.}~\bibnamefont {Sun}}, \emph {et~al.},\ }\bibfield  {title} {\bibinfo {title} {Evidence of high-temperature exciton condensation in a two-dimensional semimetal},\ }\href@noop {} {\bibfield  {journal} {\bibinfo  {journal} {Nature Communications}\ }\textbf {\bibinfo {volume} {14}},\ \bibinfo {pages} {994} (\bibinfo {year} {2023})}\BibitemShut {NoStop}%
\bibitem [{\citenamefont {Gao}\ \emph {et~al.}(2024)\citenamefont {Gao}, \citenamefont {Chan}, \citenamefont {Jiao}, \citenamefont {Chen}, \citenamefont {Yin}, \citenamefont {Tangprapha}, \citenamefont {Yang}, \citenamefont {Li}, \citenamefont {Liu}, \citenamefont {Shen} \emph {et~al.}}]{gao2024observation}%
  \BibitemOpen
  \bibfield  {author} {\bibinfo {author} {\bibfnamefont {Q.}~\bibnamefont {Gao}}, \bibinfo {author} {\bibfnamefont {Y.-h.}\ \bibnamefont {Chan}}, \bibinfo {author} {\bibfnamefont {P.}~\bibnamefont {Jiao}}, \bibinfo {author} {\bibfnamefont {H.}~\bibnamefont {Chen}}, \bibinfo {author} {\bibfnamefont {S.}~\bibnamefont {Yin}}, \bibinfo {author} {\bibfnamefont {K.}~\bibnamefont {Tangprapha}}, \bibinfo {author} {\bibfnamefont {Y.}~\bibnamefont {Yang}}, \bibinfo {author} {\bibfnamefont {X.}~\bibnamefont {Li}}, \bibinfo {author} {\bibfnamefont {Z.}~\bibnamefont {Liu}}, \bibinfo {author} {\bibfnamefont {D.}~\bibnamefont {Shen}}, \emph {et~al.},\ }\bibfield  {title} {\bibinfo {title} {Observation of possible excitonic charge density waves and metal--insulator transitions in atomically thin semimetals},\ }\href@noop {} {\bibfield  {journal} {\bibinfo  {journal} {Nature Physics}\ ,\ \bibinfo {pages} {1}} (\bibinfo {year} {2024})}\BibitemShut {NoStop}%
\bibitem [{\citenamefont {Kwan}\ \emph {et~al.}(2021)\citenamefont {Kwan}, \citenamefont {Devakul}, \citenamefont {Sondhi},\ and\ \citenamefont {Parameswaran}}]{kwan2021theory}%
  \BibitemOpen
  \bibfield  {author} {\bibinfo {author} {\bibfnamefont {Y.~H.}\ \bibnamefont {Kwan}}, \bibinfo {author} {\bibfnamefont {T.}~\bibnamefont {Devakul}}, \bibinfo {author} {\bibfnamefont {S.}~\bibnamefont {Sondhi}},\ and\ \bibinfo {author} {\bibfnamefont {S.}~\bibnamefont {Parameswaran}},\ }\bibfield  {title} {\bibinfo {title} {Theory of competing excitonic orders in insulating \uppercase{WT}e$_2$ monolayers},\ }\href@noop {} {\bibfield  {journal} {\bibinfo  {journal} {Physical Review B}\ }\textbf {\bibinfo {volume} {104}},\ \bibinfo {pages} {125133} (\bibinfo {year} {2021})}\BibitemShut {NoStop}%
\bibitem [{\citenamefont {Jia}\ \emph {et~al.}(2022)\citenamefont {Jia}, \citenamefont {Wang}, \citenamefont {Chiu}, \citenamefont {Song}, \citenamefont {Yu}, \citenamefont {J{\"a}ck}, \citenamefont {Lei}, \citenamefont {Klemenz}, \citenamefont {Cevallos}, \citenamefont {Onyszczak} \emph {et~al.}}]{jia2022evidence}%
  \BibitemOpen
  \bibfield  {author} {\bibinfo {author} {\bibfnamefont {Y.}~\bibnamefont {Jia}}, \bibinfo {author} {\bibfnamefont {P.}~\bibnamefont {Wang}}, \bibinfo {author} {\bibfnamefont {C.-L.}\ \bibnamefont {Chiu}}, \bibinfo {author} {\bibfnamefont {Z.}~\bibnamefont {Song}}, \bibinfo {author} {\bibfnamefont {G.}~\bibnamefont {Yu}}, \bibinfo {author} {\bibfnamefont {B.}~\bibnamefont {J{\"a}ck}}, \bibinfo {author} {\bibfnamefont {S.}~\bibnamefont {Lei}}, \bibinfo {author} {\bibfnamefont {S.}~\bibnamefont {Klemenz}}, \bibinfo {author} {\bibfnamefont {F.~A.}\ \bibnamefont {Cevallos}}, \bibinfo {author} {\bibfnamefont {M.}~\bibnamefont {Onyszczak}}, \emph {et~al.},\ }\bibfield  {title} {\bibinfo {title} {Evidence for a monolayer excitonic insulator},\ }\href@noop {} {\bibfield  {journal} {\bibinfo  {journal} {Nature Physics}\ }\textbf {\bibinfo {volume} {18}},\ \bibinfo {pages} {87} (\bibinfo {year} {2022})}\BibitemShut {NoStop}%
\bibitem [{\citenamefont {Sun}\ \emph {et~al.}(2022)\citenamefont {Sun}, \citenamefont {Zhao}, \citenamefont {Palomaki}, \citenamefont {Fei}, \citenamefont {Runburg}, \citenamefont {Malinowski}, \citenamefont {Huang}, \citenamefont {Cenker}, \citenamefont {Cui}, \citenamefont {Chu} \emph {et~al.}}]{sun2022evidence}%
  \BibitemOpen
  \bibfield  {author} {\bibinfo {author} {\bibfnamefont {B.}~\bibnamefont {Sun}}, \bibinfo {author} {\bibfnamefont {W.}~\bibnamefont {Zhao}}, \bibinfo {author} {\bibfnamefont {T.}~\bibnamefont {Palomaki}}, \bibinfo {author} {\bibfnamefont {Z.}~\bibnamefont {Fei}}, \bibinfo {author} {\bibfnamefont {E.}~\bibnamefont {Runburg}}, \bibinfo {author} {\bibfnamefont {P.}~\bibnamefont {Malinowski}}, \bibinfo {author} {\bibfnamefont {X.}~\bibnamefont {Huang}}, \bibinfo {author} {\bibfnamefont {J.}~\bibnamefont {Cenker}}, \bibinfo {author} {\bibfnamefont {Y.-T.}\ \bibnamefont {Cui}}, \bibinfo {author} {\bibfnamefont {J.-H.}\ \bibnamefont {Chu}}, \emph {et~al.},\ }\bibfield  {title} {\bibinfo {title} {Evidence for equilibrium exciton condensation in monolayer \uppercase{WT}e$_2$},\ }\href@noop {} {\bibfield  {journal} {\bibinfo  {journal} {Nature Physics}\ }\textbf {\bibinfo {volume} {18}},\ \bibinfo {pages} {94} (\bibinfo {year} {2022})}\BibitemShut {NoStop}%
\bibitem [{\citenamefont {Que}\ \emph {et~al.}(2024)\citenamefont {Que}, \citenamefont {Chan}, \citenamefont {Jia}, \citenamefont {Das}, \citenamefont {Tong}, \citenamefont {Chang}, \citenamefont {Cui}, \citenamefont {Kumar}, \citenamefont {Singh}, \citenamefont {Mukherjee} \emph {et~al.}}]{que2024gate}%
  \BibitemOpen
  \bibfield  {author} {\bibinfo {author} {\bibfnamefont {Y.}~\bibnamefont {Que}}, \bibinfo {author} {\bibfnamefont {Y.-H.}\ \bibnamefont {Chan}}, \bibinfo {author} {\bibfnamefont {J.}~\bibnamefont {Jia}}, \bibinfo {author} {\bibfnamefont {A.}~\bibnamefont {Das}}, \bibinfo {author} {\bibfnamefont {Z.}~\bibnamefont {Tong}}, \bibinfo {author} {\bibfnamefont {Y.-T.}\ \bibnamefont {Chang}}, \bibinfo {author} {\bibfnamefont {Z.}~\bibnamefont {Cui}}, \bibinfo {author} {\bibfnamefont {A.}~\bibnamefont {Kumar}}, \bibinfo {author} {\bibfnamefont {G.}~\bibnamefont {Singh}}, \bibinfo {author} {\bibfnamefont {S.}~\bibnamefont {Mukherjee}}, \emph {et~al.},\ }\bibfield  {title} {\bibinfo {title} {A gate-tunable ambipolar quantum phase transition in a topological excitonic insulator},\ }\href@noop {} {\bibfield  {journal} {\bibinfo  {journal} {Advanced Materials}\ }\textbf {\bibinfo {volume} {36}},\ \bibinfo {pages} {2309356} (\bibinfo {year} {2024})}\BibitemShut {NoStop}%
\bibitem [{\citenamefont {Rossnagel}\ \emph {et~al.}(2002)\citenamefont {Rossnagel}, \citenamefont {Kipp},\ and\ \citenamefont {Skibowski}}]{rossnagel2002charge}%
  \BibitemOpen
  \bibfield  {author} {\bibinfo {author} {\bibfnamefont {K.}~\bibnamefont {Rossnagel}}, \bibinfo {author} {\bibfnamefont {L.}~\bibnamefont {Kipp}},\ and\ \bibinfo {author} {\bibfnamefont {M.}~\bibnamefont {Skibowski}},\ }\bibfield  {title} {\bibinfo {title} {Charge-density-wave phase transition in 1\uppercase{T}-\uppercase{T}i\uppercase{S}e$_2$: Excitonic insulator versus band-type \uppercase{J}ahn-\uppercase{T}eller mechanism},\ }\href@noop {} {\bibfield  {journal} {\bibinfo  {journal} {Physical Review B}\ }\textbf {\bibinfo {volume} {65}},\ \bibinfo {pages} {235101} (\bibinfo {year} {2002})}\BibitemShut {NoStop}%
\bibitem [{\citenamefont {Porer}\ \emph {et~al.}(2014)\citenamefont {Porer}, \citenamefont {Leierseder}, \citenamefont {M{\'e}nard}, \citenamefont {Dachraoui}, \citenamefont {Mouchliadis}, \citenamefont {Perakis}, \citenamefont {Heinzmann}, \citenamefont {Demsar}, \citenamefont {Rossnagel},\ and\ \citenamefont {Huber}}]{porer2014non}%
  \BibitemOpen
  \bibfield  {author} {\bibinfo {author} {\bibfnamefont {M.}~\bibnamefont {Porer}}, \bibinfo {author} {\bibfnamefont {U.}~\bibnamefont {Leierseder}}, \bibinfo {author} {\bibfnamefont {J.-M.}\ \bibnamefont {M{\'e}nard}}, \bibinfo {author} {\bibfnamefont {H.}~\bibnamefont {Dachraoui}}, \bibinfo {author} {\bibfnamefont {L.}~\bibnamefont {Mouchliadis}}, \bibinfo {author} {\bibfnamefont {I.}~\bibnamefont {Perakis}}, \bibinfo {author} {\bibfnamefont {U.}~\bibnamefont {Heinzmann}}, \bibinfo {author} {\bibfnamefont {J.}~\bibnamefont {Demsar}}, \bibinfo {author} {\bibfnamefont {K.}~\bibnamefont {Rossnagel}},\ and\ \bibinfo {author} {\bibfnamefont {R.}~\bibnamefont {Huber}},\ }\bibfield  {title} {\bibinfo {title} {Non-thermal separation of electronic and structural orders in a persisting charge density wave},\ }\href@noop {} {\bibfield  {journal} {\bibinfo  {journal} {Nature materials}\ }\textbf {\bibinfo {volume} {13}},\ \bibinfo {pages} {857} (\bibinfo {year} {2014})}\BibitemShut {NoStop}%
\bibitem [{\citenamefont {Wegner}\ \emph {et~al.}(2020)\citenamefont {Wegner}, \citenamefont {Zhao}, \citenamefont {Li}, \citenamefont {Yang}, \citenamefont {Anikin}, \citenamefont {Karapetrov}, \citenamefont {Esfarjani}, \citenamefont {Louca},\ and\ \citenamefont {Chatterjee}}]{wegner2020evidence}%
  \BibitemOpen
  \bibfield  {author} {\bibinfo {author} {\bibfnamefont {A.}~\bibnamefont {Wegner}}, \bibinfo {author} {\bibfnamefont {J.}~\bibnamefont {Zhao}}, \bibinfo {author} {\bibfnamefont {J.}~\bibnamefont {Li}}, \bibinfo {author} {\bibfnamefont {J.}~\bibnamefont {Yang}}, \bibinfo {author} {\bibfnamefont {A.}~\bibnamefont {Anikin}}, \bibinfo {author} {\bibfnamefont {G.}~\bibnamefont {Karapetrov}}, \bibinfo {author} {\bibfnamefont {K.}~\bibnamefont {Esfarjani}}, \bibinfo {author} {\bibfnamefont {D.}~\bibnamefont {Louca}},\ and\ \bibinfo {author} {\bibfnamefont {U.}~\bibnamefont {Chatterjee}},\ }\bibfield  {title} {\bibinfo {title} {Evidence for pseudo--\uppercase{J}ahn-\uppercase{T}eller distortions in the charge density wave phase of 1\uppercase{T}-\uppercase{T}i\uppercase{S}e$_2$},\ }\href@noop {} {\bibfield  {journal} {\bibinfo  {journal} {Physical Review B}\ }\textbf {\bibinfo {volume} {101}},\ \bibinfo {pages} {195145} (\bibinfo {year} {2020})}\BibitemShut {NoStop}%
\bibitem [{\citenamefont {Jeong}\ \emph {et~al.}(2024)\citenamefont {Jeong}, \citenamefont {Kim}, \citenamefont {Jin}, \citenamefont {Kim},\ and\ \citenamefont {Yeom}}]{jeong2024dichotomy}%
  \BibitemOpen
  \bibfield  {author} {\bibinfo {author} {\bibfnamefont {D.}~\bibnamefont {Jeong}}, \bibinfo {author} {\bibfnamefont {J.}~\bibnamefont {Kim}}, \bibinfo {author} {\bibfnamefont {K.-H.}\ \bibnamefont {Jin}}, \bibinfo {author} {\bibfnamefont {J.}~\bibnamefont {Kim}},\ and\ \bibinfo {author} {\bibfnamefont {H.~W.}\ \bibnamefont {Yeom}},\ }\bibfield  {title} {\bibinfo {title} {Dichotomy of metallic electron density and charge density wave in 1\uppercase{T}-\uppercase{T}i\uppercase{S}e$_2$},\ }\href@noop {} {\bibfield  {journal} {\bibinfo  {journal} {Physical Review B}\ }\textbf {\bibinfo {volume} {109}},\ \bibinfo {pages} {125117} (\bibinfo {year} {2024})}\BibitemShut {NoStop}%
\bibitem [{\citenamefont {Watson}\ \emph {et~al.}(2020)\citenamefont {Watson}, \citenamefont {Markovi{\'c}}, \citenamefont {Morales}, \citenamefont {Le~F{\`e}vre}, \citenamefont {Merz}, \citenamefont {Haghighirad},\ and\ \citenamefont {King}}]{watson2020band}%
  \BibitemOpen
  \bibfield  {author} {\bibinfo {author} {\bibfnamefont {M.~D.}\ \bibnamefont {Watson}}, \bibinfo {author} {\bibfnamefont {I.}~\bibnamefont {Markovi{\'c}}}, \bibinfo {author} {\bibfnamefont {E.~A.}\ \bibnamefont {Morales}}, \bibinfo {author} {\bibfnamefont {P.}~\bibnamefont {Le~F{\`e}vre}}, \bibinfo {author} {\bibfnamefont {M.}~\bibnamefont {Merz}}, \bibinfo {author} {\bibfnamefont {A.~A.}\ \bibnamefont {Haghighirad}},\ and\ \bibinfo {author} {\bibfnamefont {P.~D.}\ \bibnamefont {King}},\ }\bibfield  {title} {\bibinfo {title} {Band hybridization at the semimetal-semiconductor transition of \uppercase{T}a$_2$\uppercase{N}i\uppercase{S}e$_5$ enabled by mirror-symmetry breaking},\ }\href@noop {} {\bibfield  {journal} {\bibinfo  {journal} {Physical Review Research}\ }\textbf {\bibinfo {volume} {2}},\ \bibinfo {pages} {013236} (\bibinfo {year} {2020})}\BibitemShut {NoStop}%
\bibitem [{\citenamefont {Windg{\"a}tter}\ \emph {et~al.}(2021)\citenamefont {Windg{\"a}tter}, \citenamefont {R{\"o}sner}, \citenamefont {Mazza}, \citenamefont {H{\"u}bener}, \citenamefont {Georges}, \citenamefont {Millis}, \citenamefont {Latini},\ and\ \citenamefont {Rubio}}]{windgatter2021common}%
  \BibitemOpen
  \bibfield  {author} {\bibinfo {author} {\bibfnamefont {L.}~\bibnamefont {Windg{\"a}tter}}, \bibinfo {author} {\bibfnamefont {M.}~\bibnamefont {R{\"o}sner}}, \bibinfo {author} {\bibfnamefont {G.}~\bibnamefont {Mazza}}, \bibinfo {author} {\bibfnamefont {H.}~\bibnamefont {H{\"u}bener}}, \bibinfo {author} {\bibfnamefont {A.}~\bibnamefont {Georges}}, \bibinfo {author} {\bibfnamefont {A.~J.}\ \bibnamefont {Millis}}, \bibinfo {author} {\bibfnamefont {S.}~\bibnamefont {Latini}},\ and\ \bibinfo {author} {\bibfnamefont {A.}~\bibnamefont {Rubio}},\ }\bibfield  {title} {\bibinfo {title} {Common \uppercase{M}icroscopic \uppercase{O}rigin of \uppercase{T}he \uppercase{P}hase \uppercase{T}ransitions in \uppercase{T}a$_2$\uppercase{N}i\uppercase{S}$_5$ and \uppercase{T}he \uppercase{E}xcitonic \uppercase{I}nsulator \uppercase{C}andidate \uppercase{T}a$_2$\uppercase{N}i\uppercase{S}e$_5$},\ }\href@noop {} {\bibfield  {journal} {\bibinfo  {journal} {npj Computational Materials}\ }\textbf {\bibinfo {volume} {7}},\ \bibinfo
  {pages} {210} (\bibinfo {year} {2021})}\BibitemShut {NoStop}%
\bibitem [{\citenamefont {Lu}\ \emph {et~al.}(2021)\citenamefont {Lu}, \citenamefont {Rossi}, \citenamefont {Kim}, \citenamefont {Yavas}, \citenamefont {Said}, \citenamefont {Nag}, \citenamefont {Garcia-Fernandez}, \citenamefont {Agrestini}, \citenamefont {Zhou}, \citenamefont {Jia} \emph {et~al.}}]{lu2021evolution}%
  \BibitemOpen
  \bibfield  {author} {\bibinfo {author} {\bibfnamefont {H.}~\bibnamefont {Lu}}, \bibinfo {author} {\bibfnamefont {M.}~\bibnamefont {Rossi}}, \bibinfo {author} {\bibfnamefont {J.-h.}\ \bibnamefont {Kim}}, \bibinfo {author} {\bibfnamefont {H.}~\bibnamefont {Yavas}}, \bibinfo {author} {\bibfnamefont {A.}~\bibnamefont {Said}}, \bibinfo {author} {\bibfnamefont {A.}~\bibnamefont {Nag}}, \bibinfo {author} {\bibfnamefont {M.}~\bibnamefont {Garcia-Fernandez}}, \bibinfo {author} {\bibfnamefont {S.}~\bibnamefont {Agrestini}}, \bibinfo {author} {\bibfnamefont {K.-J.}\ \bibnamefont {Zhou}}, \bibinfo {author} {\bibfnamefont {C.}~\bibnamefont {Jia}}, \emph {et~al.},\ }\bibfield  {title} {\bibinfo {title} {Evolution of the electronic structure in \uppercase{T}a$_2$\uppercase{N}i\uppercase{S}e$_5$ across the structural transition revealed by resonant inelastic x-ray scattering},\ }\href@noop {} {\bibfield  {journal} {\bibinfo  {journal} {Physical Review B}\ }\textbf {\bibinfo {volume} {103}},\ \bibinfo {pages} {235159}
  (\bibinfo {year} {2021})}\BibitemShut {NoStop}%
\bibitem [{\citenamefont {Kaneko}\ \emph {et~al.}(2013)\citenamefont {Kaneko}, \citenamefont {Toriyama}, \citenamefont {Konishi},\ and\ \citenamefont {Ohta}}]{kaneko2013orthorhombic}%
  \BibitemOpen
  \bibfield  {author} {\bibinfo {author} {\bibfnamefont {T.}~\bibnamefont {Kaneko}}, \bibinfo {author} {\bibfnamefont {T.}~\bibnamefont {Toriyama}}, \bibinfo {author} {\bibfnamefont {T.}~\bibnamefont {Konishi}},\ and\ \bibinfo {author} {\bibfnamefont {Y.}~\bibnamefont {Ohta}},\ }\bibfield  {title} {\bibinfo {title} {Orthorhombic-to-monoclinic phase transition of \uppercase{T}a$_2$\uppercase{N}i\uppercase{S}e$_5$ induced by the \uppercase{B}ose-\uppercase{E}instein condensation of excitons},\ }\href@noop {} {\bibfield  {journal} {\bibinfo  {journal} {Physical Review B—Condensed Matter and Materials Physics}\ }\textbf {\bibinfo {volume} {87}},\ \bibinfo {pages} {035121} (\bibinfo {year} {2013})}\BibitemShut {NoStop}%
\bibitem [{\citenamefont {Mazza}\ \emph {et~al.}(2020)\citenamefont {Mazza}, \citenamefont {R{\"o}sner}, \citenamefont {Windg{\"a}tter}, \citenamefont {Latini}, \citenamefont {H{\"u}bener}, \citenamefont {Millis}, \citenamefont {Rubio},\ and\ \citenamefont {Georges}}]{mazza2020nature}%
  \BibitemOpen
  \bibfield  {author} {\bibinfo {author} {\bibfnamefont {G.}~\bibnamefont {Mazza}}, \bibinfo {author} {\bibfnamefont {M.}~\bibnamefont {R{\"o}sner}}, \bibinfo {author} {\bibfnamefont {L.}~\bibnamefont {Windg{\"a}tter}}, \bibinfo {author} {\bibfnamefont {S.}~\bibnamefont {Latini}}, \bibinfo {author} {\bibfnamefont {H.}~\bibnamefont {H{\"u}bener}}, \bibinfo {author} {\bibfnamefont {A.~J.}\ \bibnamefont {Millis}}, \bibinfo {author} {\bibfnamefont {A.}~\bibnamefont {Rubio}},\ and\ \bibinfo {author} {\bibfnamefont {A.}~\bibnamefont {Georges}},\ }\bibfield  {title} {\bibinfo {title} {Nature of symmetry breaking at the excitonic insulator transition: \uppercase{T}a$_2$\uppercase{N}i\uppercase{S}e$_5$},\ }\href@noop {} {\bibfield  {journal} {\bibinfo  {journal} {Physical review letters}\ }\textbf {\bibinfo {volume} {124}},\ \bibinfo {pages} {197601} (\bibinfo {year} {2020})}\BibitemShut {NoStop}%
\bibitem [{\citenamefont {Scholes}\ and\ \citenamefont {Rumbles}(2006)}]{scholes2006excitons}%
  \BibitemOpen
  \bibfield  {author} {\bibinfo {author} {\bibfnamefont {G.~D.}\ \bibnamefont {Scholes}}\ and\ \bibinfo {author} {\bibfnamefont {G.}~\bibnamefont {Rumbles}},\ }\bibfield  {title} {\bibinfo {title} {Excitons in nanoscale systems},\ }\href@noop {} {\bibfield  {journal} {\bibinfo  {journal} {Nature materials}\ }\textbf {\bibinfo {volume} {5}},\ \bibinfo {pages} {683} (\bibinfo {year} {2006})}\BibitemShut {NoStop}%
\bibitem [{\citenamefont {Cui}\ \emph {et~al.}(2014)\citenamefont {Cui}, \citenamefont {Wang}, \citenamefont {Argondizzo}, \citenamefont {Garrett-Roe}, \citenamefont {Gumhalter},\ and\ \citenamefont {Petek}}]{cui2014transient}%
  \BibitemOpen
  \bibfield  {author} {\bibinfo {author} {\bibfnamefont {X.}~\bibnamefont {Cui}}, \bibinfo {author} {\bibfnamefont {C.}~\bibnamefont {Wang}}, \bibinfo {author} {\bibfnamefont {A.}~\bibnamefont {Argondizzo}}, \bibinfo {author} {\bibfnamefont {S.}~\bibnamefont {Garrett-Roe}}, \bibinfo {author} {\bibfnamefont {B.}~\bibnamefont {Gumhalter}},\ and\ \bibinfo {author} {\bibfnamefont {H.}~\bibnamefont {Petek}},\ }\bibfield  {title} {\bibinfo {title} {Transient excitons at metal surfaces},\ }\href@noop {} {\bibfield  {journal} {\bibinfo  {journal} {Nature Physics}\ }\textbf {\bibinfo {volume} {10}},\ \bibinfo {pages} {505} (\bibinfo {year} {2014})}\BibitemShut {NoStop}%
\bibitem [{\citenamefont {Byrnes}\ \emph {et~al.}(2014)\citenamefont {Byrnes}, \citenamefont {Kim},\ and\ \citenamefont {Yamamoto}}]{byrnes2014exciton}%
  \BibitemOpen
  \bibfield  {author} {\bibinfo {author} {\bibfnamefont {T.}~\bibnamefont {Byrnes}}, \bibinfo {author} {\bibfnamefont {N.~Y.}\ \bibnamefont {Kim}},\ and\ \bibinfo {author} {\bibfnamefont {Y.}~\bibnamefont {Yamamoto}},\ }\bibfield  {title} {\bibinfo {title} {Exciton--polariton condensates},\ }\href@noop {} {\bibfield  {journal} {\bibinfo  {journal} {Nature Physics}\ }\textbf {\bibinfo {volume} {10}},\ \bibinfo {pages} {803} (\bibinfo {year} {2014})}\BibitemShut {NoStop}%
\bibitem [{\citenamefont {Tanimura}\ \emph {et~al.}(2019)\citenamefont {Tanimura}, \citenamefont {Tanimura},\ and\ \citenamefont {Van~Loosdrecht}}]{tanimura2019dynamics}%
  \BibitemOpen
  \bibfield  {author} {\bibinfo {author} {\bibfnamefont {H.}~\bibnamefont {Tanimura}}, \bibinfo {author} {\bibfnamefont {K.}~\bibnamefont {Tanimura}},\ and\ \bibinfo {author} {\bibfnamefont {P.}~\bibnamefont {Van~Loosdrecht}},\ }\bibfield  {title} {\bibinfo {title} {Dynamics of incoherent exciton formation in \uppercase{C}u$_2$\uppercase{O}: Time-and angle-resolved photoemission spectroscopy},\ }\href@noop {} {\bibfield  {journal} {\bibinfo  {journal} {Physical Review B}\ }\textbf {\bibinfo {volume} {100}},\ \bibinfo {pages} {115204} (\bibinfo {year} {2019})}\BibitemShut {NoStop}%
\bibitem [{\citenamefont {Mad{\'e}o}\ \emph {et~al.}(2020)\citenamefont {Mad{\'e}o}, \citenamefont {Man}, \citenamefont {Sahoo}, \citenamefont {Campbell}, \citenamefont {Pareek}, \citenamefont {Wong}, \citenamefont {Al-Mahboob}, \citenamefont {Chan}, \citenamefont {Karmakar}, \citenamefont {Mariserla} \emph {et~al.}}]{madeo2020directly}%
  \BibitemOpen
  \bibfield  {author} {\bibinfo {author} {\bibfnamefont {J.}~\bibnamefont {Mad{\'e}o}}, \bibinfo {author} {\bibfnamefont {M.~K.}\ \bibnamefont {Man}}, \bibinfo {author} {\bibfnamefont {C.}~\bibnamefont {Sahoo}}, \bibinfo {author} {\bibfnamefont {M.}~\bibnamefont {Campbell}}, \bibinfo {author} {\bibfnamefont {V.}~\bibnamefont {Pareek}}, \bibinfo {author} {\bibfnamefont {E.~L.}\ \bibnamefont {Wong}}, \bibinfo {author} {\bibfnamefont {A.}~\bibnamefont {Al-Mahboob}}, \bibinfo {author} {\bibfnamefont {N.~S.}\ \bibnamefont {Chan}}, \bibinfo {author} {\bibfnamefont {A.}~\bibnamefont {Karmakar}}, \bibinfo {author} {\bibfnamefont {B.~M.~K.}\ \bibnamefont {Mariserla}}, \emph {et~al.},\ }\bibfield  {title} {\bibinfo {title} {Directly visualizing the momentum-forbidden dark excitons and their dynamics in atomically thin semiconductors},\ }\href@noop {} {\bibfield  {journal} {\bibinfo  {journal} {Science}\ }\textbf {\bibinfo {volume} {370}},\ \bibinfo {pages} {1199} (\bibinfo {year} {2020})}\BibitemShut {NoStop}%
\bibitem [{\citenamefont {Man}\ \emph {et~al.}(2021)\citenamefont {Man}, \citenamefont {Mad{\'e}o}, \citenamefont {Sahoo}, \citenamefont {Xie}, \citenamefont {Campbell}, \citenamefont {Pareek}, \citenamefont {Karmakar}, \citenamefont {Wong}, \citenamefont {Al-Mahboob}, \citenamefont {Chan} \emph {et~al.}}]{man2021experimental}%
  \BibitemOpen
  \bibfield  {author} {\bibinfo {author} {\bibfnamefont {M.~K.}\ \bibnamefont {Man}}, \bibinfo {author} {\bibfnamefont {J.}~\bibnamefont {Mad{\'e}o}}, \bibinfo {author} {\bibfnamefont {C.}~\bibnamefont {Sahoo}}, \bibinfo {author} {\bibfnamefont {K.}~\bibnamefont {Xie}}, \bibinfo {author} {\bibfnamefont {M.}~\bibnamefont {Campbell}}, \bibinfo {author} {\bibfnamefont {V.}~\bibnamefont {Pareek}}, \bibinfo {author} {\bibfnamefont {A.}~\bibnamefont {Karmakar}}, \bibinfo {author} {\bibfnamefont {E.~L.}\ \bibnamefont {Wong}}, \bibinfo {author} {\bibfnamefont {A.}~\bibnamefont {Al-Mahboob}}, \bibinfo {author} {\bibfnamefont {N.~S.}\ \bibnamefont {Chan}}, \emph {et~al.},\ }\bibfield  {title} {\bibinfo {title} {Experimental measurement of the intrinsic excitonic wave function},\ }\href@noop {} {\bibfield  {journal} {\bibinfo  {journal} {Science advances}\ }\textbf {\bibinfo {volume} {7}},\ \bibinfo {pages} {eabg0192} (\bibinfo {year} {2021})}\BibitemShut {NoStop}%
\bibitem [{\citenamefont {Karni}\ \emph {et~al.}(2022)\citenamefont {Karni}, \citenamefont {Barr{\'e}}, \citenamefont {Pareek}, \citenamefont {Georgaras}, \citenamefont {Man}, \citenamefont {Sahoo}, \citenamefont {Bacon}, \citenamefont {Zhu}, \citenamefont {Ribeiro}, \citenamefont {O’Beirne} \emph {et~al.}}]{karni2022structure}%
  \BibitemOpen
  \bibfield  {author} {\bibinfo {author} {\bibfnamefont {O.}~\bibnamefont {Karni}}, \bibinfo {author} {\bibfnamefont {E.}~\bibnamefont {Barr{\'e}}}, \bibinfo {author} {\bibfnamefont {V.}~\bibnamefont {Pareek}}, \bibinfo {author} {\bibfnamefont {J.~D.}\ \bibnamefont {Georgaras}}, \bibinfo {author} {\bibfnamefont {M.~K.}\ \bibnamefont {Man}}, \bibinfo {author} {\bibfnamefont {C.}~\bibnamefont {Sahoo}}, \bibinfo {author} {\bibfnamefont {D.~R.}\ \bibnamefont {Bacon}}, \bibinfo {author} {\bibfnamefont {X.}~\bibnamefont {Zhu}}, \bibinfo {author} {\bibfnamefont {H.~B.}\ \bibnamefont {Ribeiro}}, \bibinfo {author} {\bibfnamefont {A.~L.}\ \bibnamefont {O’Beirne}}, \emph {et~al.},\ }\bibfield  {title} {\bibinfo {title} {Structure of the moir{\'e} exciton captured by imaging its electron and hole},\ }\href@noop {} {\bibfield  {journal} {\bibinfo  {journal} {Nature}\ }\textbf {\bibinfo {volume} {603}},\ \bibinfo {pages} {247} (\bibinfo {year} {2022})}\BibitemShut {NoStop}%
\bibitem [{\citenamefont {Mori}\ \emph {et~al.}(2023)\citenamefont {Mori}, \citenamefont {Ciocys}, \citenamefont {Takasan}, \citenamefont {Ai}, \citenamefont {Currier}, \citenamefont {Morimoto}, \citenamefont {Moore},\ and\ \citenamefont {Lanzara}}]{mori2023spin}%
  \BibitemOpen
  \bibfield  {author} {\bibinfo {author} {\bibfnamefont {R.}~\bibnamefont {Mori}}, \bibinfo {author} {\bibfnamefont {S.}~\bibnamefont {Ciocys}}, \bibinfo {author} {\bibfnamefont {K.}~\bibnamefont {Takasan}}, \bibinfo {author} {\bibfnamefont {P.}~\bibnamefont {Ai}}, \bibinfo {author} {\bibfnamefont {K.}~\bibnamefont {Currier}}, \bibinfo {author} {\bibfnamefont {T.}~\bibnamefont {Morimoto}}, \bibinfo {author} {\bibfnamefont {J.~E.}\ \bibnamefont {Moore}},\ and\ \bibinfo {author} {\bibfnamefont {A.}~\bibnamefont {Lanzara}},\ }\bibfield  {title} {\bibinfo {title} {Spin-polarized spatially indirect excitons in a topological insulator},\ }\href@noop {} {\bibfield  {journal} {\bibinfo  {journal} {Nature}\ }\textbf {\bibinfo {volume} {614}},\ \bibinfo {pages} {249} (\bibinfo {year} {2023})}\BibitemShut {NoStop}%
\bibitem [{\citenamefont {Pareek}\ \emph {et~al.}(2024)\citenamefont {Pareek}, \citenamefont {Bacon}, \citenamefont {Zhu}, \citenamefont {Chan}, \citenamefont {Bussolotti}, \citenamefont {Chan}, \citenamefont {Urquizo}, \citenamefont {Watanabe}, \citenamefont {Taniguchi}, \citenamefont {Man} \emph {et~al.}}]{pareek2024driving}%
  \BibitemOpen
  \bibfield  {author} {\bibinfo {author} {\bibfnamefont {V.}~\bibnamefont {Pareek}}, \bibinfo {author} {\bibfnamefont {D.~R.}\ \bibnamefont {Bacon}}, \bibinfo {author} {\bibfnamefont {X.}~\bibnamefont {Zhu}}, \bibinfo {author} {\bibfnamefont {Y.-H.}\ \bibnamefont {Chan}}, \bibinfo {author} {\bibfnamefont {F.}~\bibnamefont {Bussolotti}}, \bibinfo {author} {\bibfnamefont {N.~S.}\ \bibnamefont {Chan}}, \bibinfo {author} {\bibfnamefont {J.~P.}\ \bibnamefont {Urquizo}}, \bibinfo {author} {\bibfnamefont {K.}~\bibnamefont {Watanabe}}, \bibinfo {author} {\bibfnamefont {T.}~\bibnamefont {Taniguchi}}, \bibinfo {author} {\bibfnamefont {M.~K.}\ \bibnamefont {Man}}, \emph {et~al.},\ }\href@noop {} {\bibfield  {journal} {\bibinfo  {journal} {(preprint) arXiv:2403.08725, submitted: Mar}\ } (\bibinfo {year} {2024})}\BibitemShut {NoStop}%
\bibitem [{\citenamefont {Rustagi}\ and\ \citenamefont {Kemper}(2018)}]{rustagi2018photoemission}%
  \BibitemOpen
  \bibfield  {author} {\bibinfo {author} {\bibfnamefont {A.}~\bibnamefont {Rustagi}}\ and\ \bibinfo {author} {\bibfnamefont {A.~F.}\ \bibnamefont {Kemper}},\ }\bibfield  {title} {\bibinfo {title} {Photoemission signature of excitons},\ }\href@noop {} {\bibfield  {journal} {\bibinfo  {journal} {Physical Review B}\ }\textbf {\bibinfo {volume} {97}},\ \bibinfo {pages} {235310} (\bibinfo {year} {2018})}\BibitemShut {NoStop}%
\bibitem [{\citenamefont {Fukutani}\ \emph {et~al.}(2021)\citenamefont {Fukutani}, \citenamefont {Stania}, \citenamefont {Il~Kwon}, \citenamefont {Kim}, \citenamefont {Kong}, \citenamefont {Kim},\ and\ \citenamefont {Yeom}}]{fukutani2021detecting}%
  \BibitemOpen
  \bibfield  {author} {\bibinfo {author} {\bibfnamefont {K.}~\bibnamefont {Fukutani}}, \bibinfo {author} {\bibfnamefont {R.}~\bibnamefont {Stania}}, \bibinfo {author} {\bibfnamefont {C.}~\bibnamefont {Il~Kwon}}, \bibinfo {author} {\bibfnamefont {J.~S.}\ \bibnamefont {Kim}}, \bibinfo {author} {\bibfnamefont {K.~J.}\ \bibnamefont {Kong}}, \bibinfo {author} {\bibfnamefont {J.}~\bibnamefont {Kim}},\ and\ \bibinfo {author} {\bibfnamefont {H.~W.}\ \bibnamefont {Yeom}},\ }\bibfield  {title} {\bibinfo {title} {Detecting photoelectrons from spontaneously formed excitons},\ }\href@noop {} {\bibfield  {journal} {\bibinfo  {journal} {Nature Physics}\ }\textbf {\bibinfo {volume} {17}},\ \bibinfo {pages} {1024} (\bibinfo {year} {2021})}\BibitemShut {NoStop}%
\bibitem [{\citenamefont {Hossain}\ \emph {et~al.}(2023)\citenamefont {Hossain}, \citenamefont {Cochran}, \citenamefont {Jiang}, \citenamefont {Zhang}, \citenamefont {Wu}, \citenamefont {Liu}, \citenamefont {Zheng}, \citenamefont {Kim}, \citenamefont {Cheng}, \citenamefont {Zhang} \emph {et~al.}}]{hossain2023discovery}%
  \BibitemOpen
  \bibfield  {author} {\bibinfo {author} {\bibfnamefont {M.~S.}\ \bibnamefont {Hossain}}, \bibinfo {author} {\bibfnamefont {T.~A.}\ \bibnamefont {Cochran}}, \bibinfo {author} {\bibfnamefont {Y.-X.}\ \bibnamefont {Jiang}}, \bibinfo {author} {\bibfnamefont {S.}~\bibnamefont {Zhang}}, \bibinfo {author} {\bibfnamefont {H.}~\bibnamefont {Wu}}, \bibinfo {author} {\bibfnamefont {X.}~\bibnamefont {Liu}}, \bibinfo {author} {\bibfnamefont {X.}~\bibnamefont {Zheng}}, \bibinfo {author} {\bibfnamefont {B.}~\bibnamefont {Kim}}, \bibinfo {author} {\bibfnamefont {G.}~\bibnamefont {Cheng}}, \bibinfo {author} {\bibfnamefont {Q.}~\bibnamefont {Zhang}}, \emph {et~al.},\ }\href@noop {} {\bibfield  {journal} {\bibinfo  {journal} {(preprint) arXiv:2312.15862, submitted: Dec}\ } (\bibinfo {year} {2023})}\BibitemShut {NoStop}%
\bibitem [{\citenamefont {Huang}\ \emph {et~al.}(2024)\citenamefont {Huang}, \citenamefont {Jiang}, \citenamefont {Yao}, \citenamefont {Yan}, \citenamefont {Lei}, \citenamefont {Gao}, \citenamefont {Guo}, \citenamefont {Jin}, \citenamefont {Li}, \citenamefont {Yuan} \emph {et~al.}}]{huang2024evidence}%
  \BibitemOpen
  \bibfield  {author} {\bibinfo {author} {\bibfnamefont {J.}~\bibnamefont {Huang}}, \bibinfo {author} {\bibfnamefont {B.}~\bibnamefont {Jiang}}, \bibinfo {author} {\bibfnamefont {J.}~\bibnamefont {Yao}}, \bibinfo {author} {\bibfnamefont {D.}~\bibnamefont {Yan}}, \bibinfo {author} {\bibfnamefont {X.}~\bibnamefont {Lei}}, \bibinfo {author} {\bibfnamefont {J.}~\bibnamefont {Gao}}, \bibinfo {author} {\bibfnamefont {Z.}~\bibnamefont {Guo}}, \bibinfo {author} {\bibfnamefont {F.}~\bibnamefont {Jin}}, \bibinfo {author} {\bibfnamefont {Y.}~\bibnamefont {Li}}, \bibinfo {author} {\bibfnamefont {Z.}~\bibnamefont {Yuan}}, \emph {et~al.},\ }\bibfield  {title} {\bibinfo {title} {Evidence for an excitonic insulator state in \uppercase{T}a$_2$\uppercase{P}d$_3$\uppercase{T}e$_5$},\ }\href@noop {} {\bibfield  {journal} {\bibinfo  {journal} {Physical Review X}\ }\textbf {\bibinfo {volume} {14}},\ \bibinfo {pages} {011046} (\bibinfo {year} {2024})}\BibitemShut {NoStop}%
\bibitem [{\citenamefont {Yao}\ \emph {et~al.}(2024)\citenamefont {Yao}, \citenamefont {Sheng}, \citenamefont {Zhang}, \citenamefont {Pang}, \citenamefont {Zhou}, \citenamefont {Wu}, \citenamefont {Weng}, \citenamefont {Dai}, \citenamefont {Fang},\ and\ \citenamefont {Wang}}]{yao2024excitonic}%
  \BibitemOpen
  \bibfield  {author} {\bibinfo {author} {\bibfnamefont {J.}~\bibnamefont {Yao}}, \bibinfo {author} {\bibfnamefont {H.}~\bibnamefont {Sheng}}, \bibinfo {author} {\bibfnamefont {R.}~\bibnamefont {Zhang}}, \bibinfo {author} {\bibfnamefont {R.}~\bibnamefont {Pang}}, \bibinfo {author} {\bibfnamefont {J.-J.}\ \bibnamefont {Zhou}}, \bibinfo {author} {\bibfnamefont {Q.}~\bibnamefont {Wu}}, \bibinfo {author} {\bibfnamefont {H.}~\bibnamefont {Weng}}, \bibinfo {author} {\bibfnamefont {X.}~\bibnamefont {Dai}}, \bibinfo {author} {\bibfnamefont {Z.}~\bibnamefont {Fang}},\ and\ \bibinfo {author} {\bibfnamefont {Z.}~\bibnamefont {Wang}},\ }\bibfield  {title} {\bibinfo {title} {Excitonic instability in \uppercase{T}a$_2$\uppercase{P}d$_3$\uppercase{T}e$_5$ monolayer},\ }\href@noop {} {\bibfield  {journal} {\bibinfo  {journal} {Chinese Physics Letters}\ }\textbf {\bibinfo {volume} {41}},\ \bibinfo {pages} {097101} (\bibinfo {year} {2024})}\BibitemShut {NoStop}%
\bibitem [{\citenamefont {Jiang}\ \emph {et~al.}(2024)\citenamefont {Jiang}, \citenamefont {Yao}, \citenamefont {Yan}, \citenamefont {Guo}, \citenamefont {Qu}, \citenamefont {Deng}, \citenamefont {Huang}, \citenamefont {Ding}, \citenamefont {Shi}, \citenamefont {Wang} \emph {et~al.}}]{jiang2024surface}%
  \BibitemOpen
  \bibfield  {author} {\bibinfo {author} {\bibfnamefont {B.}~\bibnamefont {Jiang}}, \bibinfo {author} {\bibfnamefont {J.}~\bibnamefont {Yao}}, \bibinfo {author} {\bibfnamefont {D.}~\bibnamefont {Yan}}, \bibinfo {author} {\bibfnamefont {Z.}~\bibnamefont {Guo}}, \bibinfo {author} {\bibfnamefont {G.}~\bibnamefont {Qu}}, \bibinfo {author} {\bibfnamefont {X.}~\bibnamefont {Deng}}, \bibinfo {author} {\bibfnamefont {Y.}~\bibnamefont {Huang}}, \bibinfo {author} {\bibfnamefont {H.}~\bibnamefont {Ding}}, \bibinfo {author} {\bibfnamefont {Y.}~\bibnamefont {Shi}}, \bibinfo {author} {\bibfnamefont {Z.}~\bibnamefont {Wang}}, \emph {et~al.},\ }\bibfield  {title} {\bibinfo {title} {Surface doping manipulation of the insulating ground states in \uppercase{T}a$_2$\uppercase{P}d$_3$\uppercase{T}e$_5$ and \uppercase{T}a$_2$\uppercase{N}i$_3$\uppercase{T}e$_5$},\ }\href@noop {} {\bibfield  {journal} {\bibinfo  {journal} {Chinese Physics B}\ } (\bibinfo {year} {2024})}\BibitemShut {NoStop}%
\bibitem [{\citenamefont {Zhang}\ \emph {et~al.}(2024)\citenamefont {Zhang}, \citenamefont {Dong}, \citenamefont {Yan}, \citenamefont {Jiang}, \citenamefont {Yang}, \citenamefont {Li}, \citenamefont {Guo}, \citenamefont {Huang}, \citenamefont {Li}, \citenamefont {Li} \emph {et~al.}}]{zhang2024spontaneous}%
  \BibitemOpen
  \bibfield  {author} {\bibinfo {author} {\bibfnamefont {P.}~\bibnamefont {Zhang}}, \bibinfo {author} {\bibfnamefont {Y.}~\bibnamefont {Dong}}, \bibinfo {author} {\bibfnamefont {D.}~\bibnamefont {Yan}}, \bibinfo {author} {\bibfnamefont {B.}~\bibnamefont {Jiang}}, \bibinfo {author} {\bibfnamefont {T.}~\bibnamefont {Yang}}, \bibinfo {author} {\bibfnamefont {J.}~\bibnamefont {Li}}, \bibinfo {author} {\bibfnamefont {Z.}~\bibnamefont {Guo}}, \bibinfo {author} {\bibfnamefont {Y.}~\bibnamefont {Huang}}, \bibinfo {author} {\bibfnamefont {Q.}~\bibnamefont {Li}}, \bibinfo {author} {\bibfnamefont {Y.}~\bibnamefont {Li}}, \emph {et~al.},\ }\bibfield  {title} {\bibinfo {title} {Spontaneous gap opening and potential excitonic states in an ideal dirac semimetal \uppercase{T}a$_2$\uppercase{P}d$_3$\uppercase{T}e$_5$},\ }\href@noop {} {\bibfield  {journal} {\bibinfo  {journal} {Physical Review X}\ }\textbf {\bibinfo {volume} {14}},\ \bibinfo {pages} {011047} (\bibinfo {year} {2024})}\BibitemShut {NoStop}%
\bibitem [{sup()}]{sup}%
  \BibitemOpen
  \href@noop {} {\emph {\bibinfo {title} {\textnormal{See Supplemental Material at http://link.aps.org/ supplemental/ for additional data and calulated desults, which includes Refs. [11,43-46,49,51,68]}}}}\BibitemShut {NoStop}%
\bibitem [{\citenamefont {Wang}\ \emph {et~al.}(2021)\citenamefont {Wang}, \citenamefont {Geng}, \citenamefont {Yan}, \citenamefont {Hu}, \citenamefont {Zhang}, \citenamefont {Yue}, \citenamefont {Sun}, \citenamefont {Kumar}, \citenamefont {Schwier}, \citenamefont {Shimada} \emph {et~al.}}]{wang2021observation}%
  \BibitemOpen
  \bibfield  {author} {\bibinfo {author} {\bibfnamefont {X.}~\bibnamefont {Wang}}, \bibinfo {author} {\bibfnamefont {D.}~\bibnamefont {Geng}}, \bibinfo {author} {\bibfnamefont {D.}~\bibnamefont {Yan}}, \bibinfo {author} {\bibfnamefont {W.}~\bibnamefont {Hu}}, \bibinfo {author} {\bibfnamefont {H.}~\bibnamefont {Zhang}}, \bibinfo {author} {\bibfnamefont {S.}~\bibnamefont {Yue}}, \bibinfo {author} {\bibfnamefont {Z.}~\bibnamefont {Sun}}, \bibinfo {author} {\bibfnamefont {S.}~\bibnamefont {Kumar}}, \bibinfo {author} {\bibfnamefont {E.~F.}\ \bibnamefont {Schwier}}, \bibinfo {author} {\bibfnamefont {K.}~\bibnamefont {Shimada}}, \emph {et~al.},\ }\bibfield  {title} {\bibinfo {title} {Observation of topological edge states in the quantum spin hall insulator \uppercase{T}a$_2$\uppercase{P}d$_3$\uppercase{T}e$_5$},\ }\href@noop {} {\bibfield  {journal} {\bibinfo  {journal} {Physical Review B}\ }\textbf {\bibinfo {volume} {104}},\ \bibinfo {pages} {L241408} (\bibinfo {year} {2021})}\BibitemShut {NoStop}%
\bibitem [{\citenamefont {Chen}\ \emph {et~al.}(2015)\citenamefont {Chen}, \citenamefont {Avila}, \citenamefont {Frantzeskakis}, \citenamefont {Levy},\ and\ \citenamefont {Asensio}}]{chen2015observation}%
  \BibitemOpen
  \bibfield  {author} {\bibinfo {author} {\bibfnamefont {C.}~\bibnamefont {Chen}}, \bibinfo {author} {\bibfnamefont {J.}~\bibnamefont {Avila}}, \bibinfo {author} {\bibfnamefont {E.}~\bibnamefont {Frantzeskakis}}, \bibinfo {author} {\bibfnamefont {A.}~\bibnamefont {Levy}},\ and\ \bibinfo {author} {\bibfnamefont {M.~C.}\ \bibnamefont {Asensio}},\ }\bibfield  {title} {\bibinfo {title} {Observation of a two-dimensional liquid of fr{\"o}hlich polarons at the bare \uppercase{S}r\uppercase{T}i\uppercase{O}$_3$ surface},\ }\href@noop {} {\bibfield  {journal} {\bibinfo  {journal} {Nature communications}\ }\textbf {\bibinfo {volume} {6}},\ \bibinfo {pages} {8585} (\bibinfo {year} {2015})}\BibitemShut {NoStop}%
\bibitem [{\citenamefont {Kang}\ \emph {et~al.}(2018)\citenamefont {Kang}, \citenamefont {Jung}, \citenamefont {Shin}, \citenamefont {Sohn}, \citenamefont {Ryu}, \citenamefont {Kim}, \citenamefont {Hoesch},\ and\ \citenamefont {Kim}}]{kang2018holstein}%
  \BibitemOpen
  \bibfield  {author} {\bibinfo {author} {\bibfnamefont {M.}~\bibnamefont {Kang}}, \bibinfo {author} {\bibfnamefont {S.~W.}\ \bibnamefont {Jung}}, \bibinfo {author} {\bibfnamefont {W.~J.}\ \bibnamefont {Shin}}, \bibinfo {author} {\bibfnamefont {Y.}~\bibnamefont {Sohn}}, \bibinfo {author} {\bibfnamefont {S.~H.}\ \bibnamefont {Ryu}}, \bibinfo {author} {\bibfnamefont {T.~K.}\ \bibnamefont {Kim}}, \bibinfo {author} {\bibfnamefont {M.}~\bibnamefont {Hoesch}},\ and\ \bibinfo {author} {\bibfnamefont {K.~S.}\ \bibnamefont {Kim}},\ }\bibfield  {title} {\bibinfo {title} {Holstein polaron in a valley-degenerate two-dimensional semiconductor},\ }\href@noop {} {\bibfield  {journal} {\bibinfo  {journal} {Nature materials}\ }\textbf {\bibinfo {volume} {17}},\ \bibinfo {pages} {676} (\bibinfo {year} {2018})}\BibitemShut {NoStop}%
\bibitem [{\citenamefont {Riley}\ \emph {et~al.}(2018)\citenamefont {Riley}, \citenamefont {Caruso}, \citenamefont {Verdi}, \citenamefont {Duffy}, \citenamefont {Watson}, \citenamefont {Bawden}, \citenamefont {Volckaert}, \citenamefont {Van Der~Laan}, \citenamefont {Hesjedal}, \citenamefont {Hoesch} \emph {et~al.}}]{riley2018crossover}%
  \BibitemOpen
  \bibfield  {author} {\bibinfo {author} {\bibfnamefont {J.~M.}\ \bibnamefont {Riley}}, \bibinfo {author} {\bibfnamefont {F.}~\bibnamefont {Caruso}}, \bibinfo {author} {\bibfnamefont {C.}~\bibnamefont {Verdi}}, \bibinfo {author} {\bibfnamefont {L.}~\bibnamefont {Duffy}}, \bibinfo {author} {\bibfnamefont {M.~D.}\ \bibnamefont {Watson}}, \bibinfo {author} {\bibfnamefont {L.}~\bibnamefont {Bawden}}, \bibinfo {author} {\bibfnamefont {K.}~\bibnamefont {Volckaert}}, \bibinfo {author} {\bibfnamefont {G.}~\bibnamefont {Van Der~Laan}}, \bibinfo {author} {\bibfnamefont {T.}~\bibnamefont {Hesjedal}}, \bibinfo {author} {\bibfnamefont {M.}~\bibnamefont {Hoesch}}, \emph {et~al.},\ }\bibfield  {title} {\bibinfo {title} {Crossover from lattice to plasmonic polarons of a spin-polarised electron gas in ferromagnetic \uppercase{E}u\uppercase{O}},\ }\href@noop {} {\bibfield  {journal} {\bibinfo  {journal} {Nature Communications}\ }\textbf {\bibinfo {volume} {9}},\ \bibinfo {pages} {2305} (\bibinfo {year} {2018})}\BibitemShut
  {NoStop}%
\bibitem [{\citenamefont {Antonius}\ \emph {et~al.}(2020)\citenamefont {Antonius}, \citenamefont {Chan},\ and\ \citenamefont {Louie}}]{antonius2020polaron}%
  \BibitemOpen
  \bibfield  {author} {\bibinfo {author} {\bibfnamefont {G.}~\bibnamefont {Antonius}}, \bibinfo {author} {\bibfnamefont {Y.-H.}\ \bibnamefont {Chan}},\ and\ \bibinfo {author} {\bibfnamefont {S.~G.}\ \bibnamefont {Louie}},\ }\bibfield  {title} {\bibinfo {title} {Polaron spectral properties in doped zno and srtio 3 from first principles},\ }\href@noop {} {\bibfield  {journal} {\bibinfo  {journal} {Physical Review Research}\ }\textbf {\bibinfo {volume} {2}},\ \bibinfo {pages} {043296} (\bibinfo {year} {2020})}\BibitemShut {NoStop}%
\bibitem [{\citenamefont {Caruso}\ \emph {et~al.}(2021)\citenamefont {Caruso}, \citenamefont {Amsalem}, \citenamefont {Ma}, \citenamefont {Aljarb}, \citenamefont {Schultz}, \citenamefont {Zacharias}, \citenamefont {Tung}, \citenamefont {Koch},\ and\ \citenamefont {Draxl}}]{caruso2021two}%
  \BibitemOpen
  \bibfield  {author} {\bibinfo {author} {\bibfnamefont {F.}~\bibnamefont {Caruso}}, \bibinfo {author} {\bibfnamefont {P.}~\bibnamefont {Amsalem}}, \bibinfo {author} {\bibfnamefont {J.}~\bibnamefont {Ma}}, \bibinfo {author} {\bibfnamefont {A.}~\bibnamefont {Aljarb}}, \bibinfo {author} {\bibfnamefont {T.}~\bibnamefont {Schultz}}, \bibinfo {author} {\bibfnamefont {M.}~\bibnamefont {Zacharias}}, \bibinfo {author} {\bibfnamefont {V.}~\bibnamefont {Tung}}, \bibinfo {author} {\bibfnamefont {N.}~\bibnamefont {Koch}},\ and\ \bibinfo {author} {\bibfnamefont {C.}~\bibnamefont {Draxl}},\ }\bibfield  {title} {\bibinfo {title} {Two-dimensional plasmonic polarons in n-doped monolayer \uppercase{M}o\uppercase{S}$_2$},\ }\href@noop {} {\bibfield  {journal} {\bibinfo  {journal} {Physical Review B}\ }\textbf {\bibinfo {volume} {103}},\ \bibinfo {pages} {205152} (\bibinfo {year} {2021})}\BibitemShut {NoStop}%
\bibitem [{\citenamefont {Kang}\ \emph {et~al.}(2021)\citenamefont {Kang}, \citenamefont {Du}, \citenamefont {Zhou}, \citenamefont {Gu}, \citenamefont {Chen}, \citenamefont {Xu}, \citenamefont {Zhang}, \citenamefont {Sun}, \citenamefont {Yin}, \citenamefont {Li} \emph {et~al.}}]{kang2021band}%
  \BibitemOpen
  \bibfield  {author} {\bibinfo {author} {\bibfnamefont {L.}~\bibnamefont {Kang}}, \bibinfo {author} {\bibfnamefont {X.}~\bibnamefont {Du}}, \bibinfo {author} {\bibfnamefont {J.}~\bibnamefont {Zhou}}, \bibinfo {author} {\bibfnamefont {X.}~\bibnamefont {Gu}}, \bibinfo {author} {\bibfnamefont {Y.}~\bibnamefont {Chen}}, \bibinfo {author} {\bibfnamefont {R.}~\bibnamefont {Xu}}, \bibinfo {author} {\bibfnamefont {Q.}~\bibnamefont {Zhang}}, \bibinfo {author} {\bibfnamefont {S.}~\bibnamefont {Sun}}, \bibinfo {author} {\bibfnamefont {Z.}~\bibnamefont {Yin}}, \bibinfo {author} {\bibfnamefont {Y.}~\bibnamefont {Li}}, \emph {et~al.},\ }\bibfield  {title} {\bibinfo {title} {Band-selective holstein polaron in \uppercase{L}uttinger liquid material \uppercase{A}$_{0.3}$\uppercase{M}o\uppercase{O}$_3$ (\uppercase{A}= \uppercase{K}, \uppercase{R}b)},\ }\href@noop {} {\bibfield  {journal} {\bibinfo  {journal} {Nature Communications}\ }\textbf {\bibinfo {volume} {12}},\ \bibinfo {pages} {6183} (\bibinfo {year}
  {2021})}\BibitemShut {NoStop}%
\bibitem [{\citenamefont {Wang}\ \emph {et~al.}(2023)\citenamefont {Wang}, \citenamefont {Li}, \citenamefont {Yang}, \citenamefont {Yan}, \citenamefont {Huang}, \citenamefont {Guo}, \citenamefont {Gao}, \citenamefont {Huang}, \citenamefont {Zeng}, \citenamefont {Qian} \emph {et~al.}}]{wang2023robust}%
  \BibitemOpen
  \bibfield  {author} {\bibinfo {author} {\bibfnamefont {A.}~\bibnamefont {Wang}}, \bibinfo {author} {\bibfnamefont {Y.}~\bibnamefont {Li}}, \bibinfo {author} {\bibfnamefont {G.}~\bibnamefont {Yang}}, \bibinfo {author} {\bibfnamefont {D.}~\bibnamefont {Yan}}, \bibinfo {author} {\bibfnamefont {Y.}~\bibnamefont {Huang}}, \bibinfo {author} {\bibfnamefont {Z.}~\bibnamefont {Guo}}, \bibinfo {author} {\bibfnamefont {J.}~\bibnamefont {Gao}}, \bibinfo {author} {\bibfnamefont {J.}~\bibnamefont {Huang}}, \bibinfo {author} {\bibfnamefont {Q.}~\bibnamefont {Zeng}}, \bibinfo {author} {\bibfnamefont {D.}~\bibnamefont {Qian}}, \emph {et~al.},\ }\bibfield  {title} {\bibinfo {title} {A robust and tunable luttinger liquid in correlated edge of transition-metal second-order topological insulator \uppercase{T}a$_2$\uppercase{P}d$_3$\uppercase{T}e$_5$},\ }\href@noop {} {\bibfield  {journal} {\bibinfo  {journal} {Nature Communications}\ }\textbf {\bibinfo {volume} {14}},\ \bibinfo {pages} {7647} (\bibinfo {year}
  {2023})}\BibitemShut {NoStop}%
\bibitem [{\citenamefont {Christiansen}\ \emph {et~al.}(2019)\citenamefont {Christiansen}, \citenamefont {Selig}, \citenamefont {Malic}, \citenamefont {Ernstorfer},\ and\ \citenamefont {Knorr}}]{christiansen2019theory}%
  \BibitemOpen
  \bibfield  {author} {\bibinfo {author} {\bibfnamefont {D.}~\bibnamefont {Christiansen}}, \bibinfo {author} {\bibfnamefont {M.}~\bibnamefont {Selig}}, \bibinfo {author} {\bibfnamefont {E.}~\bibnamefont {Malic}}, \bibinfo {author} {\bibfnamefont {R.}~\bibnamefont {Ernstorfer}},\ and\ \bibinfo {author} {\bibfnamefont {A.}~\bibnamefont {Knorr}},\ }\bibfield  {title} {\bibinfo {title} {Theory of exciton dynamics in time-resolved \uppercase{ARPES}: \uppercase{I}ntra-and intervalley scattering in two-dimensional semiconductors},\ }\href@noop {} {\bibfield  {journal} {\bibinfo  {journal} {Physical Review B}\ }\textbf {\bibinfo {volume} {100}},\ \bibinfo {pages} {205401} (\bibinfo {year} {2019})}\BibitemShut {NoStop}%
\bibitem [{\citenamefont {Ni}\ \emph {et~al.}(2017)\citenamefont {Ni}, \citenamefont {Huynh}, \citenamefont {Cheminal}, \citenamefont {Thomas}, \citenamefont {Shivanna}, \citenamefont {Hinrichsen}, \citenamefont {Ahmad}, \citenamefont {Sadhanala},\ and\ \citenamefont {Rao}}]{ni2017real}%
  \BibitemOpen
  \bibfield  {author} {\bibinfo {author} {\bibfnamefont {L.}~\bibnamefont {Ni}}, \bibinfo {author} {\bibfnamefont {U.}~\bibnamefont {Huynh}}, \bibinfo {author} {\bibfnamefont {A.}~\bibnamefont {Cheminal}}, \bibinfo {author} {\bibfnamefont {T.~H.}\ \bibnamefont {Thomas}}, \bibinfo {author} {\bibfnamefont {R.}~\bibnamefont {Shivanna}}, \bibinfo {author} {\bibfnamefont {T.~F.}\ \bibnamefont {Hinrichsen}}, \bibinfo {author} {\bibfnamefont {S.}~\bibnamefont {Ahmad}}, \bibinfo {author} {\bibfnamefont {A.}~\bibnamefont {Sadhanala}},\ and\ \bibinfo {author} {\bibfnamefont {A.}~\bibnamefont {Rao}},\ }\bibfield  {title} {\bibinfo {title} {Real-time observation of exciton--phonon coupling dynamics in self-assembled hybrid perovskite quantum wells},\ }\href@noop {} {\bibfield  {journal} {\bibinfo  {journal} {ACS nano}\ }\textbf {\bibinfo {volume} {11}},\ \bibinfo {pages} {10834} (\bibinfo {year} {2017})}\BibitemShut {NoStop}%
\bibitem [{\citenamefont {Lu}\ \emph {et~al.}(2017)\citenamefont {Lu}, \citenamefont {Kono}, \citenamefont {Larkin}, \citenamefont {Rost}, \citenamefont {Takayama}, \citenamefont {Boris}, \citenamefont {Keimer},\ and\ \citenamefont {Takagi}}]{lu2017zero}%
  \BibitemOpen
  \bibfield  {author} {\bibinfo {author} {\bibfnamefont {Y.}~\bibnamefont {Lu}}, \bibinfo {author} {\bibfnamefont {H.}~\bibnamefont {Kono}}, \bibinfo {author} {\bibfnamefont {T.}~\bibnamefont {Larkin}}, \bibinfo {author} {\bibfnamefont {A.}~\bibnamefont {Rost}}, \bibinfo {author} {\bibfnamefont {T.}~\bibnamefont {Takayama}}, \bibinfo {author} {\bibfnamefont {A.}~\bibnamefont {Boris}}, \bibinfo {author} {\bibfnamefont {B.}~\bibnamefont {Keimer}},\ and\ \bibinfo {author} {\bibfnamefont {H.}~\bibnamefont {Takagi}},\ }\bibfield  {title} {\bibinfo {title} {Zero-gap semiconductor to excitonic insulator transition in \uppercase{T}a$_2$\uppercase{N}i\uppercase{S}e$_5$},\ }\href@noop {} {\bibfield  {journal} {\bibinfo  {journal} {Nature communications}\ }\textbf {\bibinfo {volume} {8}},\ \bibinfo {pages} {14408} (\bibinfo {year} {2017})}\BibitemShut {NoStop}%
\bibitem [{\citenamefont {Okazaki}\ \emph {et~al.}(2018)\citenamefont {Okazaki}, \citenamefont {Ogawa}, \citenamefont {Suzuki}, \citenamefont {Yamamoto}, \citenamefont {Someya}, \citenamefont {Michimae}, \citenamefont {Watanabe}, \citenamefont {Lu}, \citenamefont {Nohara}, \citenamefont {Takagi} \emph {et~al.}}]{okazaki2018photo}%
  \BibitemOpen
  \bibfield  {author} {\bibinfo {author} {\bibfnamefont {K.}~\bibnamefont {Okazaki}}, \bibinfo {author} {\bibfnamefont {Y.}~\bibnamefont {Ogawa}}, \bibinfo {author} {\bibfnamefont {T.}~\bibnamefont {Suzuki}}, \bibinfo {author} {\bibfnamefont {T.}~\bibnamefont {Yamamoto}}, \bibinfo {author} {\bibfnamefont {T.}~\bibnamefont {Someya}}, \bibinfo {author} {\bibfnamefont {S.}~\bibnamefont {Michimae}}, \bibinfo {author} {\bibfnamefont {M.}~\bibnamefont {Watanabe}}, \bibinfo {author} {\bibfnamefont {Y.}~\bibnamefont {Lu}}, \bibinfo {author} {\bibfnamefont {M.}~\bibnamefont {Nohara}}, \bibinfo {author} {\bibfnamefont {H.}~\bibnamefont {Takagi}}, \emph {et~al.},\ }\bibfield  {title} {\bibinfo {title} {Photo-induced semimetallic states realised in electron--hole coupled insulators},\ }\href@noop {} {\bibfield  {journal} {\bibinfo  {journal} {Nature communications}\ }\textbf {\bibinfo {volume} {9}},\ \bibinfo {pages} {4322} (\bibinfo {year} {2018})}\BibitemShut {NoStop}%
\bibitem [{\citenamefont {Werdehausen}\ \emph {et~al.}(2018)\citenamefont {Werdehausen}, \citenamefont {Takayama}, \citenamefont {Albrecht}, \citenamefont {Lu}, \citenamefont {Takagi},\ and\ \citenamefont {Kaiser}}]{werdehausen2018photo}%
  \BibitemOpen
  \bibfield  {author} {\bibinfo {author} {\bibfnamefont {D.}~\bibnamefont {Werdehausen}}, \bibinfo {author} {\bibfnamefont {T.}~\bibnamefont {Takayama}}, \bibinfo {author} {\bibfnamefont {G.}~\bibnamefont {Albrecht}}, \bibinfo {author} {\bibfnamefont {Y.}~\bibnamefont {Lu}}, \bibinfo {author} {\bibfnamefont {H.}~\bibnamefont {Takagi}},\ and\ \bibinfo {author} {\bibfnamefont {S.}~\bibnamefont {Kaiser}},\ }\bibfield  {title} {\bibinfo {title} {Photo-excited dynamics in the excitonic insulator \uppercase{T}a$_2$\uppercase{N}i\uppercase{S}e$_5$},\ }\href@noop {} {\bibfield  {journal} {\bibinfo  {journal} {Journal of Physics: Condensed Matter}\ }\textbf {\bibinfo {volume} {30}},\ \bibinfo {pages} {305602} (\bibinfo {year} {2018})}\BibitemShut {NoStop}%
\bibitem [{\citenamefont {Chen}\ \emph {et~al.}(2023{\natexlab{a}})\citenamefont {Chen}, \citenamefont {Chen}, \citenamefont {Tang}, \citenamefont {Li}, \citenamefont {Wang}, \citenamefont {Ding}, \citenamefont {Kang}, \citenamefont {Jozwiak}, \citenamefont {Bostwick}, \citenamefont {Rotenberg} \emph {et~al.}}]{chen2023role}%
  \BibitemOpen
  \bibfield  {author} {\bibinfo {author} {\bibfnamefont {C.}~\bibnamefont {Chen}}, \bibinfo {author} {\bibfnamefont {X.}~\bibnamefont {Chen}}, \bibinfo {author} {\bibfnamefont {W.}~\bibnamefont {Tang}}, \bibinfo {author} {\bibfnamefont {Z.}~\bibnamefont {Li}}, \bibinfo {author} {\bibfnamefont {S.}~\bibnamefont {Wang}}, \bibinfo {author} {\bibfnamefont {S.}~\bibnamefont {Ding}}, \bibinfo {author} {\bibfnamefont {Z.}~\bibnamefont {Kang}}, \bibinfo {author} {\bibfnamefont {C.}~\bibnamefont {Jozwiak}}, \bibinfo {author} {\bibfnamefont {A.}~\bibnamefont {Bostwick}}, \bibinfo {author} {\bibfnamefont {E.}~\bibnamefont {Rotenberg}}, \emph {et~al.},\ }\bibfield  {title} {\bibinfo {title} {Role of electron-phonon coupling in excitonic insulator candidate \uppercase{T}a$_2$\uppercase{N}i\uppercase{S}e$_5$},\ }\href@noop {} {\bibfield  {journal} {\bibinfo  {journal} {Physical Review Research}\ }\textbf {\bibinfo {volume} {5}},\ \bibinfo {pages} {043089} (\bibinfo {year} {2023}{\natexlab{a}})}\BibitemShut {NoStop}%
\bibitem [{\citenamefont {Chen}\ \emph {et~al.}(2023{\natexlab{b}})\citenamefont {Chen}, \citenamefont {Tang}, \citenamefont {Chen}, \citenamefont {Kang}, \citenamefont {Ding}, \citenamefont {Scott}, \citenamefont {Wang}, \citenamefont {Li}, \citenamefont {Ruff}, \citenamefont {Hashimoto} \emph {et~al.}}]{chen2023anomalous}%
  \BibitemOpen
  \bibfield  {author} {\bibinfo {author} {\bibfnamefont {C.}~\bibnamefont {Chen}}, \bibinfo {author} {\bibfnamefont {W.}~\bibnamefont {Tang}}, \bibinfo {author} {\bibfnamefont {X.}~\bibnamefont {Chen}}, \bibinfo {author} {\bibfnamefont {Z.}~\bibnamefont {Kang}}, \bibinfo {author} {\bibfnamefont {S.}~\bibnamefont {Ding}}, \bibinfo {author} {\bibfnamefont {K.}~\bibnamefont {Scott}}, \bibinfo {author} {\bibfnamefont {S.}~\bibnamefont {Wang}}, \bibinfo {author} {\bibfnamefont {Z.}~\bibnamefont {Li}}, \bibinfo {author} {\bibfnamefont {J.~P.}\ \bibnamefont {Ruff}}, \bibinfo {author} {\bibfnamefont {M.}~\bibnamefont {Hashimoto}}, \emph {et~al.},\ }\bibfield  {title} {\bibinfo {title} {Anomalous excitonic phase diagram in band-gap-tuned \uppercase{T}a$_2$\uppercase{N}i(\uppercase{S}e, \uppercase{S})$_5$},\ }\href@noop {} {\bibfield  {journal} {\bibinfo  {journal} {Nature Communications}\ }\textbf {\bibinfo {volume} {14}},\ \bibinfo {pages} {7512} (\bibinfo {year} {2023}{\natexlab{b}})}\BibitemShut {NoStop}%
\bibitem [{\citenamefont {Wang}\ \emph {et~al.}(2019)\citenamefont {Wang}, \citenamefont {Erten}, \citenamefont {Wang},\ and\ \citenamefont {Xing}}]{wang2019prediction}%
  \BibitemOpen
  \bibfield  {author} {\bibinfo {author} {\bibfnamefont {R.}~\bibnamefont {Wang}}, \bibinfo {author} {\bibfnamefont {O.}~\bibnamefont {Erten}}, \bibinfo {author} {\bibfnamefont {B.}~\bibnamefont {Wang}},\ and\ \bibinfo {author} {\bibfnamefont {D.}~\bibnamefont {Xing}},\ }\bibfield  {title} {\bibinfo {title} {Prediction of a topological p+ ip excitonic insulator with parity anomaly},\ }\href@noop {} {\bibfield  {journal} {\bibinfo  {journal} {Nature communications}\ }\textbf {\bibinfo {volume} {10}},\ \bibinfo {pages} {210} (\bibinfo {year} {2019})}\BibitemShut {NoStop}%
\bibitem [{\citenamefont {Kozii}\ and\ \citenamefont {Fu}(2015)}]{kozii2015odd}%
  \BibitemOpen
  \bibfield  {author} {\bibinfo {author} {\bibfnamefont {V.}~\bibnamefont {Kozii}}\ and\ \bibinfo {author} {\bibfnamefont {L.}~\bibnamefont {Fu}},\ }\bibfield  {title} {\bibinfo {title} {Odd-parity superconductivity in the vicinity of inversion symmetry breaking in spin-orbit-coupled systems},\ }\href@noop {} {\bibfield  {journal} {\bibinfo  {journal} {Physical review letters}\ }\textbf {\bibinfo {volume} {115}},\ \bibinfo {pages} {207002} (\bibinfo {year} {2015})}\BibitemShut {NoStop}%
\bibitem [{\citenamefont {Guo}\ \emph {et~al.}(2021)\citenamefont {Guo}, \citenamefont {Yan}, \citenamefont {Sheng}, \citenamefont {Nie}, \citenamefont {Shi},\ and\ \citenamefont {Wang}}]{guo2021quantum}%
  \BibitemOpen
  \bibfield  {author} {\bibinfo {author} {\bibfnamefont {Z.}~\bibnamefont {Guo}}, \bibinfo {author} {\bibfnamefont {D.}~\bibnamefont {Yan}}, \bibinfo {author} {\bibfnamefont {H.}~\bibnamefont {Sheng}}, \bibinfo {author} {\bibfnamefont {S.}~\bibnamefont {Nie}}, \bibinfo {author} {\bibfnamefont {Y.}~\bibnamefont {Shi}},\ and\ \bibinfo {author} {\bibfnamefont {Z.}~\bibnamefont {Wang}},\ }\bibfield  {title} {\bibinfo {title} {Quantum spin \uppercase{H}all effect in \uppercase{T}a$_2$\uppercase{M}$_3$\uppercase{T}e$_5$ (\uppercase{M}= \uppercase{P}d, \uppercase{N}i)},\ }\href@noop {} {\bibfield  {journal} {\bibinfo  {journal} {Physical Review B}\ }\textbf {\bibinfo {volume} {103}},\ \bibinfo {pages} {115145} (\bibinfo {year} {2021})}\BibitemShut {NoStop}%
\bibitem [{\citenamefont {Li}\ \emph {et~al.}(2024)\citenamefont {Li}, \citenamefont {Yan}, \citenamefont {Hong}, \citenamefont {Sheng}, \citenamefont {Wang}, \citenamefont {Dou}, \citenamefont {Guo}, \citenamefont {Shi}, \citenamefont {Su}, \citenamefont {Lyu} \emph {et~al.}}]{li2024interfering}%
  \BibitemOpen
  \bibfield  {author} {\bibinfo {author} {\bibfnamefont {Y.}~\bibnamefont {Li}}, \bibinfo {author} {\bibfnamefont {D.}~\bibnamefont {Yan}}, \bibinfo {author} {\bibfnamefont {Y.}~\bibnamefont {Hong}}, \bibinfo {author} {\bibfnamefont {H.}~\bibnamefont {Sheng}}, \bibinfo {author} {\bibfnamefont {A.}~\bibnamefont {Wang}}, \bibinfo {author} {\bibfnamefont {Z.}~\bibnamefont {Dou}}, \bibinfo {author} {\bibfnamefont {X.}~\bibnamefont {Guo}}, \bibinfo {author} {\bibfnamefont {X.}~\bibnamefont {Shi}}, \bibinfo {author} {\bibfnamefont {Z.}~\bibnamefont {Su}}, \bibinfo {author} {\bibfnamefont {Z.}~\bibnamefont {Lyu}}, \emph {et~al.},\ }\bibfield  {title} {\bibinfo {title} {Interfering josephson diode effect in \uppercase{T}a$_2$\uppercase{P}d$_3$\uppercase{T}e$_5$ asymmetric edge interferometer},\ }\href@noop {} {\bibfield  {journal} {\bibinfo  {journal} {Nature Communications}\ }\textbf {\bibinfo {volume} {15}},\ \bibinfo {pages} {9031} (\bibinfo {year} {2024})}\BibitemShut {NoStop}%
\bibitem [{\citenamefont {Yu}\ \emph {et~al.}(2024)\citenamefont {Yu}, \citenamefont {Yan}, \citenamefont {Guo}, \citenamefont {Zhou}, \citenamefont {Yang}, \citenamefont {Li}, \citenamefont {Wang}, \citenamefont {Xiang}, \citenamefont {Li}, \citenamefont {Ma} \emph {et~al.}}]{yu2024observation}%
  \BibitemOpen
  \bibfield  {author} {\bibinfo {author} {\bibfnamefont {H.}~\bibnamefont {Yu}}, \bibinfo {author} {\bibfnamefont {D.}~\bibnamefont {Yan}}, \bibinfo {author} {\bibfnamefont {Z.}~\bibnamefont {Guo}}, \bibinfo {author} {\bibfnamefont {Y.}~\bibnamefont {Zhou}}, \bibinfo {author} {\bibfnamefont {X.}~\bibnamefont {Yang}}, \bibinfo {author} {\bibfnamefont {P.}~\bibnamefont {Li}}, \bibinfo {author} {\bibfnamefont {Z.}~\bibnamefont {Wang}}, \bibinfo {author} {\bibfnamefont {X.}~\bibnamefont {Xiang}}, \bibinfo {author} {\bibfnamefont {J.}~\bibnamefont {Li}}, \bibinfo {author} {\bibfnamefont {X.}~\bibnamefont {Ma}}, \emph {et~al.},\ }\bibfield  {title} {\bibinfo {title} {Observation of emergent superconductivity in the topological insulator \uppercase{T}a$_2$\uppercase{P}d$_3$\uppercase{T}e$_5$ via pressure manipulation},\ }\href@noop {} {\bibfield  {journal} {\bibinfo  {journal} {Journal of the American Chemical Society}\ }\textbf {\bibinfo {volume} {146}},\ \bibinfo {pages} {3890} (\bibinfo {year}
  {2024})}\BibitemShut {NoStop}%
\bibitem [{\citenamefont {Zhang}\ \emph {et~al.}(2011)\citenamefont {Zhang}, \citenamefont {Richard}, \citenamefont {Qian}, \citenamefont {Xu}, \citenamefont {Dai},\ and\ \citenamefont {Ding}}]{zhang2011precise}%
  \BibitemOpen
  \bibfield  {author} {\bibinfo {author} {\bibfnamefont {P.}~\bibnamefont {Zhang}}, \bibinfo {author} {\bibfnamefont {P.}~\bibnamefont {Richard}}, \bibinfo {author} {\bibfnamefont {T.}~\bibnamefont {Qian}}, \bibinfo {author} {\bibfnamefont {Y.-M.}\ \bibnamefont {Xu}}, \bibinfo {author} {\bibfnamefont {X.}~\bibnamefont {Dai}},\ and\ \bibinfo {author} {\bibfnamefont {H.}~\bibnamefont {Ding}},\ }\bibfield  {title} {\bibinfo {title} {A precise method for visualizing dispersive features in image plots},\ }\href@noop {} {\bibfield  {journal} {\bibinfo  {journal} {Review of Scientific Instruments}\ }\textbf {\bibinfo {volume} {82}} (\bibinfo {year} {2011})}\BibitemShut {NoStop}%
\bibitem [{\citenamefont {Kresse}\ and\ \citenamefont {Furthm{\"u}ller}(1996)}]{kresse1996efficient}%
  \BibitemOpen
  \bibfield  {author} {\bibinfo {author} {\bibfnamefont {G.}~\bibnamefont {Kresse}}\ and\ \bibinfo {author} {\bibfnamefont {J.}~\bibnamefont {Furthm{\"u}ller}},\ }\bibfield  {title} {\bibinfo {title} {Efficient iterative schemes for ab initio total-energy calculations using a plane-wave basis set},\ }\href@noop {} {\bibfield  {journal} {\bibinfo  {journal} {Physical review B}\ }\textbf {\bibinfo {volume} {54}},\ \bibinfo {pages} {11169} (\bibinfo {year} {1996})}\BibitemShut {NoStop}%
\bibitem [{\citenamefont {Perdew}\ \emph {et~al.}(1996)\citenamefont {Perdew}, \citenamefont {Burke},\ and\ \citenamefont {Ernzerhof}}]{perdew1996generalized}%
  \BibitemOpen
  \bibfield  {author} {\bibinfo {author} {\bibfnamefont {J.~P.}\ \bibnamefont {Perdew}}, \bibinfo {author} {\bibfnamefont {K.}~\bibnamefont {Burke}},\ and\ \bibinfo {author} {\bibfnamefont {M.}~\bibnamefont {Ernzerhof}},\ }\bibfield  {title} {\bibinfo {title} {Generalized gradient approximation made simple},\ }\href@noop {} {\bibfield  {journal} {\bibinfo  {journal} {Physical review letters}\ }\textbf {\bibinfo {volume} {77}},\ \bibinfo {pages} {3865} (\bibinfo {year} {1996})}\BibitemShut {NoStop}%
\bibitem [{\citenamefont {Kaneko}\ and\ \citenamefont {Ohta}(2025)}]{kaneko2025new}%
  \BibitemOpen
  \bibfield  {author} {\bibinfo {author} {\bibfnamefont {T.}~\bibnamefont {Kaneko}}\ and\ \bibinfo {author} {\bibfnamefont {Y.}~\bibnamefont {Ohta}},\ }\bibfield  {title} {\bibinfo {title} {A new era of excitonic insulators},\ }\href@noop {} {\bibfield  {journal} {\bibinfo  {journal} {Journal of the Physical Society of Japan}\ }\textbf {\bibinfo {volume} {94}},\ \bibinfo {pages} {012001} (\bibinfo {year} {2025})}\BibitemShut {NoStop}%
\end{thebibliography}%

\begin{figure*}
\includegraphics[width=\linewidth]{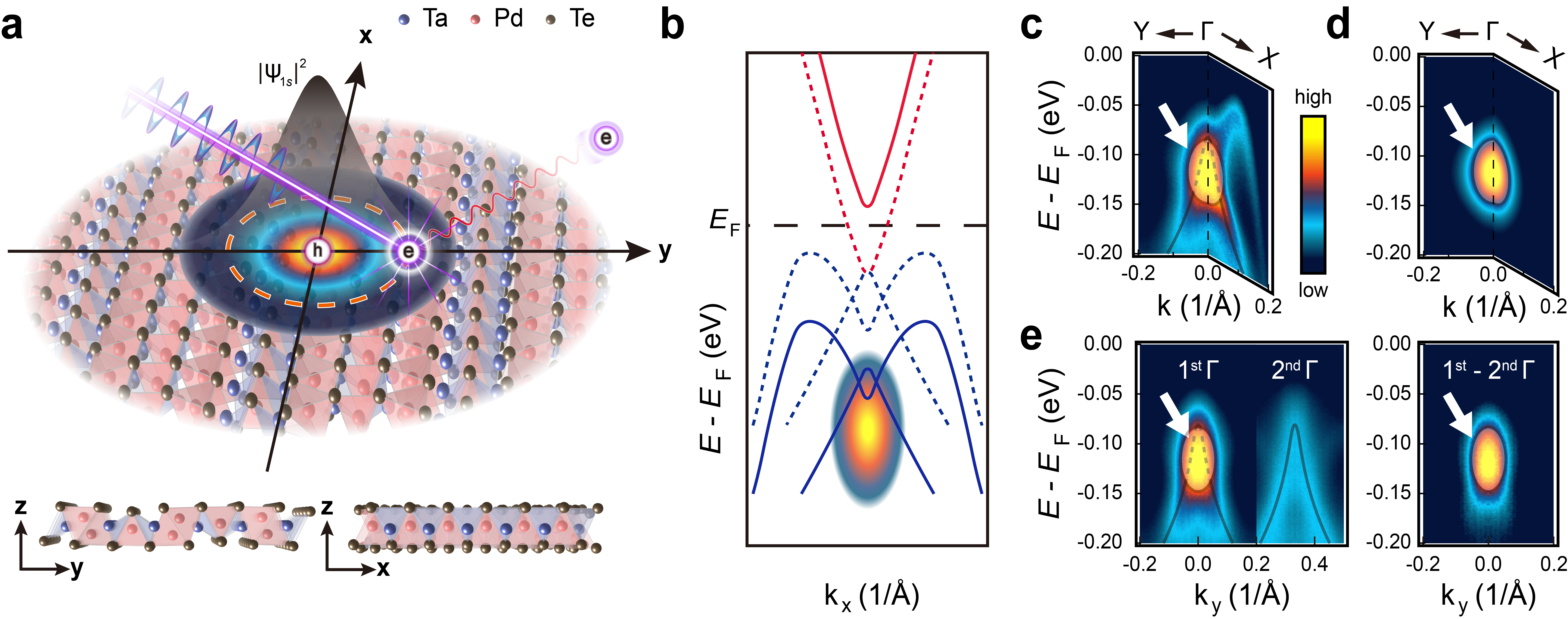}
\caption{\justifying a) An experimental setup of exciton photoemission with linearly polarized light. The squared exciton radial wave function simulated with experimental parameters are plot over the schematic model of the surface layer of Ta$_2$Pd$_3$Te$_5$. See text for the simulation procedure. Side views of the layer are given at the bottom. b) A schematic band structure of Ta$_2$Pd$_3$Te$_5$ around the Fermi energy (\textit{E}$_F$) along \textit{k}$_x$. The solid (dashed) lines represent the bands below (above) \textit{T}$_c$. The simulated momentum distribution of excitons is plot together. c) ARPES spectra of Ta$_2$Pd$_3$Te$_5$ along the $\Gamma-Y$  ($\Gamma-X$) direction. d) The exciton photoemission spectral function simulated from the ARPES spectra in (c). e) (left) ARPES spectra along $\Gamma-Y$ taken with xz-polarization at the centers of the first and the second BZs. (right) The difference of the two spectra is show clearly by the subtraction after normalization, which corresponds to the pure exciton photoemission signal.}
\end{figure*}

\begin{figure*}
\includegraphics[width=\linewidth]{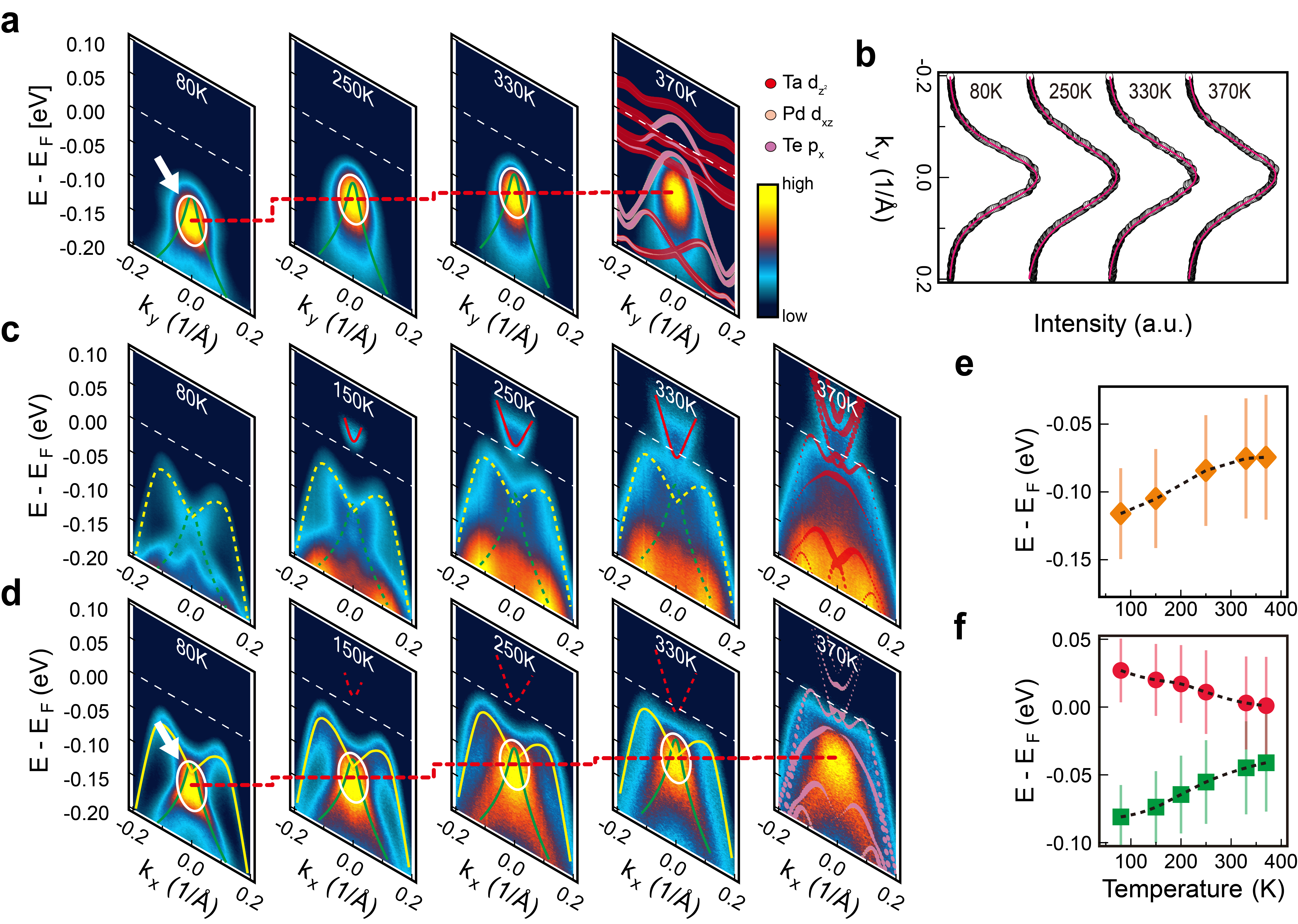}
\caption{\justifying a) ARPES spectra taken along the $\Gamma-Y$ direction exhibits the evolution of the exciton photoemission as function of temperature with the valence band photoemission. 
b) MDCs (dots) crossing the maximum spectral weights from (a) and their fits (pink lines) by the exciton wave functions. 
c) Temperature-dependent ARPES spectra (divided by Fermi Dirac functions) obtained along $\Gamma-X$ with yz polarization. The band gap evolution is very clear. 
d) Temperature-dependent ARPES spectra along $\Gamma-X$ obtained with x polarization. The calculated band structure for the normal phase are overlaid over the 370 K spectra in (a), (c) and (d) with the orbital contributions resolved.
e) The shifts of exciton photoemission peak as function of temperature, and they are extracted from $\Gamma$ point of xz-polarized ARPES results. 
f) The shifts of the conduction (red) and valence (green) band edges as function of temperature exhibiting the gradual opening of the band gap upon cooling. The conduction and valence bands are extracted from the $\Gamma$ point of yz- and xz-polarized ARPES results, respectively.}
\end{figure*}

\begin{figure*}
\includegraphics[width=\linewidth]{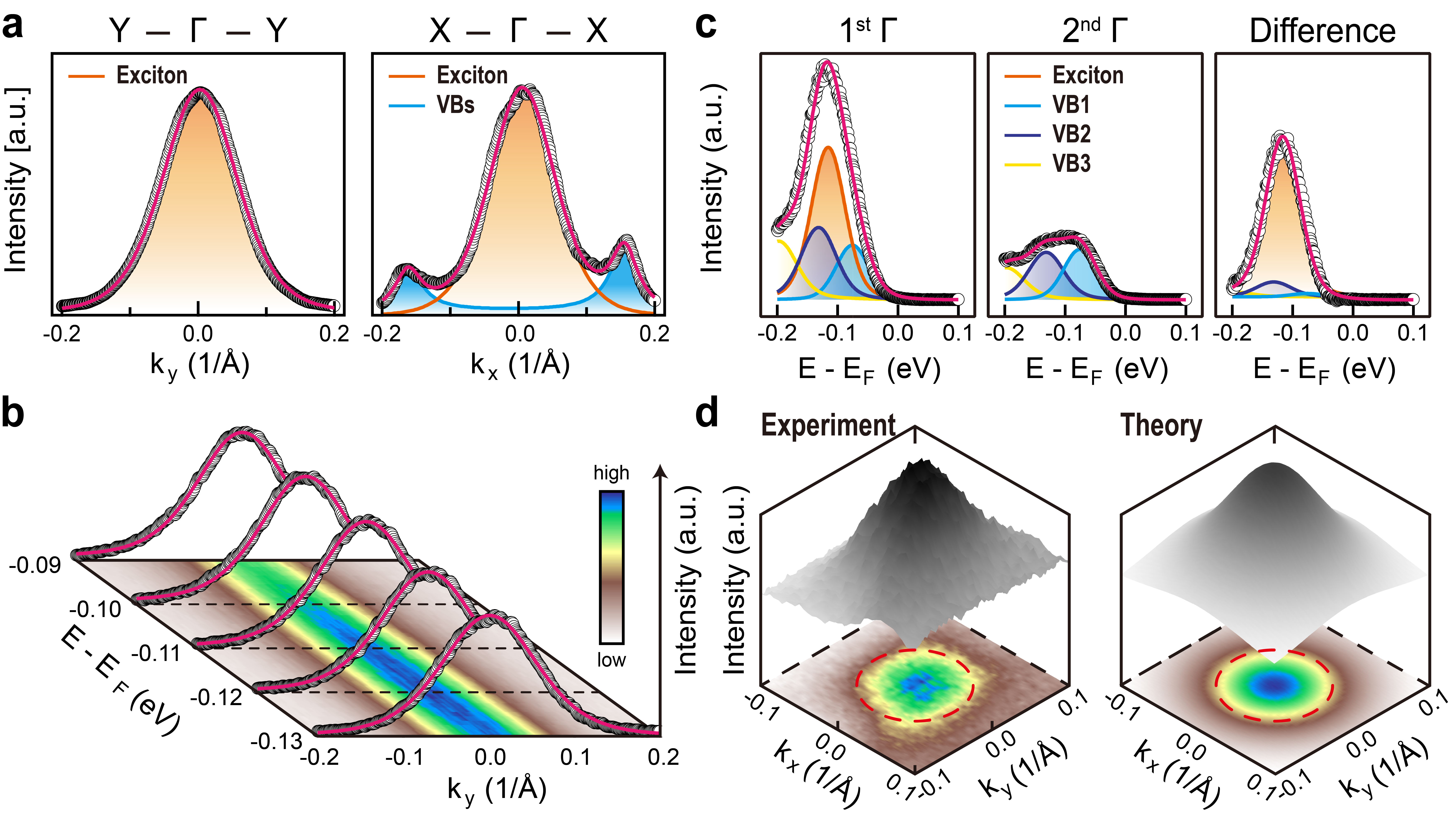}
\caption{\justifying a) MDCs crossing the maximum spectral weights from the spectra shown in Fig. 1c (circles) are fit by the exciton wave function (pink lines). 
b) Energy dependence of the fits (pink lines) of exciton wave function for the MDCs (circles) obtained for the ARPES spectrum  along $\Gamma-Y$ direction (xz-polarization) at 80K.  
c) Fits for EDCs from the spectra and its difference shown in Fig. 1e. 
d) (left) ARPES momentum distribution map obtained from x-polarized photon at the peak position of exciton photoemission (-0.115 eV). (right) 2D distribution of exciton wave function obtained from fittings of the experimental map in the left.}
\end{figure*}

\clearpage
\preprint{}
\section{Supplemental Material for \\``Exciton photoemission from a ground state of a solid Ta$_2$Pd$_3$Te$_5$''}

\author{Siwon Lee$^{1,2}$}
\author{Kyung-Hwan Jin$^{1,3}$}
\author{SeongJin Kwon$^{1,2}$}
\author{Hyunjin Jung$^{1,2}$}
\author{Choongjae Won$^{4}$}
\author{Sang-Wook Cheong$^{5,6}$}
\author{Gil Young Cho$^{1,2,}$}
\author{Jaeyoung Kim$^{1}$}
\author{Han Woong Yeom$^{1,2,}$}
\email{yeom@postech.ac.kr}
\affiliation{$^1$Center for Artificial Low Dimensional Electronic Systems, Institute for Basic Science (IBS), Pohang 37673, Republic of Korea \\ $
^2$Department of Physics, Pohang University of Science and Technology (POSTECH), Pohang 37673, Republic of Korea \\$^3$Department of Physics and Research Institute of Physics and Chemistry, Jeonbuk National University, Jeonju, 54896, Republic of Korea \\
$^4$Max Planck POSTECH/Korea Research Initiative, Center for Complex Phase of Materials, Pohang 37673, Republic of Korea
$^5$Laboratory of Pohang Emergent Materials, Pohang Accelerator Laboratory, Pohang 37673, Republic of Korea
$^6$Rutgers Center for emergent Materials and Department of Physics and Astronomy, Rutgers University, NJ, USA}

\subsection{Electronic Dominance in STM Topography Without Density Waves}
We have performed STM/STS/DFT studies on the structural details of this system. At present, we can note that the DFT-optimized structure model reproduces the fine details of the high-resolution STM topographies as shown in Fig. S1. This supports the validity of the structure model at hand but this model and the DFT calculation provide only a small energy gap of less than 10 meV. The STM images also straightforwardly exclude the emergence of any extra superstructure in the local density of states, which may be related to a charge density wave. Density waves are also excluded from the fact that the Fermi surface exists only as a point at Gamma, indicating no finite momenta for density wave couplings. A substantial electron correlation effect is not likely either since the band gap forms from a highly dispersing Dirac bands.

\subsection{Polarization dependence of exciton photoemission}
The polarization dependence in ARPES spectra is directly related to the mirror symmetry of orbitals in the involved bands through the dipole selection rule of an optical transition. Ta$_2$Pd$_3$Te$_5$ have three major orbitals that constitute valence and conduction bands. Each orbital as different mirror symmetries along yz and xz mirror planes (Figure S4). The polarization dependence of ARPES bands observed in Figure S5 aligns well with the orbital symmetry given in Figure S4d. That is, the exciton photoemission is visible for xz and x polarized photons, indicating that the exciton doesn't follow a polarization dependence of a typical 1$s$ orbital (Figure S4d).
The electron in an exciton originates from the conduction electron in the material, which is bound with a hole in the valence band. Consequently, the photoelectron from an exciton follows mainly the parity and orbital character of the conduction band (Ta \textit{d}$_{z^2}$ in Figure S4a). However, it is very clear in Figure S4d that the present exciton photoemission follows the orbital character of the valence band near the $\Gamma$ point (Pd \textit{d}$_{xz}$ or Te \textit{p}$_x$ in Figures S4b and S2c). This can only be explained by the interband hybridization, which occurs in an exciton insulator. 
For comparison, we show the orbital characters and photoemission selection rules for conduction and valence bands of Ta$_2$NiSe$_5$ in Figure S4e \cite{rustagi2018photoemission,fukutani2021detecting}. While the polarization dependence of the exciton photoemission from Ta$_2$NiSe$_5$ seems consistent with that of the valence band, it is not distinguished from that of the 1$s$-like exciton state.

\subsection{Fittings of exciton wavefunction}
We have followed the previous theoretical model and similar experimental assumptions, which were used for the ARPES work on Ta$_2$NiSe$_5$ \cite{rustagi2018photoemission,fukutani2021detecting}. The theoretical model was originally designed for excitons induced by optical pulses in time-resolved ARPES, but was also successfully applied to spontaneously formed excitons in equilibrium condition. In this model, the total photoemission intensity from excitons follows the formula:
\begin{widetext}
\begin{equation}
P\cong 2\pi \sigma ^2\vert M_{k,k'}^2\vert \Sigma_{\lambda,Q} \rho_{\lambda,Q}\vert \phi_{\lambda,w}(k'-w-\alpha Q) \vert^2\\ exp(-\sigma ^2[-\omega +E_{\lambda,Q}+\epsilon_{\nu ,k'-w-Q}]).
\end{equation}
\end{widetext}

We assumed that the matrix element ($M$) is constant, and exciton center-of-mass momentum ($Q$) is set to 0, which means the zero exciton temperature for simplicity. The momentum separation between the valence and conduction bands ($w$) is 0 for a direct gap material like Ta$_2$Pd$_3$Te$_5$. The temporal width of the probe pulse ($\sigma$) is less than 30 ps (20 ps in the previous work \cite{fukutani2021detecting}.). Therefore we can summarize the total photoemission intensity is directly related to the exciton wavefunction for the lowest energy 1$s$ exciton : $P\propto \vert\phi_{1s,w}\vert^2=1/[1+k^2a_0^2/4]^4$. The above polarization dependence is ignored in this model by the simplification on a constant matrix element.

\subsection{Two valence bands with different shapes}
The butterfly-shaped valence bands consist of two distinct components: one exhibiting an M-shaped dispersion and the other a $\Lambda$-shaped dispersion (Figure S7a and S7b). The M-shaped band is a trivial valence band that does not contribute to exciton formation. In contrast, the $\Lambda$-shaped band participates in exciton formation, as it intersects the conduction band, resulting in a Dirac-like band touching above T$_c$ \cite{zhang2024spontaneous,huang2024evidence}. The previous calculations, which neglected spin-orbit coupling (SOC), captured the $\Lambda$-shaped valence band \cite{zhang2024spontaneous,huang2024evidence}. However, when SOC was included, the $\Lambda$-shaped valence band became a M-shape, deviating from the ARPES results in details.
This is consistent with our calculations for the bulk Ta$_2$Pd$_3$Te$_5$ (Figure S6 and Figure S7c). Specifically, the $\Lambda$-shaped valence band became a M-shape when SOC was included. We performed similar calculations for a two-layer slab model, where the $\Lambda$-shaped band is closer to the experimental result even after including SOC (Figure S7d). 
However, the discrepancy in the detailed dispersion of this band between the ARPES experiments and the DFT calculations available requests further investigation. 

\subsection{Two band model calculation with an excitonic interaction}
In order to examine the excitonic correlation properly, we considered the minimal band model (two-band model) with the excitonic interaction term, as follows : 
\begin{widetext}
\begin{equation}
H=h(k) \sigma_0+M(k)\sigma_z+A_x  \sin(k_x ) \sigma_x+A_y  \sin(k_y ) \sigma_y+h_{ei},
\end{equation}
\end{widetext}
\begin{widetext}
\begin{equation}
h(k)=C_x (1+\cos(k_x ) )+C_y (1-\cos(k_y ) ), M(k)=M_0+M_x (1-\cos(k_x ) )+M_y (1-\cos(k_y ) ).
\end{equation}
\end{widetext}
$\sigma_i$  are the Pauli matrices. The first term represents the hopping interaction between neighboring atoms, while the second term governs the band inversion characteristics. The third and fourth terms account for the effective spin-orbit coupling (SOC) to deal with the topological degree of freedom of Ta$_2$Pd$_3$Te$_5$ \cite{jerome1967excitonic,hossain2023discovery}. Finally, the last term describes the Coulomb interaction between electrons in the valence and conduction bands with $h_{ei}=\Sigma_q\Sigma_{kk^\prime}V_c (q)c^\dag_{k+q,c} c^\dag_{k^\prime-q,v} c_{k^\prime,v} c_{k,c}$ where $V_c (q)=4\pi e^2/\epsilon (q) q^2$. By applying Hartree-Fock mean-field approximation and introducing the excitonic order parameter $\Delta_k=\Sigma_{k^\prime}V_c (k-k^\prime )<c^\dag_{k^\prime,c} c_{k^\prime,v}>$ similar to BCS-like approach, we can obtain one-electron excitonic insulating states. 
In the noninteracting case without any correlation or SOC effect, Figure S8a below shows the band structure in the semimetallic phase, where the valence and conduction bands overlap marginally. The calculated band dispersion highlights the metallic nature of the system, consistent with the experimentally observed high-temperature behavior. This model neglects the degeneracy of the valence band. Incorporating SOC into the calculations reveals the opening of a small topological gap, accompanied by spin-polarized edge states (Figure S8b). This phase is crucial for understanding the topological properties of the system, which was not explicitly discussed in the present work, and serves as a precursor to the correlated excitonic state. However, this state cannot explain the gradual gap opening of an order of 100 meV observed experimentally. Finally, we modeled the excitonic phase by including many-body interactions within a mean-field framework. The excitonic state arises from the coupling of electron-hole pairs, leading to a significant increase in the band gap, reflecting the binding energy of the excitons (Figure S8c). These results demonstrate how many-body interactions drive the transition from the semimetallic phase to the excitonic state, thereby validating the experimental claims of a correlated ground state.

\subsection{ARPES simulation and electron-phonon interaction}
To rule out the possibility of electron-phonon coupling as the origin of our unusual photoemission feature, we simulated the valence band spectrum using a simple Bethe-Salpeter equation model with two quadratic bands (Figure S10b) \cite{kaneko2025new}. The sum of this function with the simulated exciton contribution results in a broadened valence band feature (Figure S10f). However, this broadening does not match the characteristics of phonon-induced band broadening (Figure S10h).
Furthermore, the intensity-weighted summation of both simulations (Figure S10g) closely resembles the ARPES spectrum in the first Brillouin zone, where both the valence band and exciton photoemission features are clearly observed. In our model, electron-phonon coupling is simplified to induce band broadening by generating replica bands. This interaction is expressed by the following equation:
\begin{equation}
I(E,k)=\Sigma_{N}e^{(-AN)}I_{VB}(E-N\hbar\omega,k)
\end{equation}
where N represents the number of phonons and $\hbar\omega$ is the energy of each phonon. The simulated phonon-induced broadening in the photoemission spectra (Figure S11) does not match our ARPES spectra in the first Brillouin zone (Figure S10e) or the simulation result (Figure S10g).\\

\newpage

\begin{table}[t]
\caption*{\textbf{Table I.} Bohr radius values used in Figure 2b.}
\begin{ruledtabular}
\begin{tabular}{lc}
\textrm{Temperature} & \textrm{Bohr radius} \\
\colrule
80K & $13.56\pm0.022\text{\AA}$ \\
250K & $12.32\pm0.046\text{\AA}$ \\
330K & $12.05\pm0.049\text{\AA}$ \\
370K & $12.02\pm0.051\text{\AA}$ \\
\end{tabular}
\end{ruledtabular}
\end{table}

\begin{table}[t]
\caption*{\textbf{Table II.} Bohr radius values used in Figure 3a.}
\begin{ruledtabular}
\begin{tabular}{lc}
\textrm{Direction} & \textrm{Bohr radius} \\
\colrule
$\Gamma-X$    & $15.10\pm0.045\AA$ \\
$\Gamma-X$    & $15.10\pm0.045\AA$ \\
\end{tabular}
\end{ruledtabular}
\end{table}

\begin{table}[t]
\caption*{\textbf{Table III.} Momentum position and Lorentzian width used in Figure 3a.}
\begin{ruledtabular}
\begin{tabular}{lc}
\textrm{$k_{VB}$} & \textrm{Lorentzian width of VB} \\
\colrule
 0.150  & 0.025 \\
\end{tabular}
\end{ruledtabular}
\end{table}

\begin{table}
\caption*{\textbf{Table IV.} Bohr radius values used in Figure 3b.}
\begin{ruledtabular}
\begin{tabular}{lc}
\textrm{Binding energy} & \textrm{Bohr radius} \\
\colrule
0.09eV  & $13.89\pm0.026\text{\AA}$  \\
0.10eV  & $13.64\pm0.018\text{\AA}$  \\
0.11eV  & $13.75\pm0.022\text{\AA}$  \\
0.12eV  & $13.67\pm0.020\text{\AA}$  \\
\end{tabular}
\end{ruledtabular}
\end{table}

\begin{table*}
\caption*{\textbf{Table V.} Fitting parameters used in result from the first Brillouin zone of Figure 3c.}
\begin{ruledtabular}
\begin{tabular}{lccccc}
\textrm{Spectrum} & \textrm{Binding energy} & \textrm{FWHM} & \textrm{Width ratio (Lor./Gau.)} & \textrm{Normalized intensity} \\
\colrule
VB1  & 0.078eV & 0.060eV & 0.14 & 0.18 \\
VB2  & 0.132eV & 0.058eV & 0.46 & 0.20 \\
VB3  & 0.199eV & 0.057eV & 0.56 & 0.16 \\
Exciton  & 0.115eV & 0.058eV & 0.20 & 0.46 \\
\end{tabular}
\end{ruledtabular}
\end{table*}

\begin{table*}
\caption*{\textbf{Table VI.} Fitting parameters used in result from the second Brillouin zone of Figure 3c.}
\begin{ruledtabular}
\begin{tabular}{lccccc}
\textrm{Spectrum} & \textrm{Binding energy} & \textrm{FWHM} & \textrm{Width ratio (Lor./Gau.)} & \textrm{Normalized intensity} \\
\colrule
VB1  & 0.077eV & 0.060eV & 0.14 & 0.35 \\
VB2  & 0.132eV & 0.058eV & 0.46 & 0.38 \\
VB3  & 0.199eV & 0.057eV & 0.56 & 0.27 \\
\end{tabular}
\end{ruledtabular}
\end{table*}

\begin{table*}
\caption*{\textbf{Table VII.} Fitting parameters used in result from the difference of two Brillouin zones in Figure 3c.}
\begin{ruledtabular}
\begin{tabular}{lccccc}
\textrm{Spectrum} & \textrm{Binding energy} & \textrm{FWHM} & \textrm{Width ratio (Lor./Gau.)} & \textrm{Normalized intensity} \\
\colrule
VB1  & 0.078eV & 0.060eV & 0.14 & 0.05 \\
VB2  & 0.132eV & 0.058eV & 0.46 & 0.20 \\
VB3  & 0.199eV & 0.057eV & 0.56 & 0.04 \\
Exciton  & 0.115eV & 0.058eV & 0.20 & 0.71 \\
\end{tabular}
\end{ruledtabular}
\end{table*}

\newpage

 \begin{figure*}
\centering
\includegraphics[width=1.0\textwidth]{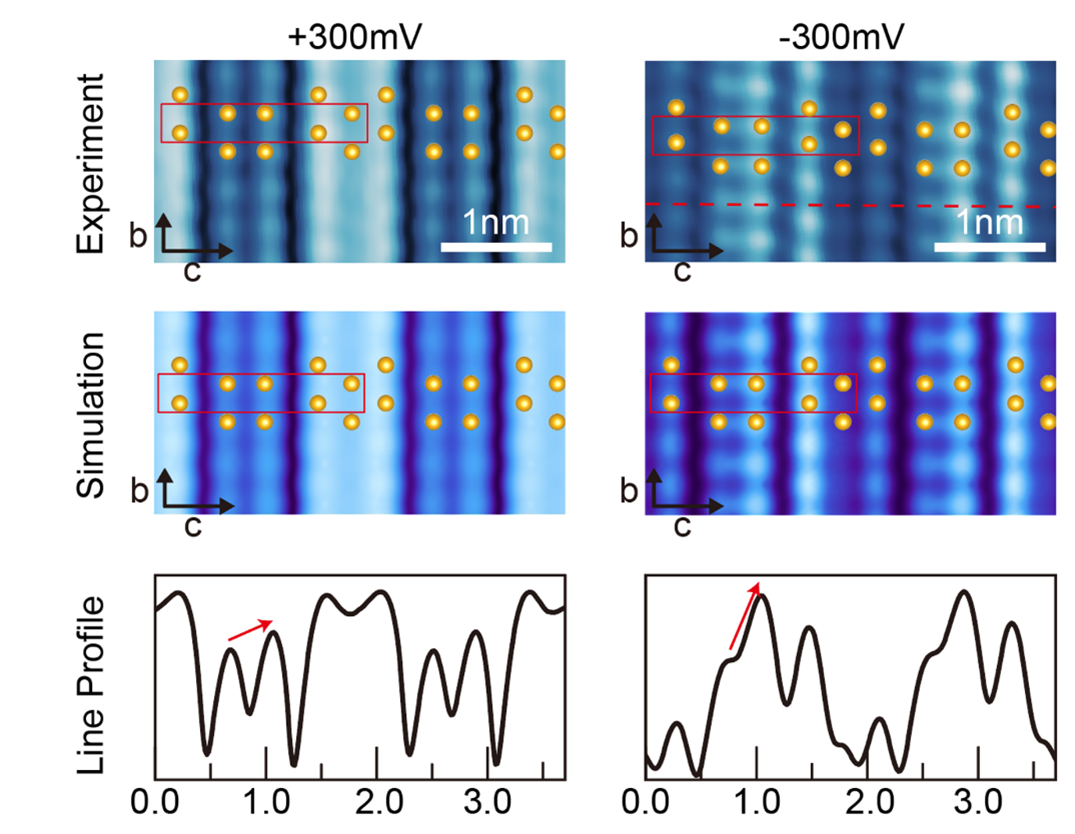}
\caption*{\justifying\textbf{Fig. S1.} Atomic resolution STM topographies (obtained at 78 K) and corresponding DFT simulations of the gapped low temperature phase. The experimental STM images and simulated results at both occupied (-300mV) and unoccupied (+300mV) states reveal a clear breaking of mirror symmetry, as evidenced by both the topographic contrast and the line profiles. The details of the STM results will be published elsewhere. 
}
\end{figure*}

 \begin{figure*}
\centering
\includegraphics[width=0.7\textwidth]{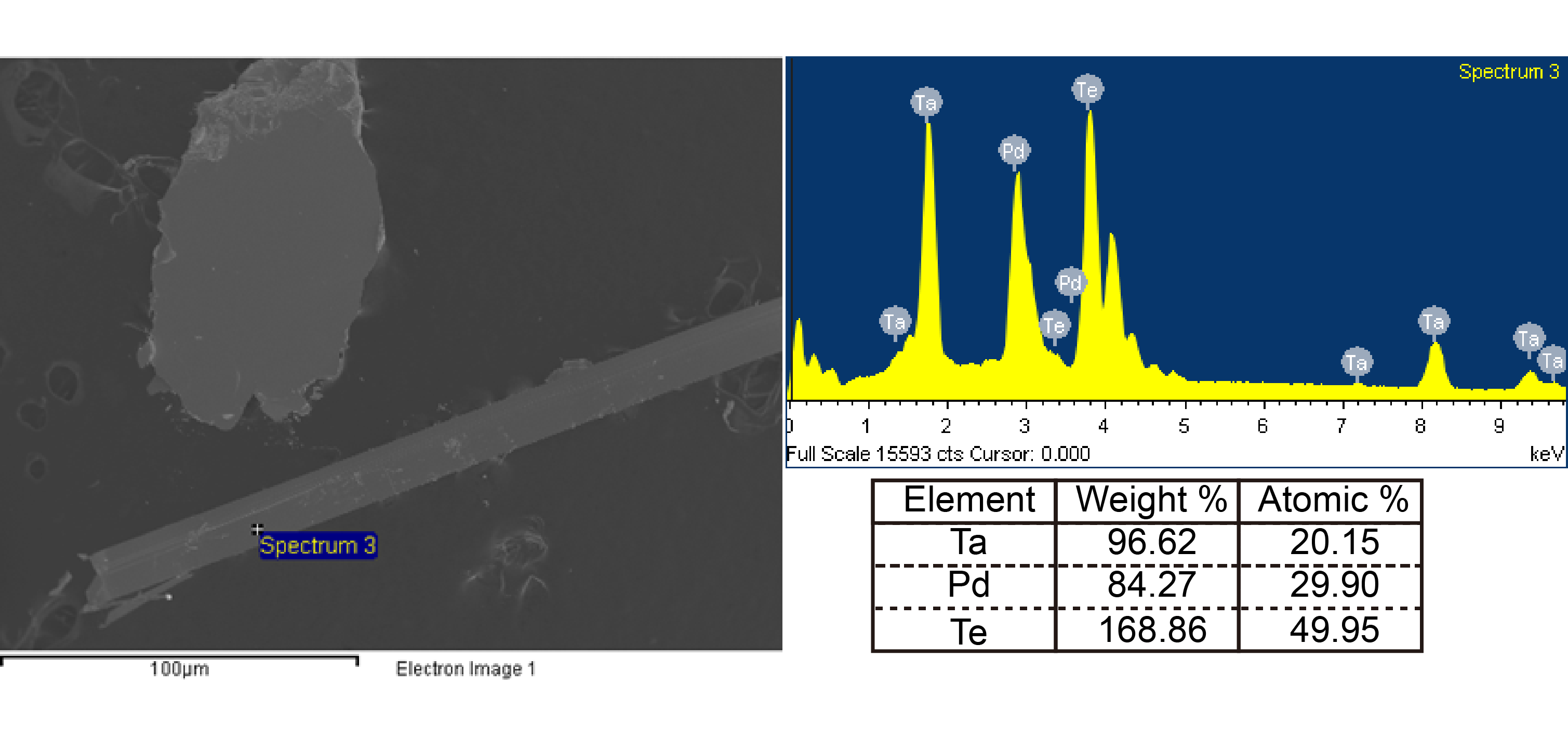}
\caption*{\justifying\textbf{Fig. S2.} Energy dispersive X-ray spectroscopy (EDS) analysis of the synthesized Ta$_2$Pd$_3$Te$_5$ crystal confirms an atomic ratio of Ta, Pd, and Te as 2:3:5, indicating successful synthesis of the crystal.
}
\end{figure*}

 \begin{figure*}
\centering
\includegraphics[width=0.5\textwidth]{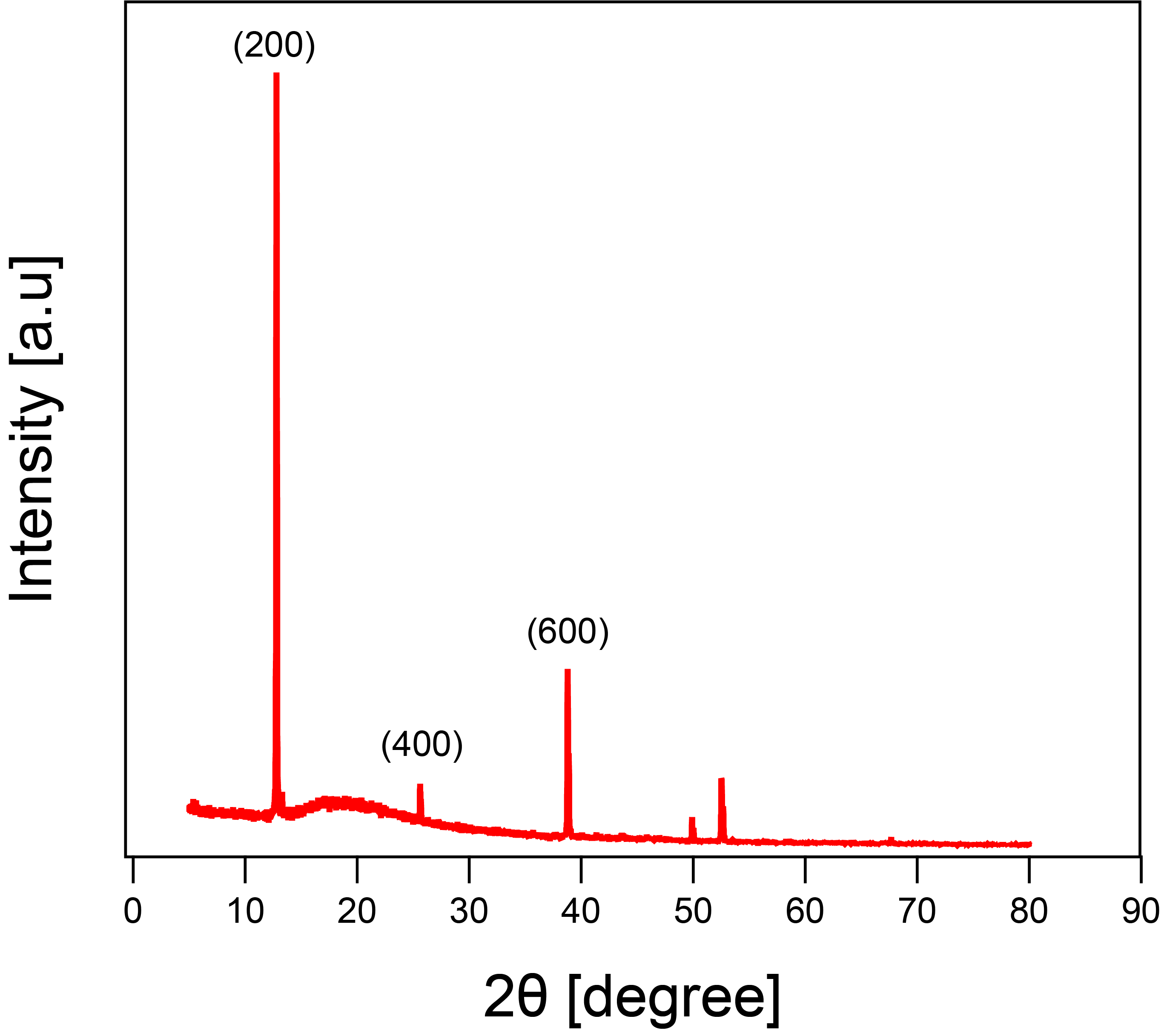}
\caption*{\justifying\textbf{Fig. S3.} The X-ray diffraction (XRD) pattern matches well with previously reported XRD data, confirming the consistency and accuracy of the synthesized crystal structure \cite{guo2021quantum,zhang2024spontaneous}.
}
\end{figure*}

 \begin{figure*}
\centering
\includegraphics[width=1.0\textwidth]{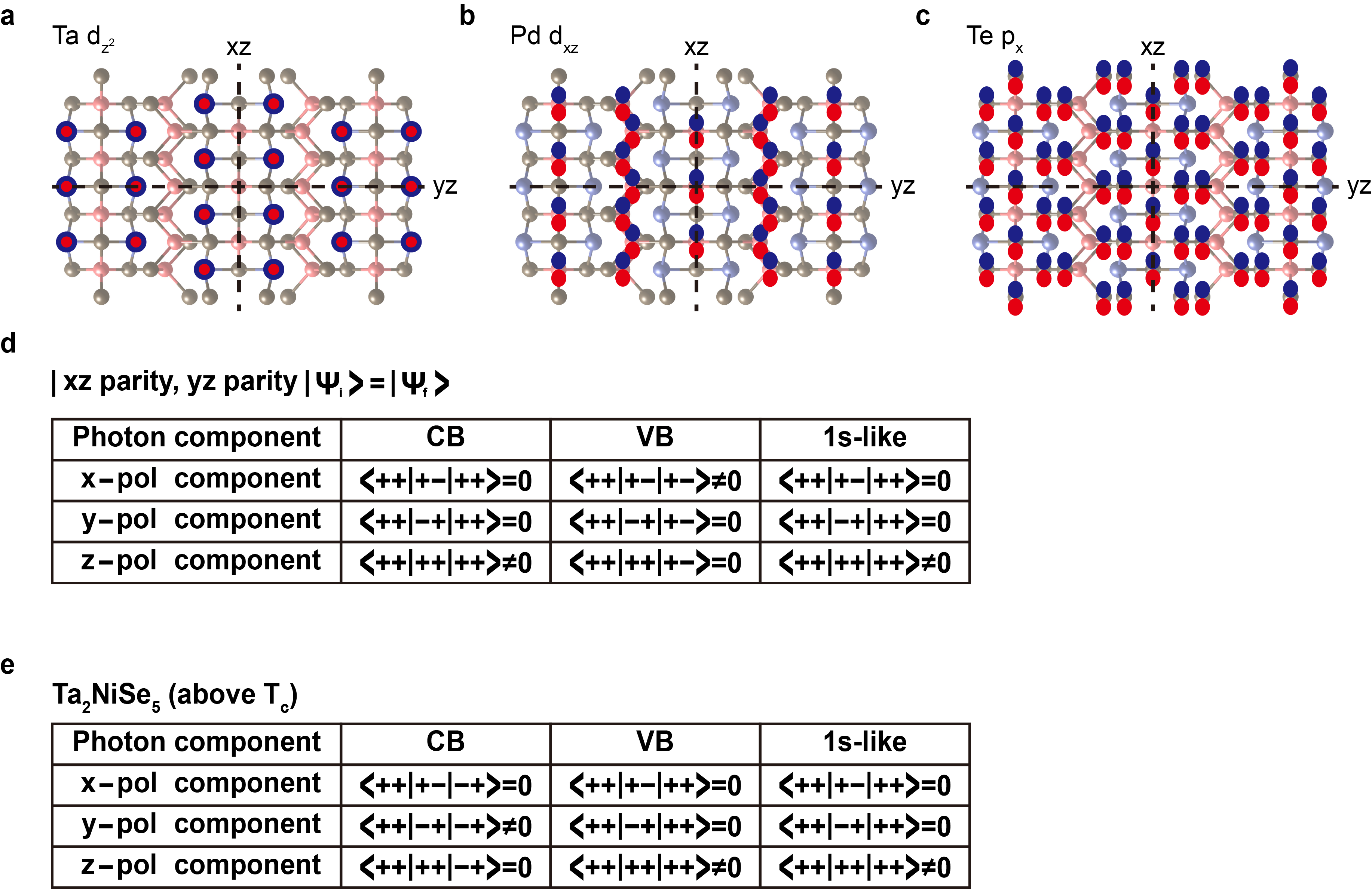}
\caption*{\justifying\textbf{Fig. S4.} Atomic and orbital structures of Ta$_2$Pd$_3$Te$_5$ highlighting mirror symmetries and orbital parities of three major bands across the band gap. (a) Ta \textit{d}$_{z^2}$ orbitals, (b) Pd \textit{d}$_{xz}$ orbitals, and (c) Te \textit{p}$_x$ orbitals, each shown on its atomic site with the xz and yz mirror planes indicated. (d) Table summarizing the xz and yz parities of each orbital type under different photon polarization components (x, y, and z) in the dipole transition, illustrating the polarization dependence of given orbital characters. The conduction band (CB) (Ta \textit{d}$_{z^2}$) is only visible for z-polarized photons and the valence band (VB) (Pd \textit{d}$_{xz}$ and Te \textit{p}$_x$) is visible for x-polarized photons. (e) Similar polarization dependence of the CB, the VB and the 1$s$-like orbital in Ta$_2$NiSe$_5$.
}
\end{figure*}

 \begin{figure*}
\centering
\includegraphics[width=1.0\textwidth]{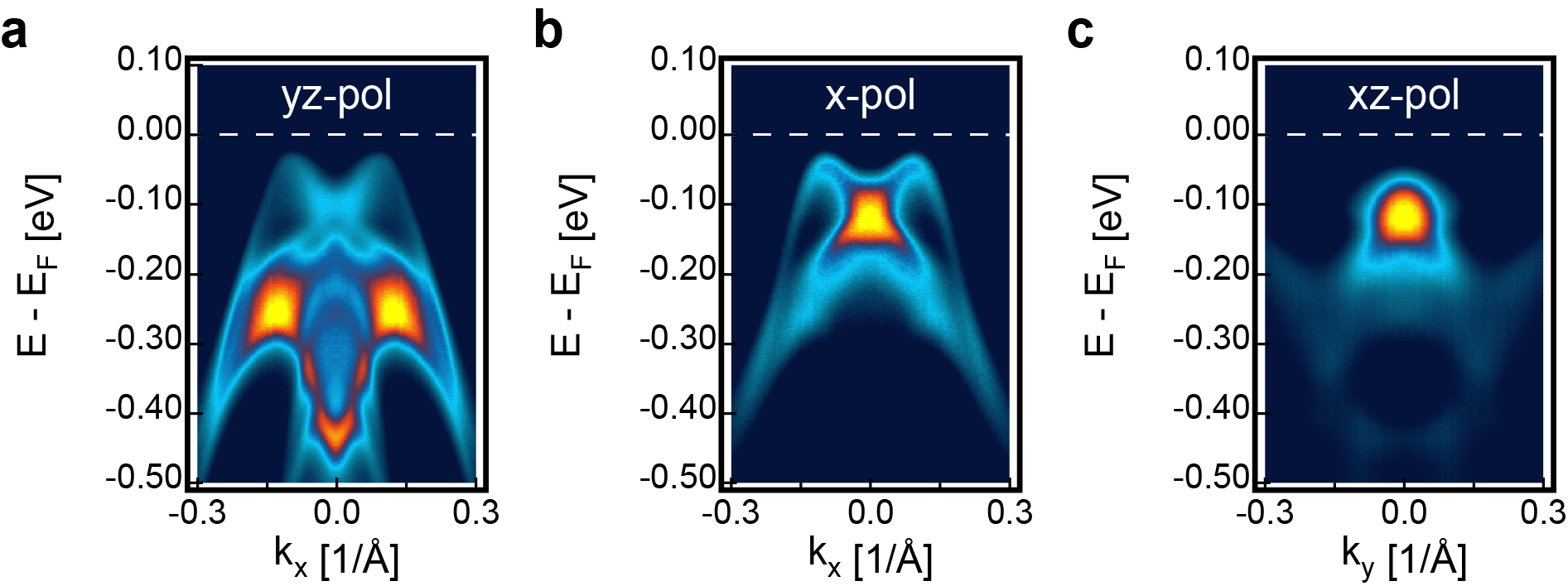}
\caption*{\justifying\textbf{Fig. S5.} ARPES results for different light polarizations. The ARPES spectra are obtained using (a) yz-, (b) x-, and (c) xz-polarized photons, respectively. Ta \textit{d}$_{z^2}$ bands are visible with z-polarized photons (a, c), while Pd \textit{d}$_{xz}$ Te \textit{p}$_x$ bands are visible with x-polarized photons (b, c).}
\end{figure*}

 \begin{figure*}
\centering
\includegraphics[width=1.0\textwidth]{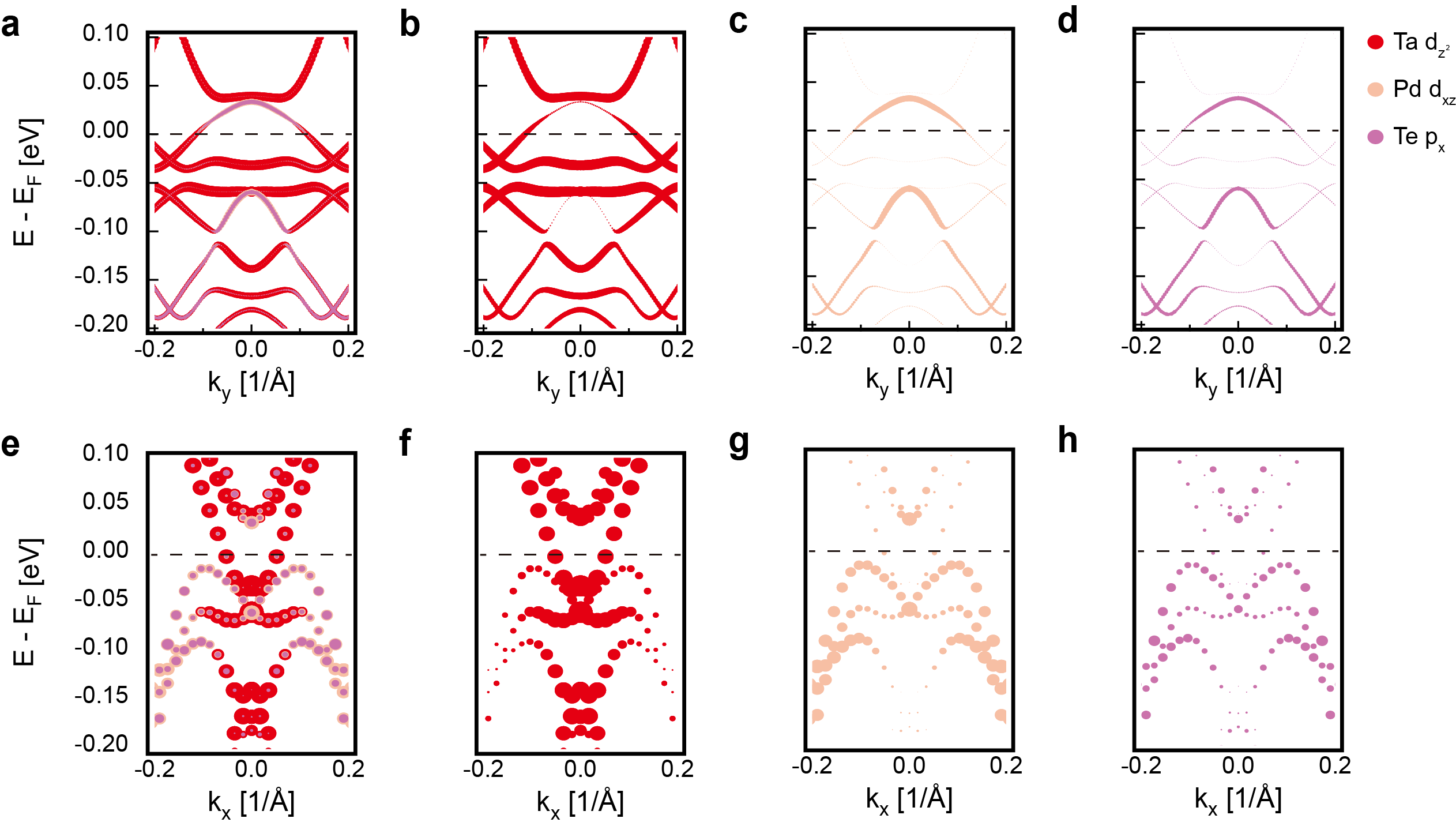}
\caption*{\justifying\textbf{Fig. S6.} Orbital-projected DFT calculations along $\Gamma-Y$ (a-d) and $\Gamma-X$ (e-h). The overall band structures along two different directions, $\Gamma-Y$ (a) and $\Gamma-X$ (b), consist of three different orbital characters, Ta $d_{z^2}$, Pd $d_{xz}$ and Te $p_x$. (b,f) are Ta $d_{z^2}$ orbital projected DFT calculations and (c,g) and (d,h) are Te $p_x$ and are Pd $d_{xz}$ orbitals, respectively.}
\end{figure*}

 \begin{figure*}
\centering
\includegraphics[width=0.6\textwidth]{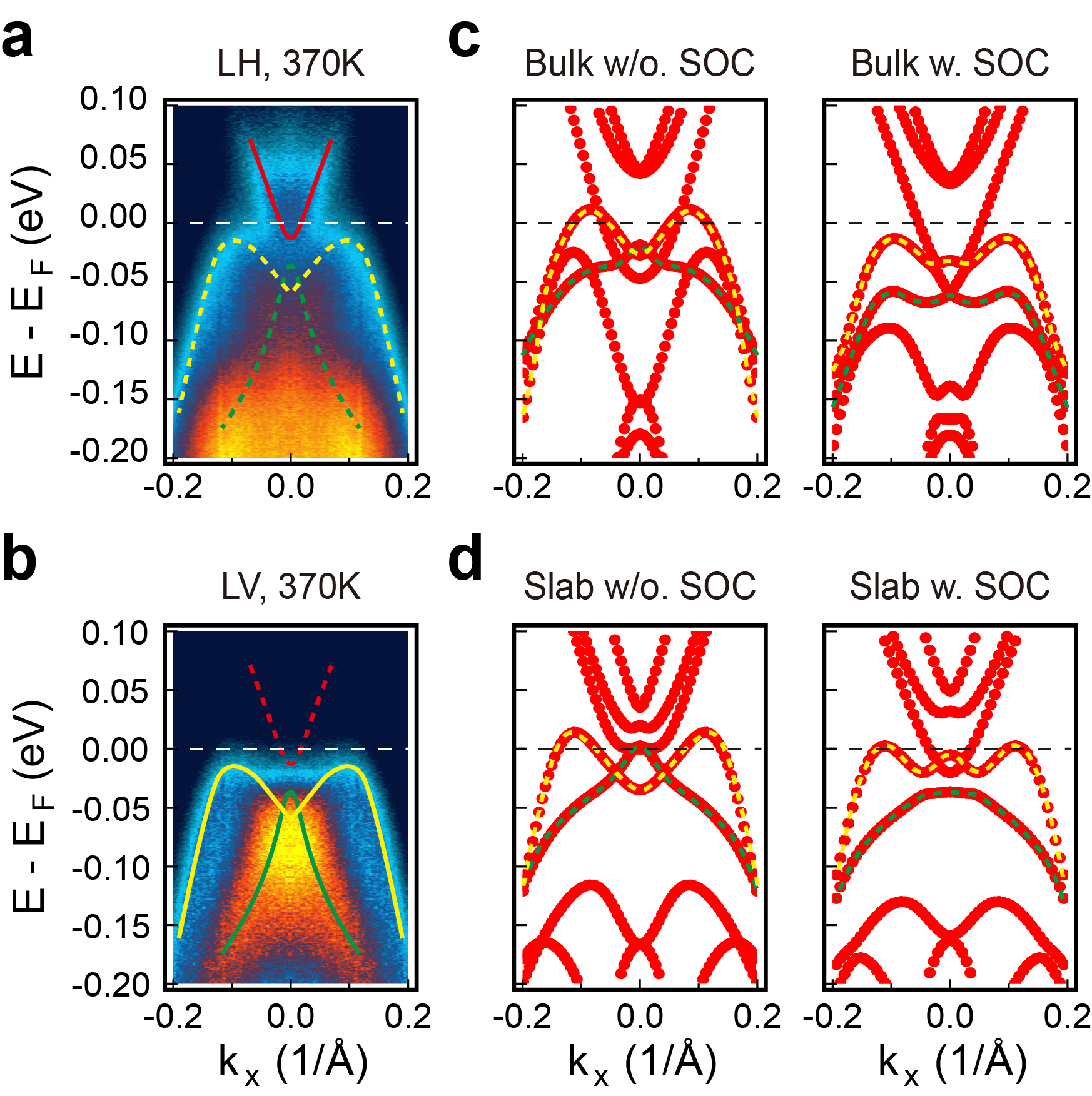}
\caption*{\justifying\textbf{Fig. S7.} ARPES spectra and calculated band structures with and without SOC.
(a-b) ARPES results obtained using LH (a) and LV (b) polarized photons at 370K.
(c) Calculated bulk band structures of Ta$_2$Pd$_3$Te$_5$ without and with SOC.
(d) Calculated band structures of a two-layer Ta$_2$Pd$_3$Te$_5$ slab without and with SOC.}
\end{figure*}

 \begin{figure*}
\centering
\includegraphics[width=1.0\textwidth]{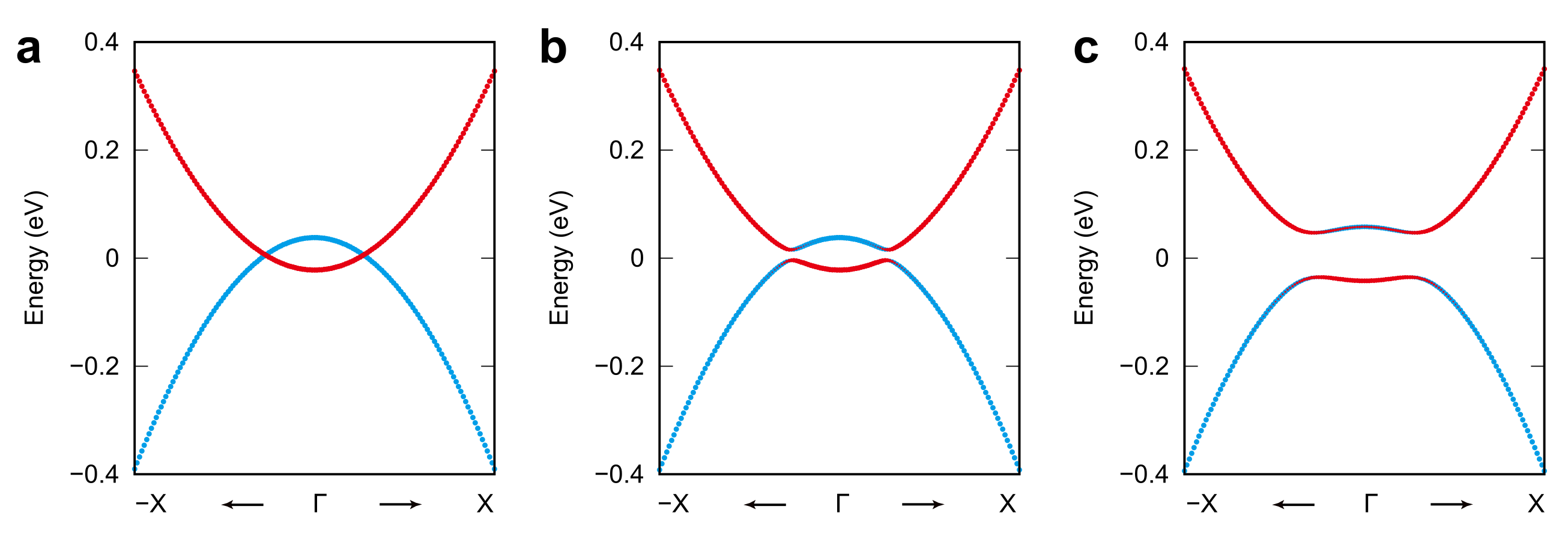}
\caption*{\justifying\textbf{Fig. S8.} Two-band exciton model. (a) Model band structure without the SOC and the exciton interaction as simplified from the DFT result. (b) Band structure after including the SOC interaction. (c) Model band structure after including the SOC and excitonic interaction with an order parameter of $\Delta$ = 40meV. We adopted the parameters as follows : M$_0$=-0.03 eV, M$_y$=-4 eV, M$_x$=0.05 eV, A$_x$=0.007 eV, A$_y$=0.09 eV, C$_x$=0.004 eV, and C$_y$=-0.3 eV.}
\end{figure*}

 \begin{figure*}
\centering
\includegraphics[width=1.0\textwidth]{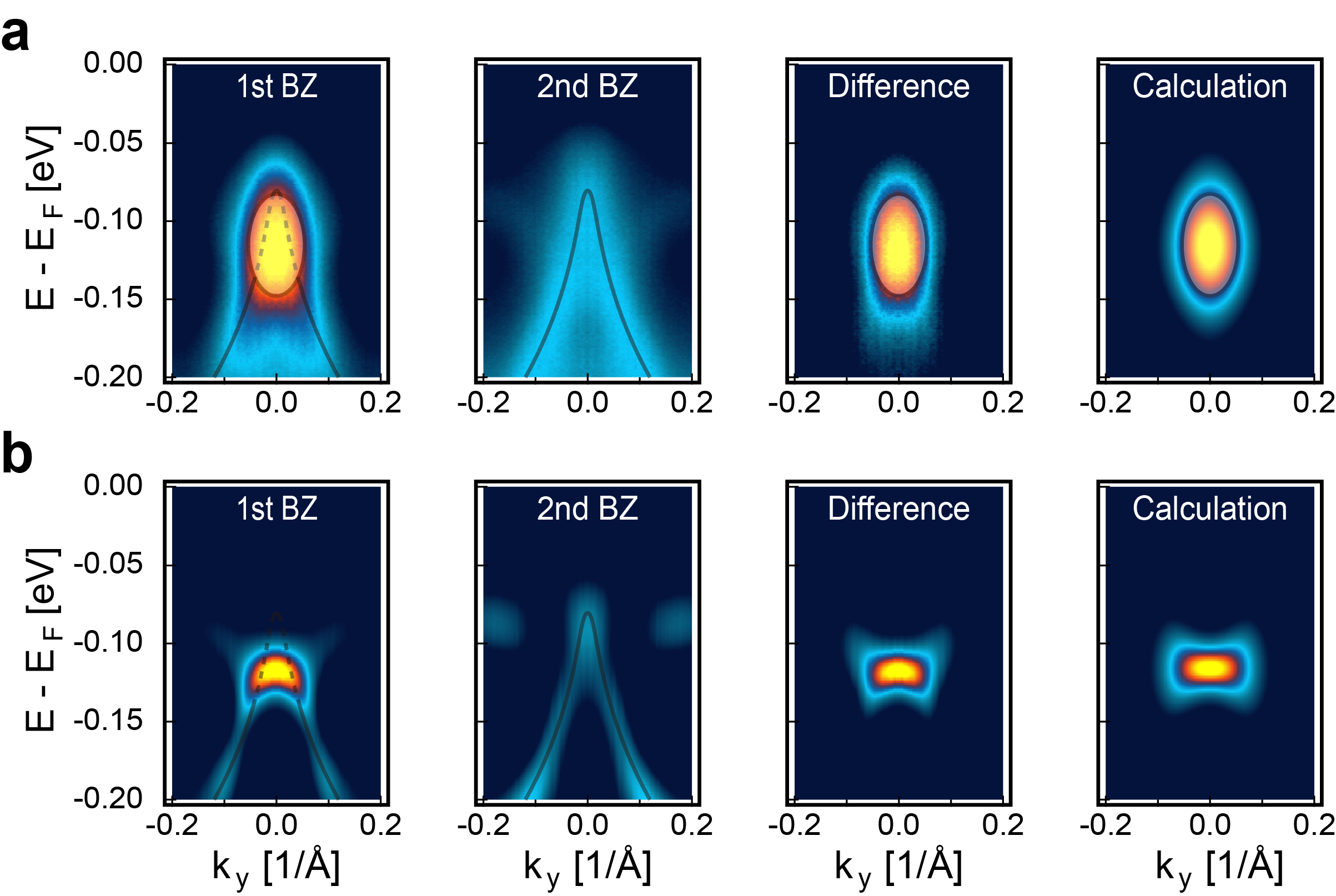}
\caption*{\justifying\textbf{Fig. S9.} a) The ARPES spectra along the $\Gamma - Y$ direction in the first and second Brillouin zones (BZs). The difference in their spectral intensities resembles closely the calculated exciton photoemission signal. b) The curvature plots to extract the dispersions more clearly, which are derived from the data in (a). The photoemission from the exciton shows a flat band, indicating no band dispersion regardless of the energy. The curvature plot is a more sophisticated method to extract band dispersions from the ARPES intensity map than a simple second-derivative method \cite{zhang2011precise}.}
\end{figure*}

\begin{figure*}
\centering
\includegraphics[width=1.0\textwidth]{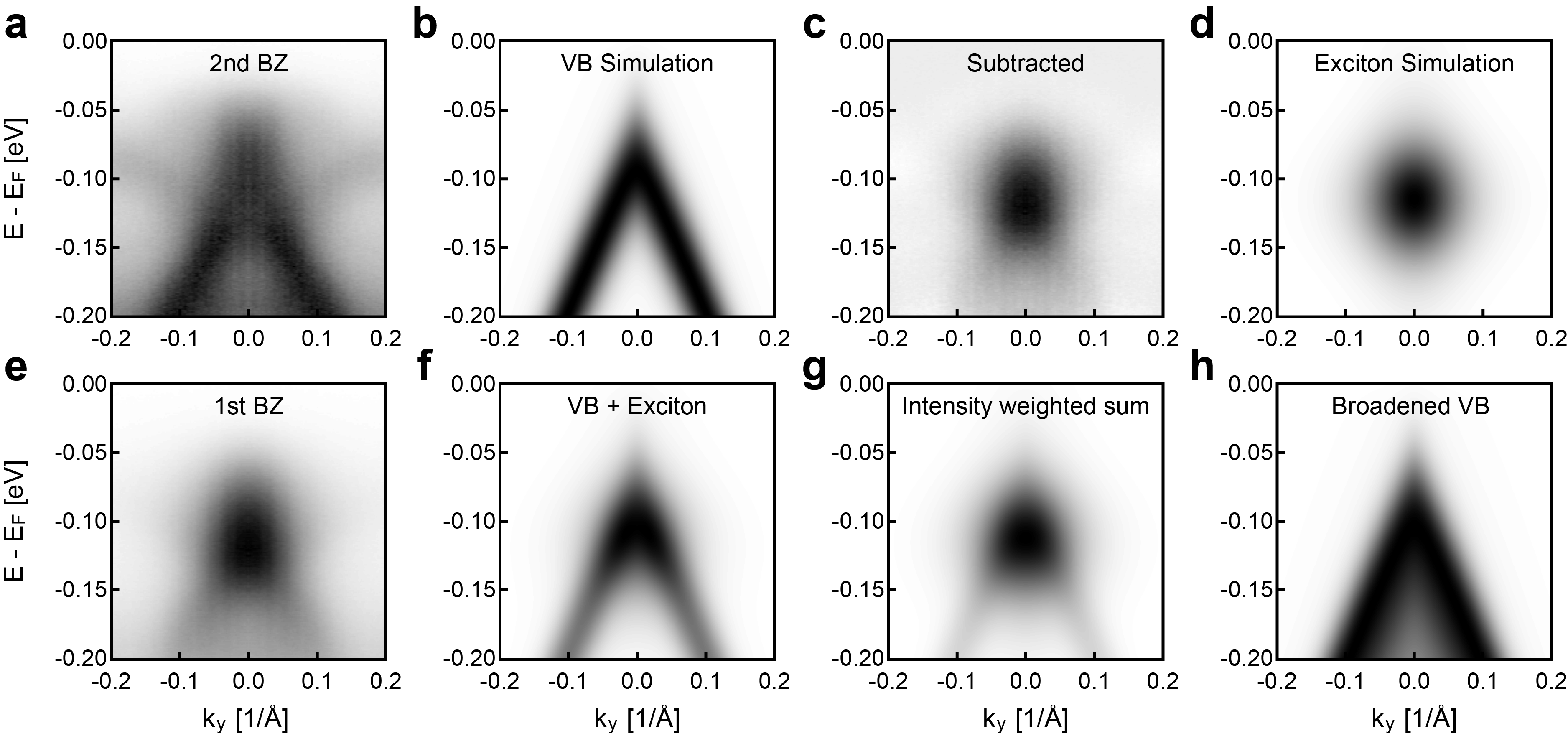}
\caption*{\justifying\textbf{Fig. S10.} a) ARPES spectrum obtained from xz-polarized photons along the $\Gamma-Y$ direction in the second Brillouin zone.
b) Simulated valence band in the excitonic insulating phase, calculated using a simple two-band Bethe-Salpeter equation for two quadratic bands.
c) Difference spectrum between the first and second Brillouin zone ARPES measurements.
d) Simulated exciton photoemission feature.
e) ARPES spectrum obtained from xz-polarized photons along the $\Gamma-Y$ direction in the first Brillouin zone.
f) Summation of the simulated valence band and exciton spectra.
g) Intensity-weighted summation of the valence band and exciton simulations.
h) Simulated valence band broadening due to phonon-induced replica bands (A=1 and $\hbar\omega=30meV$).}
\end{figure*}

\begin{figure*}
\includegraphics[width=1.0\textwidth]{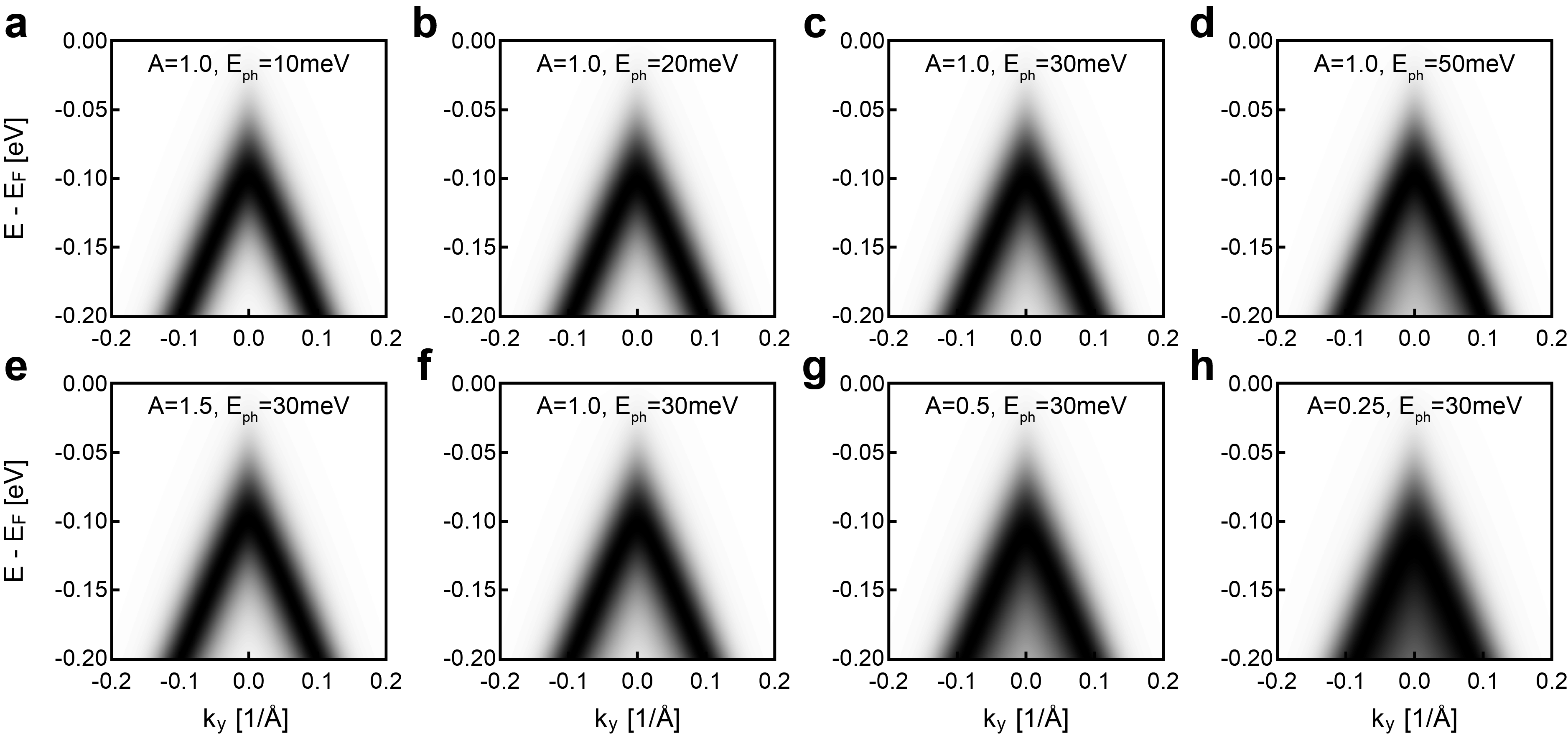}
\caption*{\justifying\textbf{Fig. S11.} Simulated phonon-induced valence band broadening features with different parameter values.}
\end{figure*}

\begin{figure*}
\centering
\includegraphics[width=1.0\textwidth]{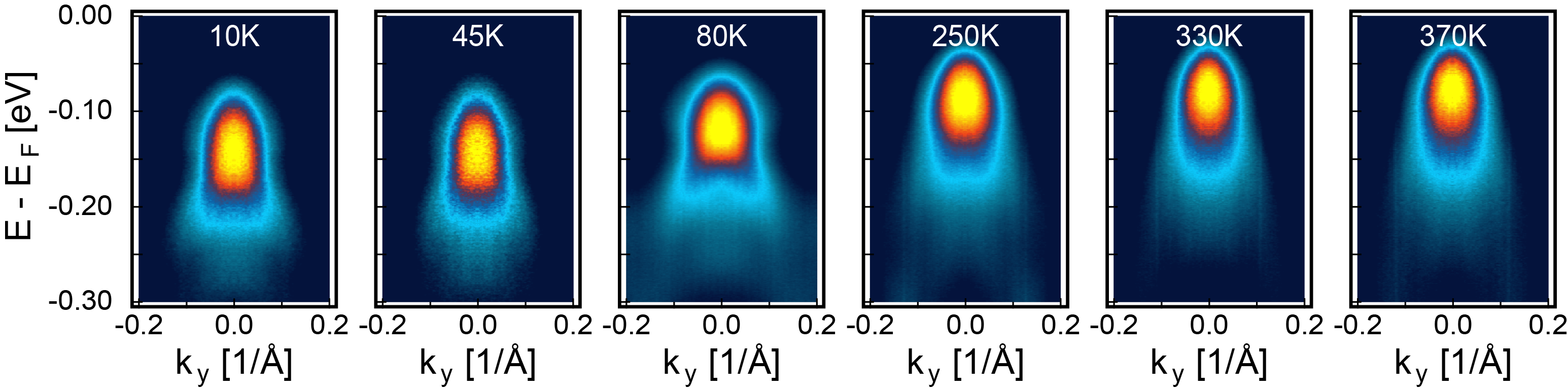}
\caption*{\justifying\textbf{Fig. S12.} Temperature-depending ARPES results along the $\Gamma-Y$ direction. The photoemission feature from the exciton survives down to 10 K.}
\end{figure*}

\begin{figure*}
\centering
\includegraphics[width=1.0\textwidth]{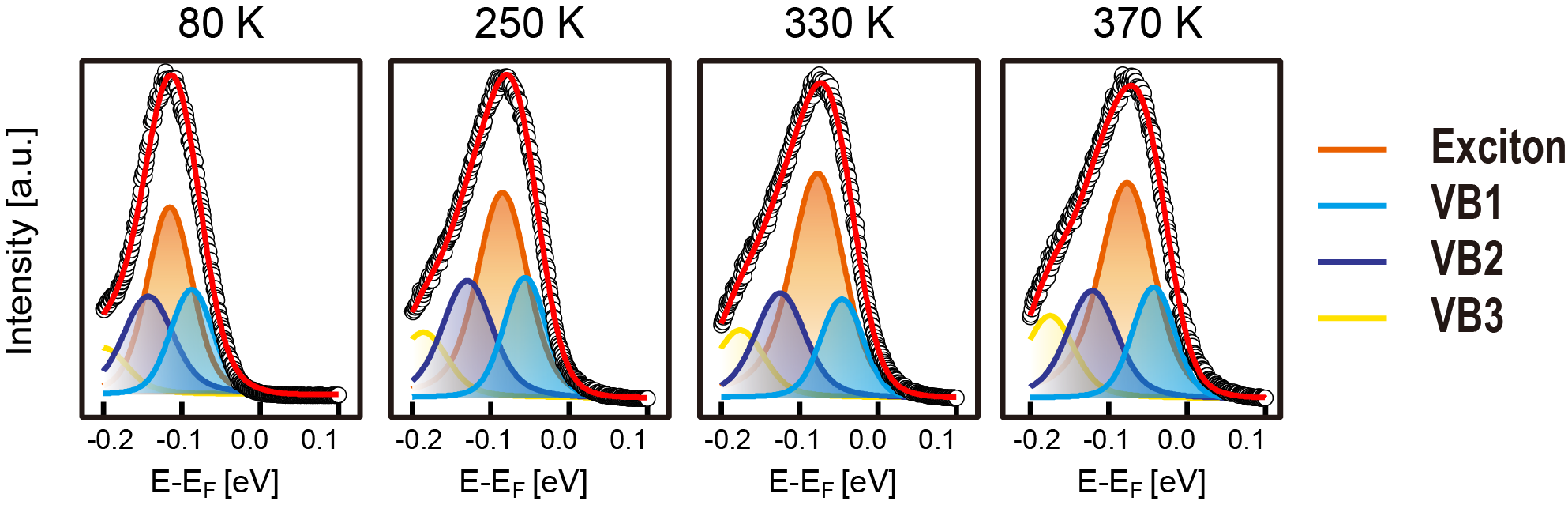}
\caption*{\justifying\textbf{Fig. S13.} Temperature-dependent ARPES EDCs at k=0 with spectral decomposition through curve fittings with Voigt functions. The orange exciton peaks broaden as the temperature increases. We set the broadening factor to values proportional to thermal energy (k$_B$T). Alongside the exciton signal, the valence band shifts to lower binding energy, indicating the decrease of the band gap and the exciton binding energy.}
\end{figure*}

 \begin{figure*}
\centering
\includegraphics[width=1.0\textwidth]{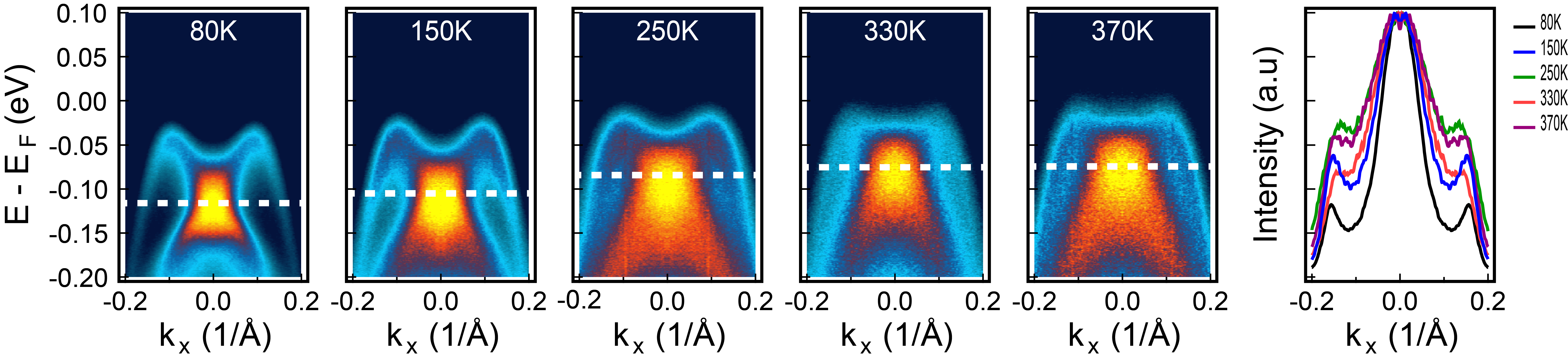}
\caption*{\justifying\textbf{Fig. S14.} The temperature-dependent spectral weight transfer between the ARPES spectra and exciton photoemission. The ARPES data are taken from Fig. 2d and the line plots on the right are the MDC’s taken crossing the centers of the exciton photoemission signal (along the dashed lines in the ARPES maps). Note that both the valence band and the exciton feature move toward a higher binding energy as the temperature decreases. The MDC’s are normalized by the maximum intensity.}
\end{figure*}

\begin{figure*}
\centering
\includegraphics[width=1.0\textwidth]{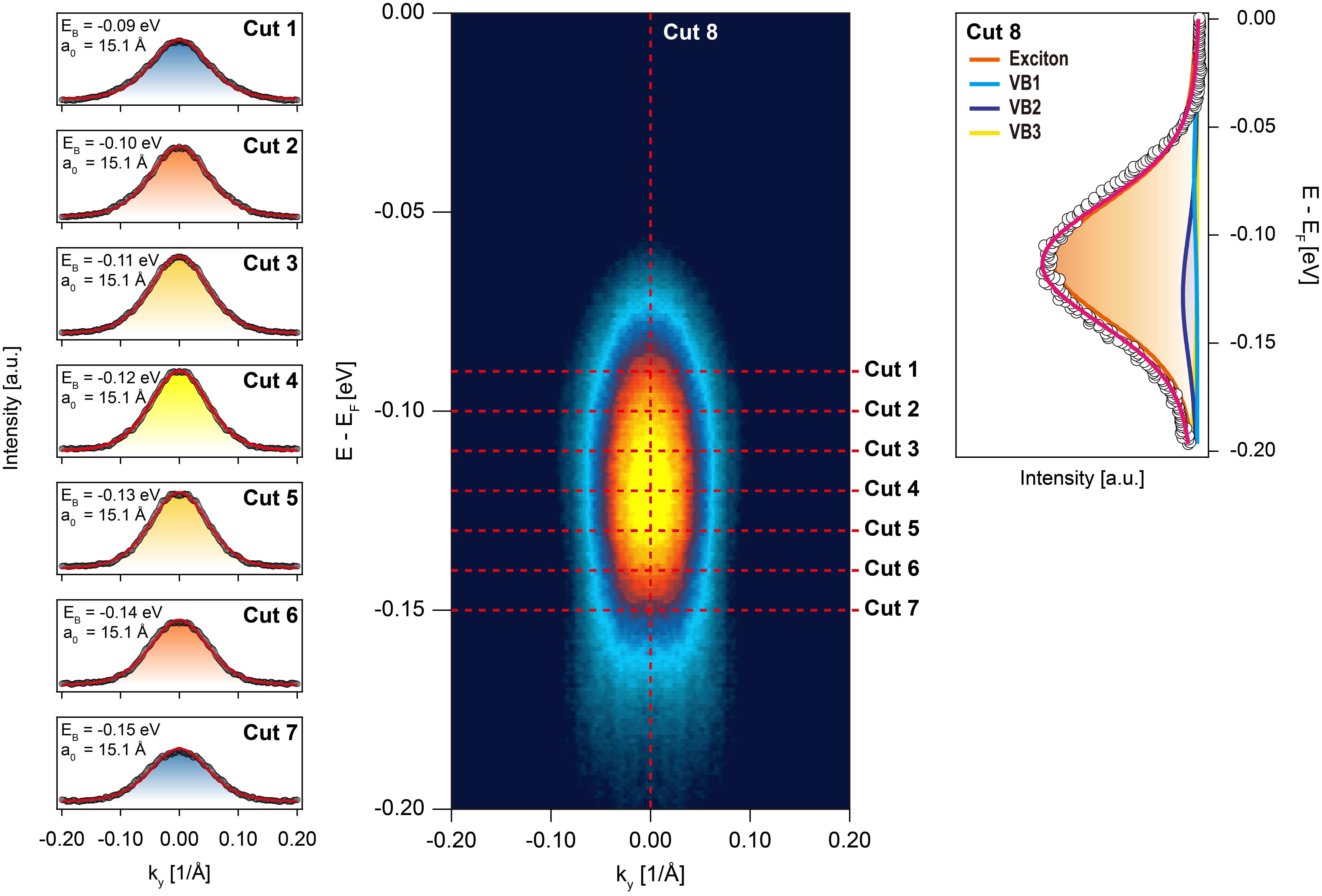}
\caption*{\justifying\textbf{Fig. S15.} a) ARPES energy and 2D momentum-space mappings obtained using yz-polarized photons. b) ARPES energy and 2D momentum-space mapping obtained using xz-polarized photons. The photoemission signal from the exciton at the zone center is suppressed for the yz-polarized photons.}
\end{figure*}

\begin{figure*}
\centering
\includegraphics[width=1.0\textwidth]{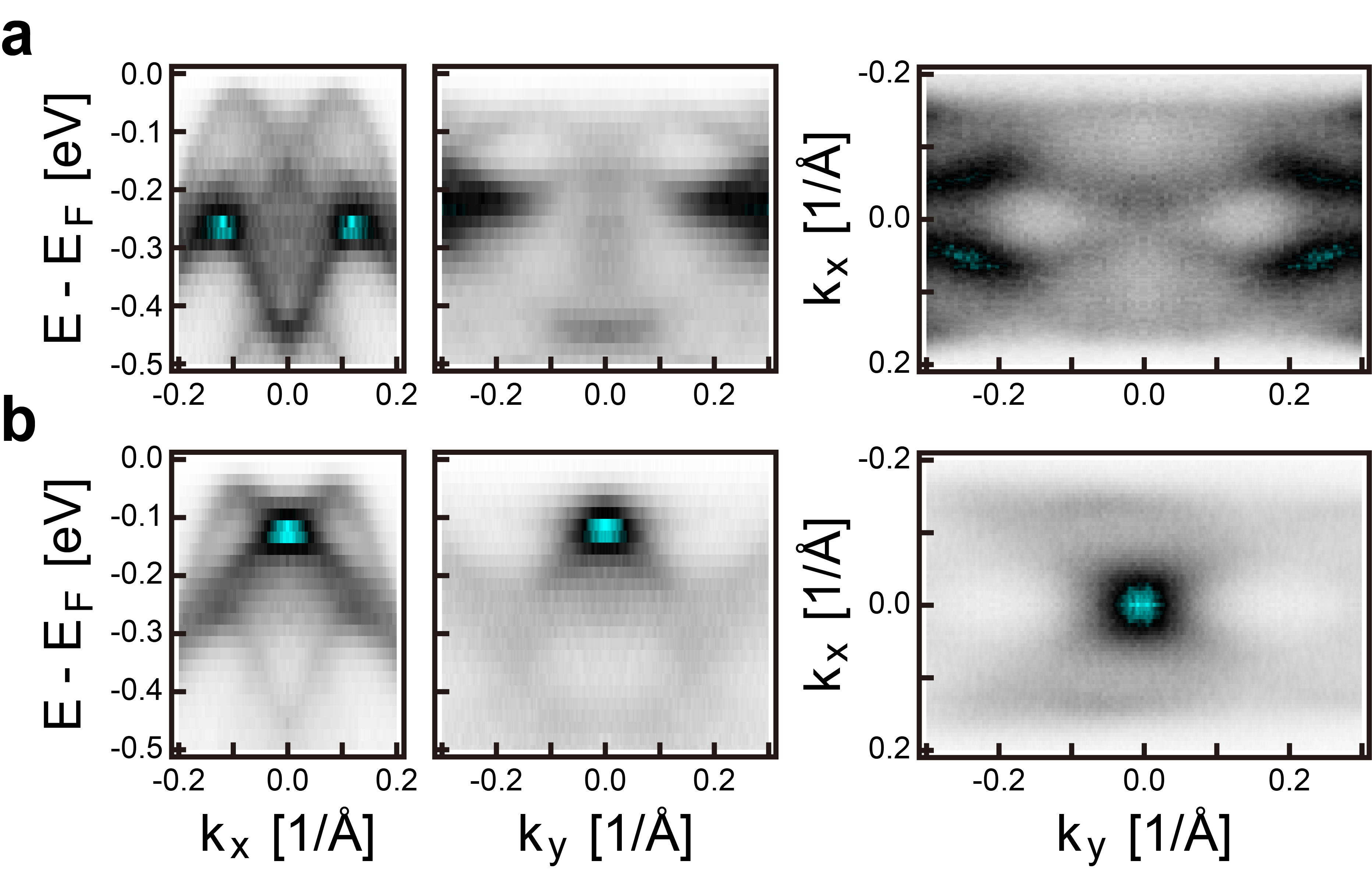}
\caption*{\justifying\textbf{Fig. S16.} Energy broadening of the exciton photoemission spectral feature. The spectral function of excitons is derived from the difference between the ARPES measurements along $\Gamma-Y$ with the exciton contribution (measured in the 1st BZ) and without it (measured in the 2nd BZ) as shown in Fig. 1e. The momentum cuts at different energies of the spectral map follow the exciton photoemission spectral function very well, exhibiting a Bohr radius of 15.1 Å and a binding energy range from 0.09 to 0.15 eV.}
\end{figure*}
\end{document}